\documentclass[numberedappendix]{emulateapj}
\usepackage{apjfonts}
\usepackage{amsmath}
\usepackage{amssymb}
\usepackage{epsfig}

\newcommand*{\Msun}{\ensuremath{\mathrm{M_\odot}}}%
\newcommand{\oii}{[\textrm{O}\,\textsc{ii}]}
\newcommand{\hii}{\textrm{H}\,\textsc{ii}}
\newcommand{\ha}{\textrm{H}\ensuremath{\alpha}}
\newcommand{\hb}{\textrm{H}\ensuremath{\beta}}
\newcommand{\msun}{\mbox{$M_{\rm \odot}$}}
\newcommand{\angstrom}{\text{\normalfont\AA}}
\newcommand{\s}{{\it Spitzer}}
\newcommand{\h}{{\it Herschel}}
\newcommand{\lir}{$L_{\rm IR}$}

\lefthead{Pannella et al.}

\slugcomment{to appear in ApJ}
\shorttitle{Star formation, dust attenuation and the FIR--radio correlation up to $z\simeq$4}
\shortauthors{Pannella et al.}

\begin{document}
 \title{GOODS-{\it HERSCHEL}~: star formation, dust attenuation and the FIR-radio correlation\\ on the Main Sequence of star-forming galaxies up to \lowercase{$z\simeq4$}\altaffilmark{*}}
 \author{M.~Pannella$\!$\altaffilmark{1,2,$\diamond$},
 D.~Elbaz$\!$\altaffilmark{1},        
 E.~Daddi$\!$\altaffilmark{1},
 M.~Dickinson$\!$\altaffilmark{3},
 H.S.~Hwang$\!$\altaffilmark{1,4},
 C.~Schreiber$\!$\altaffilmark{1},
 V.~Strazzullo$\!$\altaffilmark{1,5},\\
H.~Aussel$\!$\altaffilmark{1},
M.~Bethermin$\!$\altaffilmark{1,6},
V.~Buat$\!$\altaffilmark{7},
V.~Charmandaris$\!$\altaffilmark{8,9},
A.~Cibinel$\!$\altaffilmark{1,10},
S.~Juneau$\!$\altaffilmark{1},
R.J.~Ivison$\!$\altaffilmark{11,6},\\
D.~Le Borgne$\!$\altaffilmark{12,2},
E.~Le~Floc'h$\!$\altaffilmark{1},
R.~Leiton$\!$\altaffilmark{1,13},
L.~Lin$\!$\altaffilmark{14},
G.~Magdis$\!$\altaffilmark{15,8},
G.E.~Morrison\altaffilmark{16,17},\\
J.~Mullaney$\!$\altaffilmark{1,18},
M.~Onodera$\!$\altaffilmark{19},
A.~Renzini$\!$\altaffilmark{20},
S.~Salim$\!$\altaffilmark{21},\\
M.~T.~Sargent$\!$\altaffilmark{1,10},
D.~Scott$\!$\altaffilmark{22},
X.~Shu$\!$\altaffilmark{1,23},
T,~Wang$\!$\altaffilmark{1,24}
}

\altaffiltext{*}{Based on observations collected at the \h, \s, Keck, NRAO-VLA, Subaru, KPNO and CFHT observatories. \h~is an European Space Agency Cornerstone Mission with science instruments provided by European-led Principal Investigator consortia and with significant participation by NASA. The National Radio Astronomy Observatory is a facility of the National Science Foundation operated under cooperative agreement by Associated Universities, Inc.}

\altaffiltext{1}{Laboratoire AIM-Paris-Saclay, CEA/DSM/Irfu - CNRS - Universit\'e Paris Diderot, CEA-Saclay, F-91191 Gif-sur-Yvette, France}
\altaffiltext{2}{Institut d'Astrophysique de Paris, UMR 7095, CNRS, 98bis boulevard Arago, F-75005 Paris, France}
\altaffiltext{$\diamond$}{Present address: Max-Planck-Institut f{\"u}r Extraterrestrische Physik, Giessenbachstrasse 1, D-85748 Garching, Germany}
\altaffiltext{3}{National Optical Astronomy Observatory, 950 North Cherry Avenue, Tucson, AZ 85719, USA}
\altaffiltext{4}{School of Physics, Korea Institute for Advanced Study, 85 Hoegiro, Dongdaemun-Gu, 130-722 Seoul, Korea}
\altaffiltext{5}{Ludwig-Maximilians-Universit\"{a}t, Department of Physics, Scheinerstr.\ 1, 81679 M\"{u}nchen, Germany}
\altaffiltext{6}{European Southern Observatory, Karl-Schwarzschild-Strasse 2, D-85748 Garching, Germany}
\altaffiltext{7}{Aix-Marseille Universit\'e, CNRS, LAM (Laboratoire d'Astrophysique de Marseille) UMR7326, 13388, Marseille, France}
\altaffiltext{8}{Institute for Astronomy, Astrophysics, Space Applications \& Remote Sensing, National Observatory of Athens, 15236, Penteli, Greece}
\altaffiltext{9}{Department of Physics, University of Crete, 71003, Heraklion, Greece}
\altaffiltext{10}{Astronomy Centre, Dept. of Physics \& Astronomy, University of Sussex, Brighton BN1 9QH, UK}
\altaffiltext{11}{Institute for Astronomy, University of Edinburgh, Blackford Hill, Edinburgh EH9 3HJ, UK}
\altaffiltext{12}{Sorbonne Universit\'es, UPMC Univ Paris 06, UMR 7095, Institut d'Astrophysique de Paris, F-75005, Paris, France }
\altaffiltext{13}{Astronomy Department, Universidad de Concepci\'on, Casilla 160-C, Concepci\'on, Chile}
\altaffiltext{14}{Institute of Astronomy \& Astrophysics, Academia Sinica, Taipei 106, Taiwan (R.O.C.)}
\altaffiltext{15}{Department of Physics, University of Oxford, Keble Road, Oxford OX1 3RH}
\altaffiltext{16}{Institute for Astronomy, University of Hawaii, Honolulu,  Hawaii, 96822, USA}
\altaffiltext{17}{Canada-France-Hawaii Telescope, Kamuela, Hawaii, 96743, USA}
\altaffiltext{18}{Department of Physics \& Astronomy, University of Sheffield, Sheffield, S3 7RH, UK}
\altaffiltext{19}{Institute for Astronomy, ETH Zürich, Wolfgang-Pauli-strasse 27, 8093 Zürich, Switzerland}
\altaffiltext{20}{INAF-Osservatorio Astronomico di Padova, Vicolo dell'Osservatorio 5, I-35122 Padova, Italy}
\altaffiltext{21}{Indiana University, Department of Astronomy, Swain Hall West 319, Bloomington, IN 47405-7105, USA}
\altaffiltext{22}{Department of Physics and Astronomy, University of British Columbia, Vancouver, BC V6T 1Z1, Canada}
\altaffiltext{23}{Department of Physics, Anhui Normal University, Wuhu, Anhui, 241000, China}
\altaffiltext{24}{School of Astronomy and Space Sciences, Nanjing University, Nanjing, 210093, China}

\begin{abstract}
We use deep panchromatic datasets in the GOODS-N field,  from {\it GALEX}  to the deepest Herschel far-infrared and VLA radio continuum imaging, to explore, using mass-complete samples, the evolution of the star formation activity and dust attenuation of star-forming galaxies to $z\simeq$4.  Our main results can be summarized as follows: i)  the slope of the SFR--$M_*$ correlation is consistent with being constant $\simeq$0.8 at least to $z\simeq$1.5, while its normalization keeps increasing with redshift; ii) for the first time here we are able to explore the FIR--Radio correlation for a mass-selected sample of star-forming galaxies: the correlation does not evolve up to $z\simeq$4; iii) we confirm that galaxy stellar mass is a robust proxy for UV dust attenuation in star-forming galaxies, with more massive galaxies being more dust attenuated, strikingly we find that this attenuation relation evolves very weakly with redshift, the amount of dust attenuation increasing by less than 0.3 magnitudes over the redshift range [0.5--4] for a fixed stellar mass, as opposed to a tenfold increase of star formation rate; iv) the correlation between dust attenuation and the UV spectral slope evolves in redshift, with the median UV spectral slope of star-forming galaxies becoming bluer with redshift. By $z\simeq$3, typical UV slopes are inconsistent, given the measured dust attenuation, with the predictions of commonly used empirical laws. Finally, building on existing results, we show that gas reddening is marginally larger (by a factor of around 1.3) than stellar reddening at all redshifts probed, and also that the amount of dust attenuation at a fixed ISM metallicity increases with redshift. We speculate that our results support evolving ISM conditions of typical star-forming galaxies such that at $z\ge$1.5 Main Sequence galaxies have ISM conditions getting closer to those of local starbursts.

\end{abstract}


%__________________________________________________________________
\section{Introduction}
\setcounter{footnote}{0} 

Our understanding of galaxy formation and evolution has made substantial progress in recent years. Having reached a robust measurement, at least up to $z\simeq$4, of the stellar mass growth and star formation rate (SFR) density evolution over cosmic time \citep[e.g.,][]{dickinson2003,drory2005,P06,font06,mmm09,pgp08,P09,bouwens09,ilbert10,hop06,karim11,muzzin13,ilbert13,burga13,mad14}, we must now try to understand which are the main processes that drive the star formation histories in galaxies and, more specifically, the time scales of the mass growth in star-forming galaxies and the main reasons for the downsizing pattern observed in the passive galaxy population \citep[e.g.,][]{thomas2005,cim06,PP09,renzini09,peng10,Elbaz11}. 

A preferred tool for investigating this topic is to study the tight correlation between the galaxy star formation rate~(SFR) and stellar mass~($M_*$) content, which is present at all explored redshifts, and also known as the ``main sequence'' (MS) of star-forming galaxies \citep[e.g.,][]{brinchmann04,salim07,noeske07,elbaz07,daddi071,PP09,magdis10,Elbaz11,salmi12,whita12}. The slope and scatter of this correlation, the evolution of its normalization with cosmic time, but also the detailed dissection of its demographics, i.e. the exact percentages of objects that live above it (starburst galaxies, SB) or below it (quiescent galaxies) contain crucial, and still poorly known, pieces of the galaxy evolution puzzle \citep[e.g.,][]{karim11,rod11,wuyts11,sargent12}. 

While photometric redshift and SED-fitting techniques have become common and are robust ways to estimate galaxy distances and stellar masses, we are still heading down an unpaved way toward obtaining accurate star formation rate measurements for cosmologically relevant galaxy samples. Substantial progress has been made thanks to tracers of star formation such as the mid infrared \s, radio continuum (VLA) and, more recently, far-infrared (FIR) \h~surveys which are not subject to dust attenuation corrections. Due to sensitivity limits, however, the bulk of the star-forming galaxy population is not detected in these kind of data at redshift greater than 1, i.e. when most of the stellar mass growth in the Universe took place. The only way to populate and study the SFR--$M_*$ plane with individual detections at z$>$1 is to use other, dust-attenuated tracers of star formation activity. The most common and easily accessible tracer in most multi-wavelength databases is the UV continuum light emitted by young massive stars. A viable, though more expensive alternative to the rest-frame UV light, is to use line emission, such as \oii~or \ha~which are also good tracers of  star formation rate (e.g., Kennicutt 1998). These can be probed through narrow-band imaging \citep[e.g.,][]{garn2010,sobral12} or by spectroscopic surveys \citep[e.g.,][]{gilbank10,kashino13}.

The main issue with estimating star formation from the emerging UV light, as well as from \oii~or \ha~line emission, is the need to correct these, at least in a statistical sense, for the intervening dust attenuation. A common approach is to use the correlation between the slope of the UV spectrum of galaxies and dust attenuation \citep[e.g.,][]{meurer99,Calzetti00,daddi2004,overzier11}. The effectiveness of such correlations has been questioned in the local Universe \citep{kong2004} and is still debated in the literature \citep[e.g.,][]{buat05,reddy2010,oteo13}. Dust attenuation affecting line measurements can in principle be derived by measuring  the Balmer decrement  (i.e. \ha/\hb) in galaxy spectra \citep[e.g.,][]{brinchmann04,garn2010,zahid13} but this information is rarely available in the distant Universe, due to the scarcity of NIR spectroscopy and, of course, cannot be determined using narrow-band imaging data. For this reason, line emission studies at high redshift often rely on an indirect way to correct for dust attenuation, such as the comparison with a robust SFR tracer like the galaxy FIR or radio continuum emission \citep[see, e.g.,][]{garn2010} or a SFR derived through dust-corrected UV/SED-fitting \citep[e.g.,][]{fs09,mancini11}.

By using a sample of BzK selected star-forming galaxies \citet{PP09} showed that star-forming galaxies at $z\simeq$1.7 were growing in mass in a self-similar, exponential way. Consequently galaxies cannot have lived on the Main Sequence for their entire life, as this would imply a dramatic overproduction of mass compared to the measured evolution of the stellar mass density over cosmic time (see also e.g., Heinis et al. 2014 for similar conclusions). Pannella et al.~(2009a) also found that the correlation between UV slope and UV dust attenuation allowed them to retrieve accurate - again, in a statistical sense - galaxy star formation rates and that dust attenuation is a strong function of the galaxy stellar mass. Finally they were also able to show that the measured emerging UV light is poorly correlated (or perhaps even anti-correlated) with the actual star formation present in a galaxy, as measured by the 1.4~GHz radio continuum luminosity.

The main aim of this paper is to extend the work presented in \citet{PP09} by focusing on a mass-complete sample of star-forming galaxies up to $z\simeq$4. Accurate SFR measurements are obtained from the deepest \h~FIR and radio 1.4~GHz continuum imaging available to date, permitting an unbiased derivation of the Main Sequence evolution with cosmic time. For the first time, in this paper, we study the radio-FIR correlation from a pure stellar mass selection perspective. Finally, we are able to thoroughly study the dust attenuation properties of star-forming galaxies over a vast range of cosmic time by comparing the derived SFR to the measured UV light.  

This paper is organized as follows: in Section 2 we describe the selection and stellar population properties of the star-forming galaxy sample used in this work. In Section 3 we describe the stacking analysis performed on the GOODS-\h~FIR and VLA radio continuum images. In Section 4 we show our results on the Main Sequence of star-forming galaxies, their dust attenuation properties and the FIR-radio correlation. In Section 5 we discuss our results and show some implications of the extra attenuation suffered by line emission compared to the stellar continuum, and the correlation between gas phase metallicity and dust attenuation. We close in Section 6 by summarizing our main results.

Throughout this paper we use AB magnitudes, a Salpeter~(1955) initial mass function (IMF) and adopt a $\Lambda$CDM cosmo$\log$y with \mbox{$\Omega_M=0.3$}, \mbox{$\Omega_\Lambda=0.7$} and \mbox{$H_0=70 \; \mathrm{km} \, \mathrm{s}^{-1} \, \mathrm{Mpc}^{-1}$}. As a matter of notation, we refer to the rest-frame {\it GALEX} FUV bandpass and to the total integrated IR light in the range 8--1000\,$\mu$m when using the subscripts ``UV'' and ``IR'', respectively.

\begin{figure*}
\begin{center}
  \includegraphics[height=.460\textwidth]{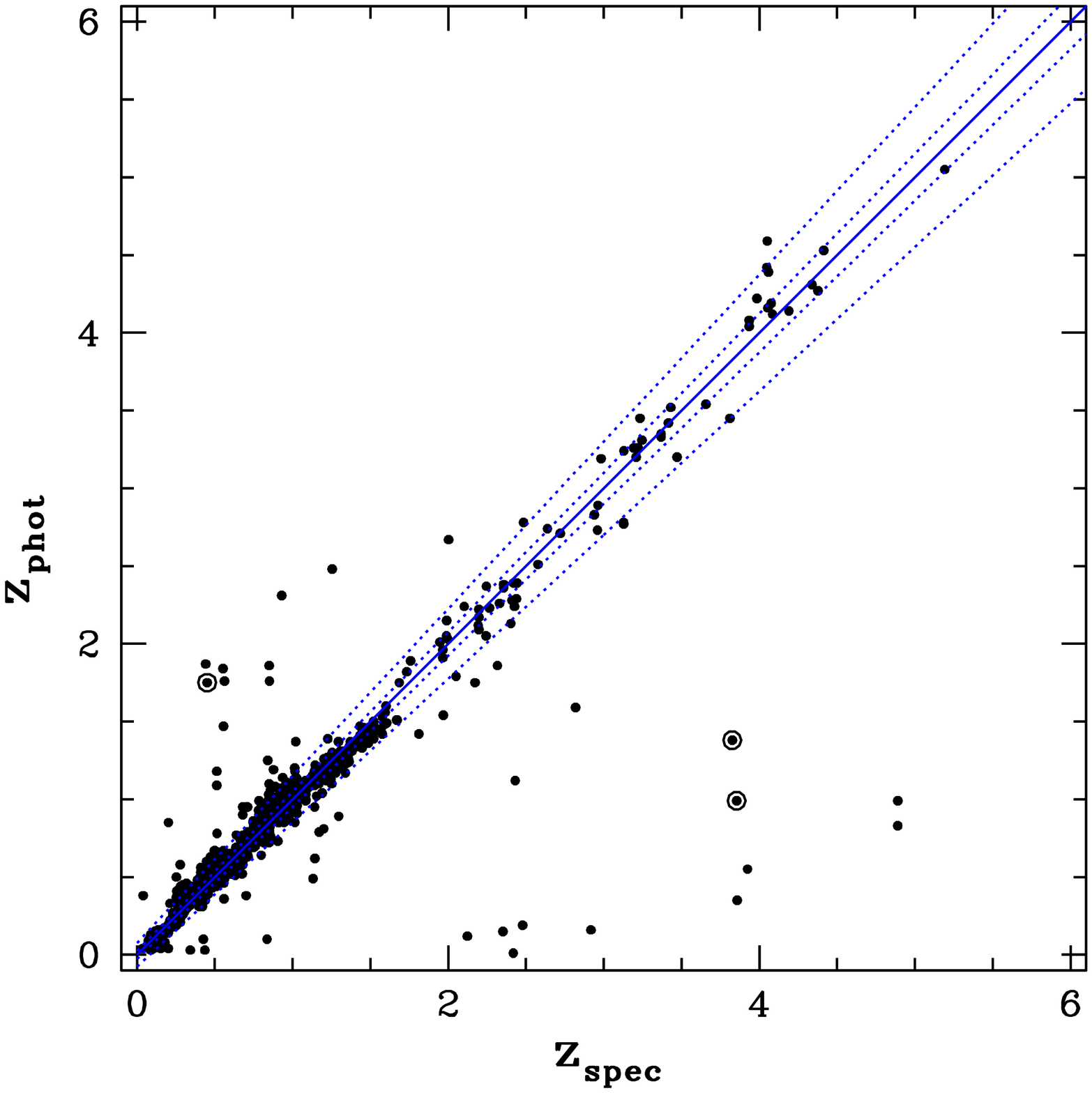}%
  \includegraphics[height=.460\textwidth]{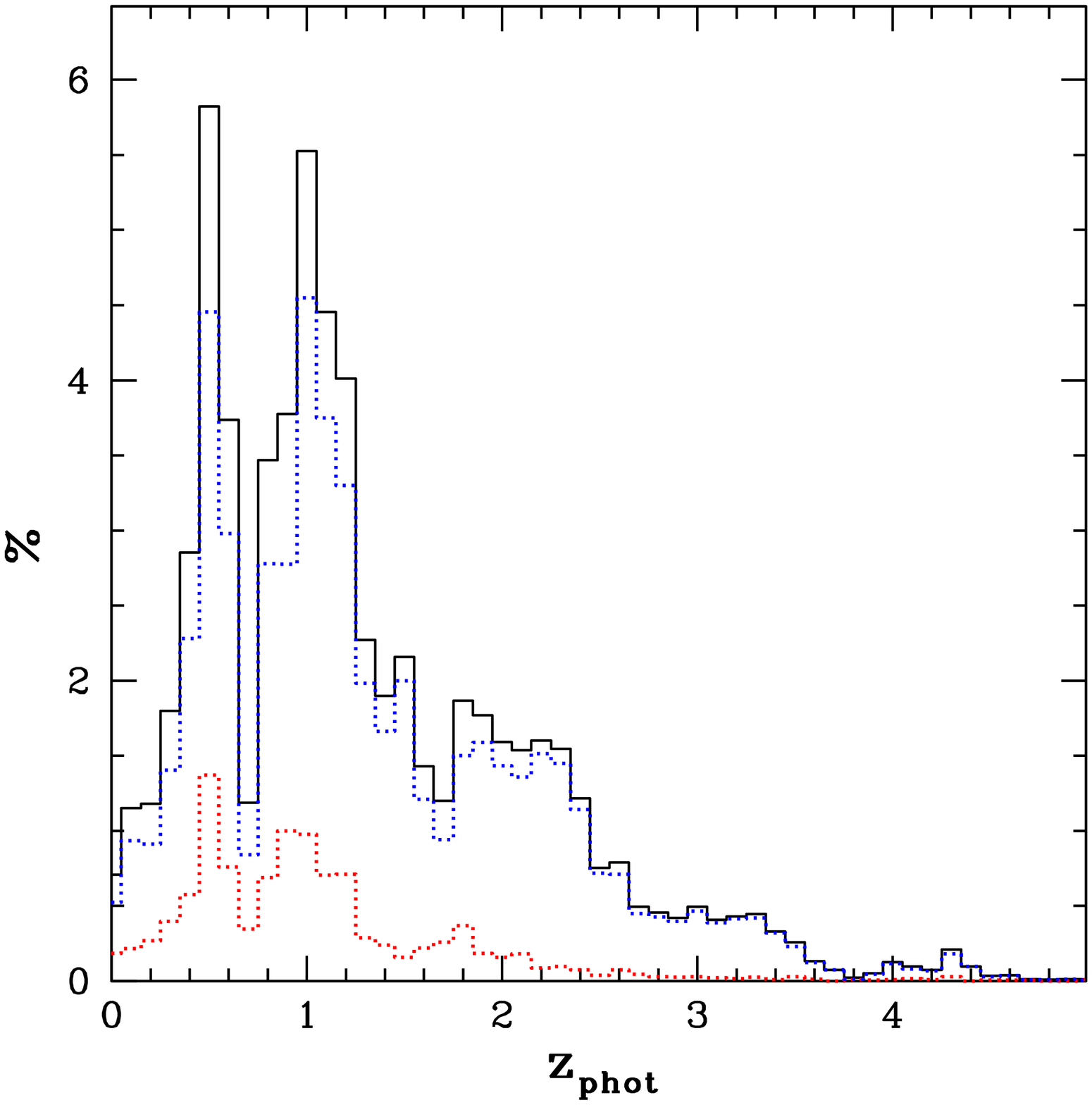}
  \caption{{\bf Left:}~Comparison between spectroscopic and photometric redshifts used in this work. The solid line is the bisector, while the dotted lines show the 3\% ($\pm$1$\sigma$) and 9\% ($\pm$3$\sigma$) bands. The three circled dots are galaxies for which two different spectroscopic determinations are available. For all three, the second determination agrees with our photometric redshift estimate. {\bf Right:}~Redshift distribution of our catalog in the GOODS-N field down to $K_{\rm AB}=$24.5, the 5$\sigma$ limiting magnitude. The solid line shows the actual sample of 18,416 galaxy sample used in this work, i.e., restricted to the FIR deep imaging area. Dashed blue (red) lines show the sub-sample of star-forming (quiescent) galaxies. We note that the two main redshift peaks at $z\simeq0.5$ and $z\simeq1$, and as well as the underdensity at $z\simeq0.75$, are real large scale structure features as they are present also in the spectroscopic redshifts distribution.}
\label{uno}
\end{center}
\end{figure*}

\section{The GOODS-N star-forming galaxy sample:\\ physical properties and mass-completeness}

\subsection{Multi-wavelength data and catalog production}

The galaxy sample we use in this work is drawn from a {$K_{\rm S}$}-band selected multi-wavelength catalog in the GOODS-North field, spanning 19 passbands from {\it GALEX} NUV to IRAC 8\,$\mu$m (namely, {\it GALEX} NUV, KPNO Mosaic U,Subaru Suprime-Cam B-V-IA624-R-I-z-Y, CFHT WIRCam J-H-Ks, Subaru MOIRCS J-H-Ks and \s~IRAC 3.6-4.5-5.8 and 8\,$\mu$m).  The {\it GALEX} data are part of the {\it GALEX} GR6 data release\footnote{Publicly available at http://galex.stsci.edu/GR6/}.  The optical, NIR and IRAC data are described in \citet{Capak2004}, \citet{ouchi09}, \citet{Wang10}, \citet{lin12}, \citet{kajisawa11}, Lin et al.~(in preparation), Onodera et al.~(in preparation) and Dickinson et al.~(in preparation). 

We first register all images to a common pixel grid and then use {\sc
SExtractor} \citep{bertin} in dual image mode, to measure photometry. The {$K_{\rm S}$}-band CFHT WIRCAM image has been adopted as the primary detection image,
because it represents the best compromise, among all available bands, between
the need for a robust tracer of galaxy stellar mass and an angular resolution
($\simeq$0.8") matching that of most of the other bands, which simplifies
catalog assembly and photometry measurements. The whole catalog contains 56,144 objects over the CFHT WIRCAM {$K_{\rm S}$} image field of approximately 900 arcmin$^2$ and down to an AB magnitude of 24.5 (i.e., the image 5$\sigma$ limiting magnitude, see Wang et al.~2010 for details).  

The images used here have very different resolutions. Rather than convolving all images to the worse resolution, which would result in a significant loss of information, we account for this in the estimate of aperture colors by applying PSF-matching corrections based on the growth-curve of point-like sources. To limit uncertainties in such corrections we use 2" diameter apertures to sample the galaxy spectral energy distributions (SEDs).
% In order to match the resolution of the different images we decided, rather than convolving all images to the worst measured seeing,  to work directly in the catalog space by applying PSF-matching corrections based on the growth curve of ``bona-fide'' point-like sources.}  Aperture magnitudes (2" diameter) are used to sample the galaxy spectral energy distributions (SED). 

Band-merging with IRAC photometry was done with a cross-match of the NUV--optical--NIR catalog with the {$K_{\rm S}$}--IRAC catalog released by Wang et al.~2010. This latter study used a ``real-cleaning'' procedure to perform PSF modeling in the IRAC images by assuming as priors the positions of sources detected in the same {$K_{\rm S}$} CFHT WIRCAM image we have used. We refer the reader to Wang et al.~(2010) for a more detailed description and assessment of the extracted IRAC photometry. Here we want to stress that because we use the same {$K_{\rm S}$} image publicly released by Wang and collaborators our cross-matching procedure turns out to be a very robust way of associating IRAC fluxes to our {$K_{\rm S}$}-based multi-wavelength catalog. 

The whole catalog contains 56,144 objects over the CFHT WIRCAM {$K_{\rm S}$} image field of approximately 900 arcmin$^2$ and down to an AB magnitude of 24.5 (i.e., the image 5$\sigma$ limiting magnitude, see Wang et al.~2010 for details).

We classified 2,072 objects as stars, and as such exclude them from the sample, according to their {\sc SExtractor} stellarity index, at bright ({$K_{\rm S} < 20$}) magnitudes, and to their position in the $BzK$ diagram as in Daddi et al.~(2004).   

The morphological selection of stars is also effective in removing those Type 1 (i.e. optically unobscured) AGN whose emission from the active nucleus dominates over the host galaxy in almost all bands. In any case, these objects are extremely
difficult to deal with in terms of both the photometric redshift derivation and stellar mass estimate \citep[e.g.,][]{salvato09,cisternas11,bongiorno12}, so it is preferable to remove them from our sample.  However, we do not attempt to remove
from the star-forming galaxy sample those X-ray sources detected in the 2-Ms {\it Chandra} data \citep{alexander03,bauer04} whose optical-to-NIR emission is dominated by the host galaxy. A number of recent studies (see, e.g., Mullaney et al.
2012, \citealt{santini12}, \citealt{juneau13}, \citealt{rosario13}) have shown that these AGNs are mostly Type-II obscured AGN and LINERS which have similar star-formation rates as normal (i.e., non-AGN) star-forming and passive galaxies, respectively, thus their inclusion has no net effect on our results. We will discuss later in Section 3 our approach to minimize the contribution of the AGNs present in our sample to the IR budget.

\subsection{Photometric redshifts}

Photometric redshifts were estimated by running {\it EAZY}\footnote{Publicly available at http://code.google.com/p/EAZY-photoz/} \citep{brammer08} on the multi-wavelength catalog. We have used {\it EAZY}  in its standard set-up and modeled the observed galaxy SED with a linear combination of the seven standard {\it EAZY}  galaxy templates \citep{whita11} in order to maximize the likelihood as a function of the galaxy redshift. Following a common procedure for producing accurate photometric redshifts \citep[see, e.g.,][]{capak2007,gab2008}, we allowed for a photometric offset in each measured band. For each band in our catalog we computed a photometric zero-point shift by iteratively running the code on a subsample of high fidelity spectroscopic redshifts. For each iteration and for each band we computed the median of the ratio between the measured flux density in the catalog and the carefully computed model value, i.e. the integral of the best fit solution template through the theoretical response curve, and apply these median offsets before starting the next iteration. Computing median offsets, instead of mean ones, is an effective way to filter out catastrophic events such as badly measured SEDs or wrong spectroscopic redshifts determinations. The procedure described here usually converges to a robust solution within a limited number of iterations.

Photometric offsets between observed and modeled data may be attributed to  many different effects, like actual zero-point errors, differences of the actual system response curves with respect to the ones adopted and inaccurate PSF-matching corrections, but also uncertainties and limitations in the template library adopted.

In the following sections, we use the multi-wavelength catalog which has been corrected for these systematic offsets. In all bands  these amount to less than 20\% of the measured flux so that, while substantially improving the photometric redshift accuracy, applying these flux corrections does not have any major effect on the derived galaxy properties, nor in general on the main results of this study. We will discuss in the relevant sections when notable differences arise from the use of the corrected vs. the uncorrected photometry.
 
When comparing to the spectroscopic sample of \citet{barger08} and Stern et al.~(in preparation),  we reach a relative (i.e., $\Delta z = (z_{\rm phot}-z_{\rm spec})/(1+z_{\rm spec})$) accuracy of 3\% (see Figure~\ref{uno}), with less than 3\% catastrophic outliers (i.e., objects with $\Delta z > 0.2$).

\subsection{Derivation of stellar masses}

Stellar masses were determined with {\it FAST}\footnote{Publicly available at http://astro.berkeley.edu/$\sim$mariska/FAST.html} \citep{fast} on the
$U$ to 4.5\,$\mu$m PSF-matched aperture photometry, using \citet[]{bc03}
delayed exponentially declining star formation histories (SFHs,
$\psi(t) \propto \frac{t}{\tau^2} \exp(-t/\tau)$) with 0.01$<\tau<$10
Gyr, solar metallicities (Z$_\odot$ = 0.02), Salpeter initial mass function (IMF), and the \citet{Calzetti00}
reddening law with A$_V$ up to 4 magnitudes. 

The choice of a fixed solar metallicity has essentially no impact on the derived stellar masses, as the age-metallicity degeneracy leaves mass-to-light ratios basically unchanged. Moreover, solar metallicities are still a fair assumption for our sample at high redshift given 
that, due to the increasing mass completeness with redshift, it contains at all cosmic epochs the most massive and most metal-rich systems.

%that the stellar mass completeness evolves with redshift so to contain only the most metal-rich systems at any cosmic epoch explored.}

%The choice of a fixed solar metallicity has essentially no impact on the derived stellar masses, as the age-metallicity degeneracy leaves mass-to-light ratios basically unchanged. Moreover, solar metallicities are a fair assumption for our sample given its redshift and stellar mass range. 

%{\bf The choice of a fixed, solar, metallicity, mainly adopteded to optimize computational time, has almost no impact on the derived stellar masses because of the age--metallicity degeneracy which does compensate, in the SED-ftting results, physical differences in metallicities with numerical differences in age by leaving the galaxy mass-to-light ratio basically unchanged. Moreover, our approximation of solar metallicities represents a fair description of our galaxy sample that going at high redshift, and due to the increasing mass completeness limit, contains only the most metal-rich systems.}

Derived (aperture) masses were extrapolated to ``total'' masses using the ratio between the 
total ({\it FLUX\_AUTO}) and aperture flux in the $Ks$-band detection image. 
While this approach corrects for the bulk of the flux loss, it is based on 
one band only and thus neglects any color gradient within the galaxy.

With respect to the choice of the SFH, we note that it has been shown that
other forms of SFHs are more appropriate for star-forming
galaxies at high redshift \citep{PP09,maraston10,papovich11}. For what
directly concerns this work, rising or constant (possibly truncated)
SFHs would change the stellar masses of our star-forming galaxy sample by an amount which is well within the median estimated accuracy of about 0.2~dex~\citep[see e.g.,][and the results presented in Appendix \ref{appmass}]{papovich06,papovich11,debarros14,buat14}. 

As a further accuracy test, we have compared stellar mass estimates derived with {\it FAST} with the SED-fitting
code described in \citet{drorymass,drory09}. The results from runs with the different codes and different adopted SFHs, show no systematic differences in stellar mass estimates, except for the highest stellar masses ($M_* \gtrsim 2\times10^{11}\,\Msun$), where the two-component fitting from \citet{drorymass} predicts masses which are larger by 0.2~dex, with a global scatter of about 0.2~dex. We provide more details on the estimated stellar mass errors and on the comparison between the two codes in Appendix \ref{appmass}.

\begin{figure*}
  \includegraphics[height=.460\textwidth, bb= 20 170 590 720, clip]{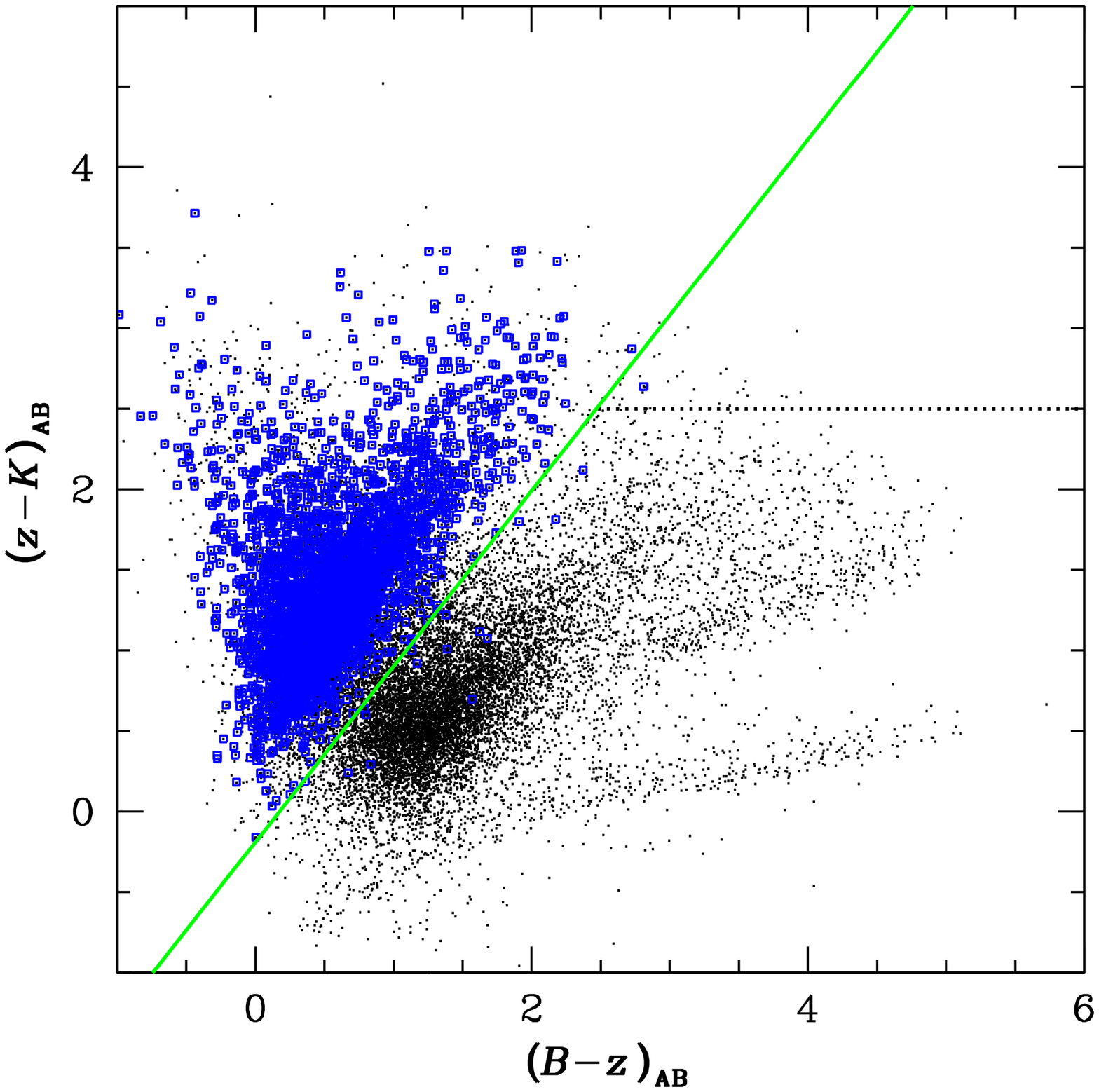}%
  \includegraphics[height=.460\textwidth, bb= 20 170 590 720, clip]{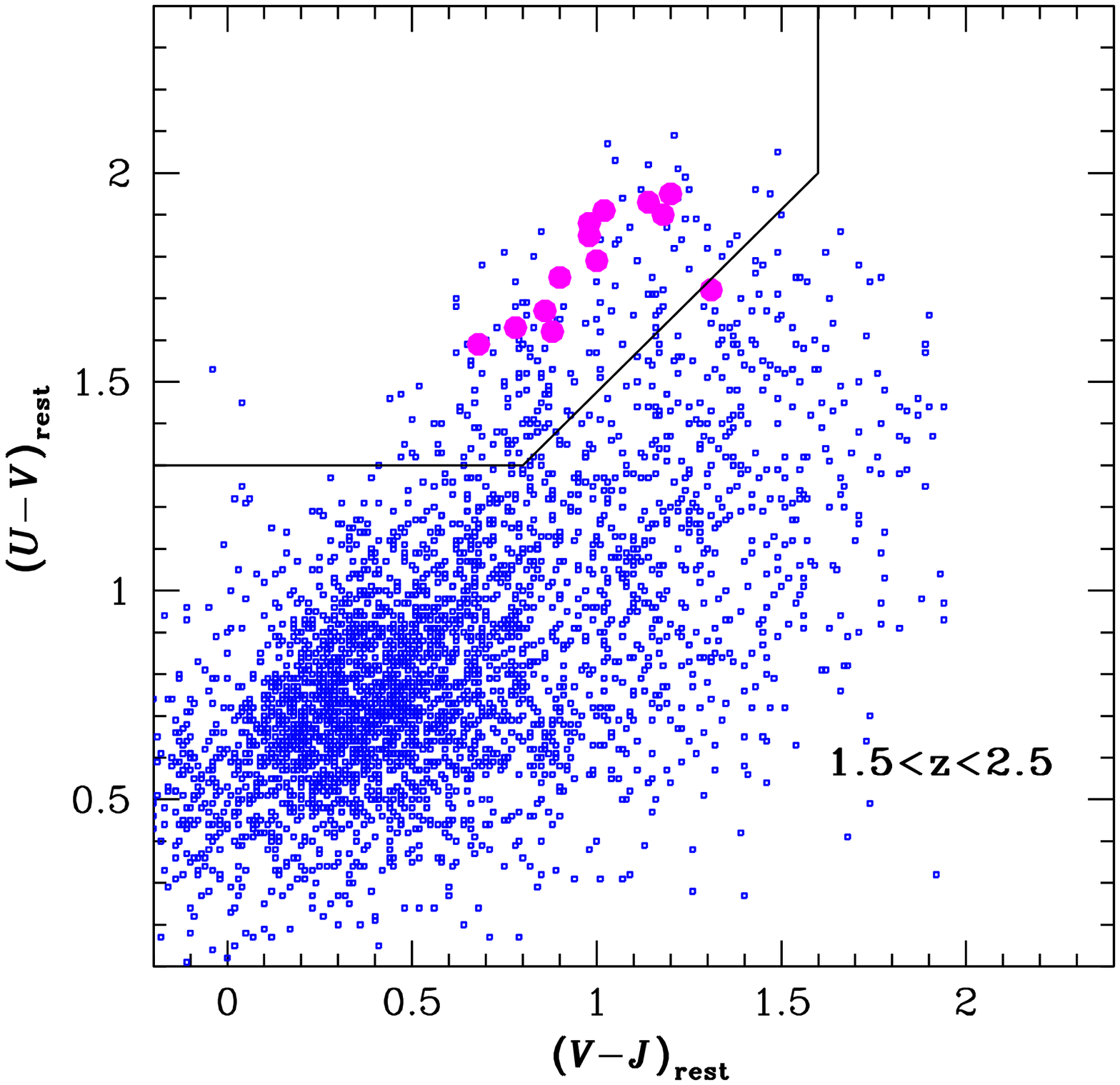}%
  \caption{{\bf Left:}~{\it BzK} diagram used for the selection of star-forming galaxies in the redshift range 1.5--2.5 according to Daddi et al.~(2004). The solid line divide the regions populated by star-forming galaxies at $z\simeq2$ ({\it sBzK}, above the solid line) from the lower redshift objects (below the solid line and below the dotted line) and the passive galaxies at  $z\simeq2$ ({\it pBzK}, below the solid line and above the dotted line). Black points show all the galaxies in our catalog, while open squares represent the {\it UVJ}-selected star-forming galaxies in the redshift range 1.5--2.5.~{\bf Right:}~{\it UVJ} diagram used for the selection of the sample of star-forming galaxies used in this work. The solid lines identify the two regions used to divide quiescent (upper-left) from star-forming (lower-right) galaxies. Here we show the galaxies in the redshift range 1.5--2.5 which would have been selected as star-forming {\it sBzK} (blue squares) and passive quiescent {\it pBzK} (magenta points).} 
\label{tre}
\end{figure*}

\subsection{The star-forming galaxy sample:\\ UVJ selection, $\beta$ slope and mass-completeness}

The main aim of this study is to characterize the relation between star formation, UV dust attenuation and the stellar mass content of galaxies over cosmic time. To do so, it is important not to mix galaxies which might still have some residual ongoing star formation (but are substantially quenched), with galaxies that are instead genuinely star-forming and reddened by dust attenuation. A selection based on direct star formation rate indicators, such as UV or line emission, inevitably mixes the two populations. Ideally one would like to remove galaxies with extremely low specific star formation rates (SSFRs) from the sample, but this is not easily achieved for individual galaxies as it requires accurate dust attenuation corrections or alternatively FIR/radio derived star formation rate estimates. Following established procedures, we have excluded quiescent galaxies from the sample by using the $U-V$ vs.~$V-J$ rest-frame color plot \citep{wuyts07,williams09}. This approach builds on two main ingredients: 1) the intrinsic rest-frame $U-V$ color is a good proxy for the SSFR of a galaxy (e.g., \citealt{salim2005}, Pannella et al.~2009b); 2) the color-color plot allows us to break the age-dust degeneracy and hence efficiently splits the sample into two classes of galaxy, which are either essentially quiescent, or still actively forming stars. We have selected star-forming galaxies at all redshifts as,
$$(U-V) < 1.3, (V-J) > 1.6, (U-V) < (V-J)\times0.88 + 0.59.$$
These $UVJ$ selection limits were originally defined by \citet{williams09} in order to maximize the difference in SSFRs between the regions. However, rest-frame color distributions might be slightly different than \citet{williams09}
due to photometric coverage, band selection but also the specific redshift where the analysis is performed (rest-frame color of quiescent galaxies are becoming redder with decreasing redshift). Because of this reason different studies have used slightly different dividing lines, sometimes changing with redshift~\citep[e.g.,][]{cardamone10,whita11,Brammer11,strazzullo13,viero13,muzzin13}. The differences between these studies are always lower than 0.2 magnitudes in colors and often comparable with the same accuracy to which the rest-frame colors are known. In order to be conservative and minimize the contamination of quiescent galaxies in our sample, we have choosen to use the same selection region at all redshifts. We have also checked that slightly changing our assumptions on the selection limits, i.e., applying $\pm$0.1 magnitudes shifts on the rest-frame colors, would not change quantitavely our results.

We checked our selection against the {\it BzK} selection of star-forming galaxies at $z\simeq$2 \citep{daddi2004} (see Figure~\ref{tre}), and the results are in good agreement over the common redshift range. Finally, we have tested that all the results of this paper would remain the same if instead of using the $UVJ$ selection to separate active from passive/quenched galaxies we would use a cut in SSFR, namely selecting as star-forming galaxies all the objects with SSFR$\ge$10$^{-11}$\,yr$^{-1}$  as derived from the FAST SED-fitting output.

Hereafter, we will concentrate on the {\it UVJ}-selected subsample of 14,483 star-forming galaxies that fall within the GOODS-\h~area, the deepest  FIR imaging available in the northern sky~(Elbaz et al. 2011).

\begin{figure}
\begin{center}
  \includegraphics[height=.18\textwidth, bb= 50 180 580 390, clip]{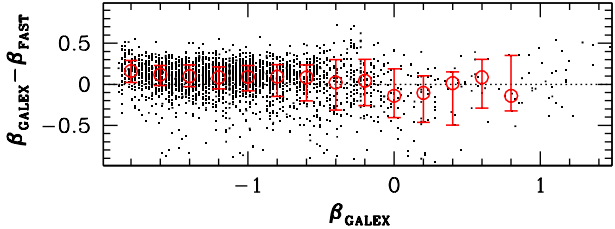}%
  \caption{Comparison between the $\beta$ values derived from the rest-frame {\it GALEX} bands and the one derived from a fit, in the rest-frame range 1250\,\angstrom~and~2600\,\angstrom, to the best-model SEDs of {\it FAST}. Red open circles show median values of the difference and error bars represent 16th and 84th percentiles of the distribution. The $\beta_{GALEX}$ values are only slightly redder, by around 0.1, than the SED-derived ones with a median uncertainty value of 0.2.}
\label{tred}
\end{center}
\end{figure}

Following \citet{Brammer11}, rest-frame magnitudes are computed with {\it EAZY} for all objects in the catalog from the best-fit SED model integrated through the theoretical filters. The $\beta$ slope of the UV continuum was calculated as
\begin{equation}
 \beta = \frac{ \ \log(f_{\rm \lambda_1}/f_{\rm \lambda_2}) }{ \ \log(\lambda_1/\lambda_2) } = \frac{ 0.4(m_2-m_1) }{ \ \log(\lambda_1/\lambda_2) }  -  2,
\label{betaeq}
\end{equation}
where $m_1$ and $m_2$ are the magnitudes at wavelengths $\lambda_1$ and $\lambda_2$, respectively (see, e.g., Overzier et al.~2010, \citealt{nord13}).
In this study we adopt the {\it GALEX}
  FUV ($\lambda_{\rm c}$$\simeq$1530\angstrom) and NUV
  ($\lambda_{\rm c}$$\simeq$2315\angstrom) response curves to compute the photometric measure of $\beta$, but we also checked that the results obtained would remain largely unchanged by using different rest-frame bands sampling the slope, or by obtaining a direct fit of the form $f_\lambda \propto \lambda^\beta$ to the best-fit SED in the rest-frame range 1250\,\angstrom~and~2600\,\angstrom. In Figure~\ref{tred} we show the results of this comparison using the best-fit SEDs of {\it FAST}. The {\it GALEX}-derived estimate of $\beta$ provides an accurate description of the UV slope, with a median offset of 0.1 and a scatter of around 0.2. This result is in good agreement with the finding reported in Finkelstein et al.~(2012). %{\bf We notice here that the pattern in Figure~\ref{tred} is due to the coarse sampling of the stellar population model parameters used for the FAST run. We have tested that running FAST on a finer grid of stellar population parameters would completely erase the pattern.}  

 An important issue, which is often not given careful consideration in high redshift studies dealing with star-forming galaxies, is the assessment of the sample's mass-completeness.
This is because the amount of dust attenuation plays an important role in the selection function: low mass galaxies (and the definition of {\it low} depends of course on the selection band and detection limit of the specific dataset) can only be detected if they suffer from low dust attenuation. In other words, as redshift increases magnitude-limited samples become more and more biased against objects with lower masses and/or high M$_*$/L ratios, such as galaxies that are not currently forming stars, or which are highly dust obscured. Because we are interested in dissecting dust attenuation properties, we want to make sure this issue has no impact on our results. 
 Following \citet{rod10}, we estimated our completeness mass by using a stellar population with a 1 Gyr old constant SFH and highly attenuated (E[B-V]=0.6).  As a function of redshift, there is a one-to-one correspondence between the observed Ks magnitude and the stellar mass built up by this model. We hence define as the completeness mass at a given redshift the mass corresponding to the completeness magnitude ($\simeq$23.8 AB mag) of our Ks selected catalog. All galaxies more massive than our completeness mass will have a brighter flux at Ks and hence will be surely detected in our survey and the same is true for galaxies suffering less dust attenuation. Obviously, this definition of mass completeness is dependent on the specific stellar population parameters adopted and can only apply to galaxies that can be likely represented by such a star formation history, i.e. star-forming galaxies. Figure A.1 of Rodighiero et al. (2010) illustrates how the actual mass completeness varies by adopting different ages and attenuations for the stellar population models. However we notice that our modeling is very conservative and the quoted mass completenesses (see Table \ref{ttmas}) should be regarded as safe and robust, although theoretical, values. As stated above our sample, being a sample of a priori selected star-forming galaxies, is also cut in SSFR (at approximately 10$^{-11}$\,$yr^{-1}$, see discussion above), meaning that, even if we are complete down to a certain mass and able to robustly trace the SFR-M* made by the bulk of the star-forming galaxy population, we would still be able to detect galaxies with extremely low SSFR if they suffer low dust attenuation. This obviously can be translated in a SFR limit at the completeness mass that we also quote in Table 1.  We want to stress here that, being our primary selection a mass selection,  the secondary limits in SFR has no influence on the paper results because it applies to galaxies well below the Main Sequence locus.

% Following \citet{rod10}, we here adopt a conservative approach and estimate the ($\simeq$90\%) mass-completeness at a given redshift as the mass of a stellar population with a constant SFH of 1\,Gyr, attenuated with an $A_{\rm UV}$ = 6 mag (i.e., assuming a Calzetti~(2000) law this corresponds to an $E[B-V]\simeq$0.6) and flux equal to the completeness magnitude of our catalog, $K_{\rm AB}$= 23.8 mag. Obviously, this definition of mass completeness is dependent on the specific stellar population parameters adopted, in Figure~~A.1 of Rodighiero et al.~(2010) one can appreciate how the mass completeness varies by adopting different stellar population models. In Table \ref{ttmas} we provide the mass-completeness values as a function of redshift adopted in our analysis.

\begin{table}
\begin{center}
\caption{Stellar mass and star formation rate~\\completeness versus redshift}         
\label{ttmas}      
\begin{tabular}{c c c c c c c}     
\hline \hline                
{\it z} & 0.7 & 1 & 1.3 & 1.7 & 2.3 & 3.3\\   
\hline                       
$\log$~$M_*$/M$_\odot$& 9.0&  9.5   & 10.0 & 10.3&  10.5  & 10.7\\
 SFR [M$_\odot$yr$^{-1}$]& 1&  2   & 10 & 20 &  30  & 90\\
\hline \hline                                    
\end{tabular}
\end{center}
\end{table}

~\\
\section{The stacking analysis}
\label{Stacking}

We study the FIR properties of star-forming galaxies at high redshift using the deepest \h~imaging available to date. Obviously any analysis based on  FIR detected sources only would be a SFR-biased view of the Universe at all redshifts. This is best exemplified by the fact that only 1095 sources are detected at more than 3$\sigma$ in the deep GOODS-\h~images (see Elbaz et al.~2011), while more than 11000 star-forming galaxies are present in our sample over the same area: at all redshifts the flux density limit of the \h~images allows us to explore only the highest values of star formation rates as opposed to a purely mass-selected sample, where UV-based star formation rates can reach down to extremely low SFR levels. Up to $z\simeq$1 the FIR data are deep enough to probe the Main Sequence down to relatively low stellar masses (i.e., $M_* \ge$10$^9$M$_\odot$). At higher redshifts only the most extreme events of star formation and the high mass end of the mass function can be detected in the GOODS-\h~data. In order to reach the FIR flux densities typical of normal MS galaxies, and to gain a comprehensive picture of the physics driving star-forming galaxies at high redshift, it is therefore mandatory to adopt a stacking approach.      

The GOODS-\h~images were generated at different pixel scales in order to allow a fair sampling ($\simeq$FWHM/5) of the beam. For each of the sources in our star-forming galaxy sample we produced a cutout, centered on the nearest pixel to the sky position of the source, in the PACS (100, 160\,$\mu$m) and SPIRE (250, 350, 500\,$\mu$m) images of 60$\times$60 pixels, which corresponds to an angular scale of about 10 times the image beam. 
The cutouts were then stacked to create median images in selected bins of redshift and stellar mass. Median stacking is more robust than the mean against the tails of the distribution (mainly due to the relatively few detections but also to possible photo-{\it z} catastrophic outliers), while the rms still goes down by approximately $\sqrt N$ and image statistics are well preserved. Total fluxes are retrieved by performing a PSF plus background modeling with the {\it GALFIT} code~\citep{peng2002} on the stacked images. In order to calculate realistic errors on the retrieved flux densities we use a bootstrapping approach. In each redshift and mass bin we randomly select (with replacement) $1/N$ of the sample within that bin and re-estimate the median flux density of this sub-sample. This is performed 1000 times for each bin and the error on the median value is calculated by taking $\sqrt{1/N}$ times the standard deviation of the results. Running this procedure with {\it N}=1,2,3 and 4 yields consistent results.

%For each \h~band, we derived total IR luminosities (\lir) by using the MS template defined in \citet{Elbaz11}. Despite the very different resolutions and data reduction procedures for the PACS and SPIRE data, the total \lir~obtained from the different bands are in good agreement (within 0.2~dex) for all the bands, at all redshifts and in all mass bins (see Appendix \ref{applir} for more details). This was reassuring for the (still poorly explored) technical issues which might be affecting the \h~data, and in particular concerning stacking analyses, like the high-pass filtering for the PACS data or the flux boosting due to sources angular correlation for SPIRE\citep[see, e.g.,][]{bethermin10,popesso12,viero13}. 
%Our results suggest that these effects should have a limited impact on the derived \lir, giving a final uncertainty of certainly less than 30\% . They also reinforce the result of Elbaz et al.~(2011) on the actual self-similarity of the FIR spectral shape of MS galaxies as a relatively uniform population. 

We corrected the estimated \h~flux densities for the bias due to clustering in a statistical way. The clustering is expected to introduce an extra signal in the PSF modeling of the stacking results due to the positive correlation of the faint sources present in the field and the stacked sample. Different formalisms have been proposed in the last years in order to estimate the impact of source clustering on the stacking results~\citep[see, e.g.,][]{bethermin10,charyandpope10,Kurcz10, bourne11,bethermin12,viero13}.    A natural expectation is that the bias is directly proportional to the beam size of the image, i.e., the worse is the image angular resolution and the higher is the expected bias, while the dependence on stellar mass and redshift is less clear.

The novel technique adopted to estimate the flux boosting due to clustering will be thoroughly detailed in \citet{schreiber15}. Here for the sake of clarity we will briefly outline the main steps involved in the procedure. A simulated map in each band is created by using the sky position, redshift and stellar mass for all galaxies in our sample. The analytical model described in~\citet{sargent12} is then used to associate a SFR, and hence an IR luminosity, to every galaxy. Finally \h~flux densities are estimated by adopting a library of FIR templates and injected, with PSFs, in a pure instrumental noise map. The stacking procedure is then run on the simulated map in the same bins of mass and redshift as for the real data. The  correction factor is derived from the comparison between the median input flux and output stacked result. 

We report in Table \ref{spirecorr} the median corrections adopted for the SPIRE bands. These values are in nice agreement with values already reported in the literature for similar experiments in the same bands~\citep{viero13} or in similar angular resolution data~\citep{bourne11}. Within the uncertainties we are not able to see any dependence of flux boosting on stellar mass and redshift and hence adopt a fix scaling in all bins. No flux correction has been applied to the PACS photometry. Here the estimated boosting is less than 10\% and is approximately counterbalancing the loss of flux due to the high-pass filtering in the data reduction process~\citep{popesso12}.

\begin{table}
\begin{center}
\caption{Corrections applied to SPIRE fluxes}         
\label{spirecorr}      
\begin{tabular}{c c c c}     
\hline \hline                
 band & 250\,$\mu$m & 350\,$\mu$m & 500\,$\mu$m\\   
\hline                       
scaling factor& 0.8&  0.7   & 0.6\\
\hline \hline                                    
\end{tabular}
\end{center}
\end{table}

After de-boosting the SPIRE flux densities, we derived the total IR luminosity by fitting the observed IR SED from 100 to 500\,$\mu$m with the \citet{ce01} library models.
  The derived IR luminosities are deemed more accurate than methods based on bolometric correction extrapolation of single IR band measurements because they use the more extended photometric information that can also account for the slight evolution in redshift of the median IR SED~\citep{magdis12,beth12}. In Appendix~\ref{applir} we compare our results to IR luminosities estimated  by using bolometric correction techniques and test the fidelity and relative accuracy of these latter.
 Finally, we want to remind the reader that as mentioned earlier we have choosen to retain Type II AGNs in our sample since they do not prevent the accurate determination of photometric redshifts and stellar masses of their host galaxies. Still, the AGN itself could dominate the mid-IR total budget. In order to minimize the impact of AGN emission to the IR stacked SEDs and on our results, we discarded from the IR SED fitting all the points below 30\,$\mu$m rest-frame. Indeed as shown by, e.g., Mullaney et al.~(2011) above this wavelength the contribution of the AGN to the measured SED becomes negligible.

To compute star formation rates from IR and UV luminosities we adopt the conversions of \citet{kennicutt98} and Daddi et al.~(2004), respectively:
\begin{equation}
 \frac{SFR_{\rm IR}}{\rm M_\odot yr^{-1}} = 1.71 \times 10^{-10} \frac{L_{\rm IR}}{L_\odot}~~~;
\end{equation}
\begin{equation}
 \frac{SFR_{\rm UV}}{\rm M_\odot yr^{-1}} = 1.13 \times 10^{-28} \frac{L_{\rm UV}}{{\rm erg\,s^{-1}\,Hz^{-1}}}~~~.
\end{equation}

The detection of UV emission from star-forming galaxies clearly indicates that at least part of the UV radiation is not absorbed by dust (and reprocessed into the infrared).  Therefore, following e.g, Bell et al.~(2003), to derive the total SFRs we use: $SFR_{\rm total} = SFR_{\rm IR} + SFR_{\rm UV}$, where $SFR_{\rm UV}$ is computed using the observed UV luminosity in equation 3.

\section{Results:\\ the properties of star-forming galaxies up to $z\simeq$4}
In the next sections we will present our results on the star formation and dust attenuation properties of galaxies in the GOODS-N field up to $z\simeq$4.

\subsection{The SSFR-$M_*$ correlation:\\ evolution of slope and normalization}

\begin{figure}
\begin{center}
\includegraphics[height=.460\textwidth]{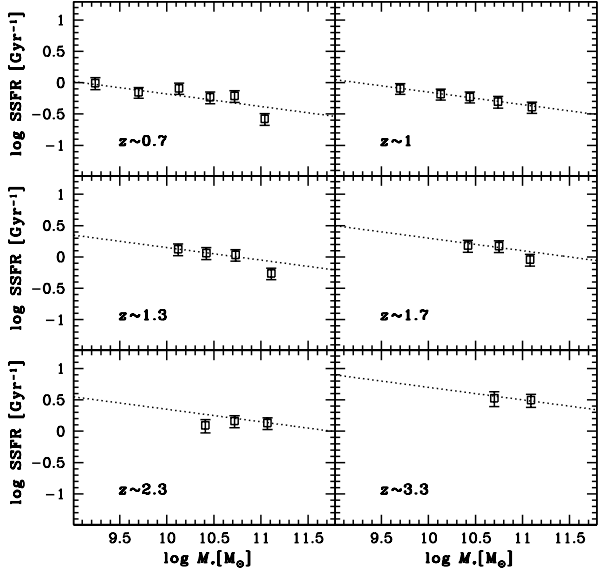}
\caption{Evolution of the specific star formation rate ($SSFR$) vs. stellar mass ($M_*$) at different reshifts from our FIR stacking analysis. We find a slope of 0.8 up to $z\simeq$1.5. For higher redshifts, we are simply assuming that the slope is the same and allow for an evolution in normalization.}
\label{MS}
\end{center}
\end{figure}

The main goal of this work is to define in a consistent way the locus of the Main Sequence of star-forming galaxies over a significant range of cosmic time. Many studies have been published on this topic in the recent years (e.g., Rodighiero et al.~2010, Karim et al.~2011). We use data probing the full FIR SED (using all \h~bands) thus not relying on the \lir~derived from the bolometric correction of a single band. In  Figure~\ref{MS} we show the Main Sequence of star-forming galaxies, i.e., the correlation between the galaxy stellar mass and specific star formation rate, in different redshift bins. Only results from mass-complete samples are shown. We fit a linear trend in the first three redshift bins and find a slope consistent with -0.2$\pm$0.08, corresponding to a slope of 0.8$\pm$0.08 for the proper log$M_*$--log$SFR$ correlation, up to redshift {\it z}$\simeq$1.5. This value of the slope is consistent with a number of previous studies (e.g., \citealt{santini09}, Rodighiero et al.~2011, Lin et al.~2012, Heinis et al.~2014, \citealt{rodighiero14}). We then assume the same slope for the higher redshift bins and fit only the normalization of the MS up to $z\simeq$4. 

\begin{figure}
\begin{center}
\includegraphics[height=.49\textwidth, bb= 0 120 580 750, clip]{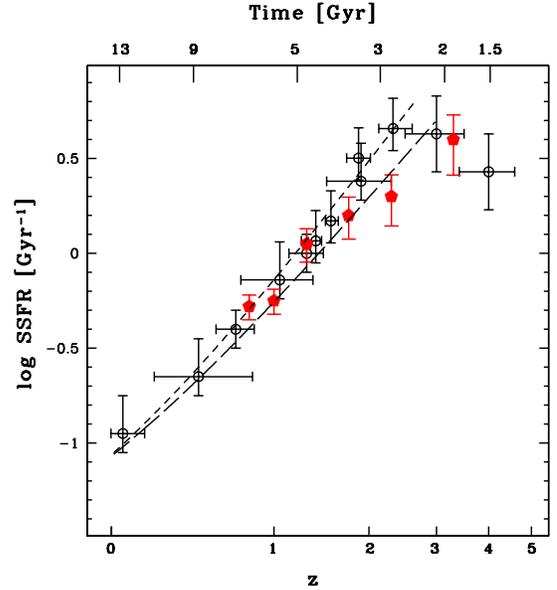}%
\caption{FIR-derived SSFRs from this work (red pentagons) for star-forming galaxies with $M_* \simeq$3$\times$10$^{10}$M$_\odot$  as a function of redshift. A sample of previously published estimates from Brinchmann et al.~(2004), Noeske et al.~(2007), Elbaz et al.~(2007), Daddi et al.~(2007), Pannella et al.~(2009a), Daddi et al.~(2009) and Magdis et al.~(2010) are plotted as empty circles. The dashed and long--dashed lines show the SSFR as a function of redshift according to the relations published in Pannella et al.~(2009a) and Elbaz et al.~(2011).}
\label{MS_norm}
\end{center}
\end{figure}

In Figure~\ref{MS_norm} we show the evolution of the Main Sequence normalization for a stellar mass of $\log$($M_*$/M$_\odot$)=10.5 as obtained from this work, together with some previously published results (Brinchmann et al.~2004, Noeske et al.~2007, Elbaz et al.~2007, Daddi et al.~2007 and 2009, Pannella et al.~2009a, Magdis et al.~2010) and the analytical descriptions of the Main Sequence evolution from Pannella et al.~(2009a) and Elbaz et al.~(2011)\footnote{ Here we plot the original derivation of Elbaz et al.~(2011) that is indeed a good match to our stacking results. However, we note that after taking into account a possible stellar mass systematic offset of 0.19~dex due to the different stellar population models used in the Elbaz et al.~(2011) study and in the one presented here~(see also e.g., \citealt{bell12}), the relation of Elbaz et al.~(2011) would shift upward and almost perfectly overlap the Pannella et al.~(2009a) result. We will investigate this issue in more detail in a future work.}

Our median SSFR values are slightly lower than  previous studies at $z\ge$1.5.
In principle this could be explained by the fact that we are dealing here with a mass-complete sample while previous studies were biased toward brigther ({$K_{\rm S} < 23$} in Pannella et al.~2009a) or star formation selected (24\,$\mu$m detected sources in Elbaz et al.~2011) samples that almost by construction are bound to overshoot the median value of the whole star-forming population. At the same time we cannot exclude the possibility that at least part of the offset could be due to some mis-classified galaxies that enter the {\it UVJ}-selected star-forming sample. 

Finally, it is worth noticing how the SSFR tends to keep increasing up to the highest redshift, without any evidence of the possible flattening at z$\ge$2 obtained in previous studies \citep[e.g.,][]{daddi2009,stark09,gonzales10}, which were based on star formation rate estimates derived from dust-attenuation-corrected UV luminosities. We will discuss this in more detail in the following sections, but based on our results this might be related to a change in the dust and/or stellar population properties of star-forming galaxies with cosmic time.

\subsection{An unbiased analysis of the radio-FIR correlation:\\
the contribution of old stellar populations to the FIR}

Taking advantage of the deep 1.4~GHz map available in the GOODS-N field \citep{morrison00}, we perform a stacking analysis of the radio continuum in the same bins of mass and redshift as done in the FIR \h~bands in order to study the evolution of the radio-FIR correlation for a uniformly selected, mass-complete sample of star-forming galaxies over cosmic time. \citet{sargent10} discussed the impact of sample selection, specifically the FIR versus radio selection, in explaining the controversial results obtained in literature studies on the evolution of the FIR-radio correlation. Based on {\s} 24\,$\mu$m-derived SFR estimates, they showed how the correlation was not evolving up to redshift 1 and possibly even at higher redshift, although this was plagued by the uncertain extrapolation from the measured 24\,$\mu$m to total IR luminosity at $z\ge$1.5. Here we have the advantage of being able to stack mass-complete samples.

Some recent \h-FIR based studies have claimed a redshift evolution of the correlation~(e.g., \citealt{ivison10,magnelli12,casey12}, but see also e.g., \citealt{strazzullo10,bourne11,barger12,delmoro13,barger14} for different conclusions), with increasing radio luminosities for the same IR luminosity  at higher redshift.

On the other hand, simple theoretical arguments predict the opposite trend \citep[e.g.,][]{condon92,carilli08}, because of the increased inverse Compton cooling of the relativistic electron population due to scattering off the increasing cosmic microwave background at high redshift, $U_{\rm CMB} \propto (1+z)^4$.
 
We derive 1.4~GHz monochromatic luminosities from the measured radio flux densities assuming a synchrotron spectral index, $\alpha$, of --0.8 and the median redshift of each stacked subsample, as
\begin{equation}
{\sl L}_{\rm 1.4} = 1.19\times10^{14}{\rm DM}^2\times{\sl S}_{\rm 1.4}\times(1+z_{\rm med})^{-(\alpha+1)} \; ({\rm W Hz^{-1}}),
\end{equation}
\noindent 
where {\rm DM} is the distance in Mpc at the median redshift $z_{\rm med}$ of the stacked subsample, and {\sl S}$_{\rm 1.4}$ is the measured flux density in the stack in units of $\mu$Jy.  We notice that our assumption of a non-evolving spectral index is justified by most observational studies dealing with statistical samples of star-forming galaxies~(see e.g., \citealt{ibar09,bourne11}). We adopt the \citet{yun2001} relation to convert such luminosities in star formation rates:
\begin{equation}
{\sl SFR}_{\rm 1.4} = 5.9\times10^{-22}~{\sl \it L}_{\rm 1.4} \; ({\rm M_\odot yr^{-1}})~.
\end{equation}
\noindent 

\begin{figure}
\begin{center}
\includegraphics[height=.460\textwidth]{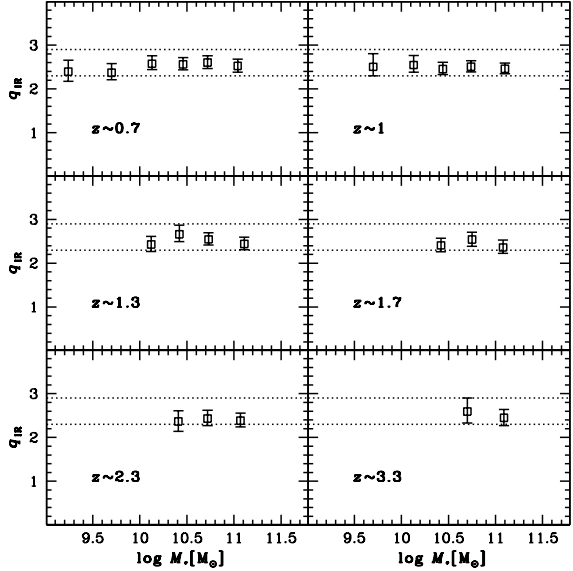}
\caption{The luminosity ratio $q_{\rm IR}$=$\log$($L_{\rm IR}/3.75\times10^{12}$) -$\log$($L_{\rm 1.4}$) as a function of redshift and for the different mass bins from stacking of mass-complete star-forming samples. The dashed lines encompass the $\pm$1$\sigma$ scatter of the local FIR-radio correlation.  We do not see any evolution of the median $q_{\rm IR}$ up to the highest redshifts probed.}
\label{qIR}
\end{center}
\end{figure}

In order to study the evolution of the correlation we compute, following \citet{helou85}, the luminosity ratio:
\begin{equation}
q_{\rm IR}=\log(L_{\rm IR}/(3.75\times10^{12}{\rm W})) - \log(L_{\rm 1.4}/{\rm W Hz^{-1}})
\end{equation}
\noindent 
for all redshift and mass bins where we have performed our stacking analysis. We show the results in Figure~\ref{qIR} together with the $\pm$1-$\sigma$ confidence locus of the local Universe $q_{\rm IR}$ value~\citep[e.g.,][]{yun2001,bell03}. We find striking agreement with the local correlation at all redshifts and masses explored. 

We stress that our result is the first so far obtained for a mass-complete uniformly selected sample of star-forming galaxies and hence is expected to be significantly more robust against selection bias, as compared to all previous analyses.    

This finding has some non-trivial implications.
Firstly, it shows that a FIR-radio correlation is well defined up to $z\simeq$4 with basically the same $q_{\rm IR}$ value measured in the local Universe, hence that radio continuum data are an ideal tool for estimating dust unbiased star formation rates at high redshift. 

Secondly, since radio luminosity is only contributed by young stellar populations, our result suggests that the IR emission of star-forming galaxies in the redshift range 0.5$< z <$4 can only have, in a statistical sense, a marginal contribution from old stellar populations and likely smaller than the 20\% estimated in the local Universe~\citep[see e.g.,][]{law11}.%Thirdly, since the timescales associated to the radio and IR SFRs are of the order of $\simeq 10$ and $\simeq 100$\,Myr~\citep[see, e.g.,][and references therein]{murphy11}, respectively, our result suggests that the star formation histories of star-forming galaxies are not strongly varying over 10\,Myr timescales in the entire redshift range 0.5--4. 

%\begin{figure}
%\begin{center}
%  \includegraphics[height=.460\textwidth]{paperaccl_13}%
%  \caption{$A_{\rm UV}$ vs. SFR$_{\rm tot}$ = SFR$_{\rm IR}$+SFR$_{\rm UV}$. The dotted line is the best fit at the $z\simeq$0.7 correlation and is only meant to guide the eye in tracing the evolution of the correlation between dust attenuation and total star formation. Confirming previous results we show here, and for the first time in a consistent way at different redshifts, that galaxies with the same star formation rate suffer less dust extinction at higher redshift.}
%\label{dust}
%\end{center}
%\end{figure}

%\begin{figure}
%\begin{center}
%  \includegraphics[height=.460\textwidth]{paperaccl_14}%
%  \caption{$A_{\rm UV}$ vs. SFR$_{\rm UV}$ without any dust correction (i.e. versus the measured rest-frame UV luminosity). Galaxies are systematically increasing their UV output with increasing redshift, according to the %fact that their star formation is increasing and the dust attenuation is decreasing. Yet, there is basically no  correlation between dust attenuation and the measured UV light.}
%\label{dust}
%\end{center}
%\end{figure}

\subsection{UV dust attenuation over cosmic time}

In order to study the redshift evolution of the UV dust attenuation, we define as an effective measurement of dust attenuation the quantity: 
\begin{equation}
A_{\rm UV} = 2.5\times \log(SFR_{\rm IR}/SFR_{\rm UV}^{\rm obs} +1)~,
\end{equation}
\noindent
i.e., the magnitude correction to the measured UV emission. This means that, by construction, we obtain\\

 $SFR_{\rm UV}^{\rm corr}$ = $SFR_{\rm total}$ = $SFR_{\rm IR}$+$SFR_{\rm UV}^{\rm obs}$.\\

For each subsample stacked in each redshift and mass bin, we estimate the median FUV emission and UV spectral slope (see equation~\ref{betaeq}).

Assuming the Calzetti et al.~(2000) empirical correlation between UV dust attenuation and slope $\beta$:
\begin{equation}
A_{\rm UV}^{\rm \beta} = 4.85 + 2.31\times \beta~,
\end{equation}
\noindent
we then define the quantity
\begin{equation}
 SFR_{\rm UV_{\rm}^{\beta}} = SFR_{\rm total}^{\rm \beta} = SFR_{\rm UV}^{\rm obs} \times 10^{A_{\rm UV}^{\rm \beta}/2.5}
\end{equation}
\noindent
which provides an estimate of the total star formation rate by correcting the observed UV luminosity assuming the measured $\beta$ slope and the Calzetti et al.~(2000) law, a standard prescription adopted for high redshift studies. 

In the following we will discuss how the measured UV dust attenuation, $A_{\rm UV}$, depends on galaxy properties, such as the stellar mass and the UV spectral slope over the redshift range 0.5--4.

%To derive median rest-frame UV properties, namely the FUV emission and the UV spectral slope, of our galaxy sample we stack, in consistent way for the same mass and redshift bins used for the FIR data, at all bands in our dataset i.e. from {\it GALEX} NUV to the CFHT WIRCAM K band data which allows to exactly map the median UV restrame galaxy properties over the redshift range used in this study. In this way median SEDs have been built for each bin of mass and redshift.  We follow the approach outlined above for the stacking procedure and extract total fluxes in all bands using the SExtractor software (Bertin\&Arnouts 1996). We then run{\it EAZY}  to consistently derive rest-frame magnitudes in the FUV and NUV {\it GALEX} bands and hence obtain an estimate of both the total FUV luminosity and the UV spectral slope index, $\beta$. Finally we run the FAST code on the median SEDs with a constant star formation history library and allowing a varying dust attenuation according to the Calzetti et al.~(2000) recipe to derive the stellar continuum dust reddening. The SED-fitting of a dust attenuated constant SFH has been widely used in the last years (e.g., Daddi et al.~2004, Wuyts et al.~2010) to obtain more reliable estimates of the ongoing star formation rates in star-forming galaxies from a standard SED-fitting approach as compared to more standard exponentially declining SFH and in general in all cases where any dust unbiased tracer is not available.

\begin{figure*}
\begin{center}
  \includegraphics[height=.460\textwidth]{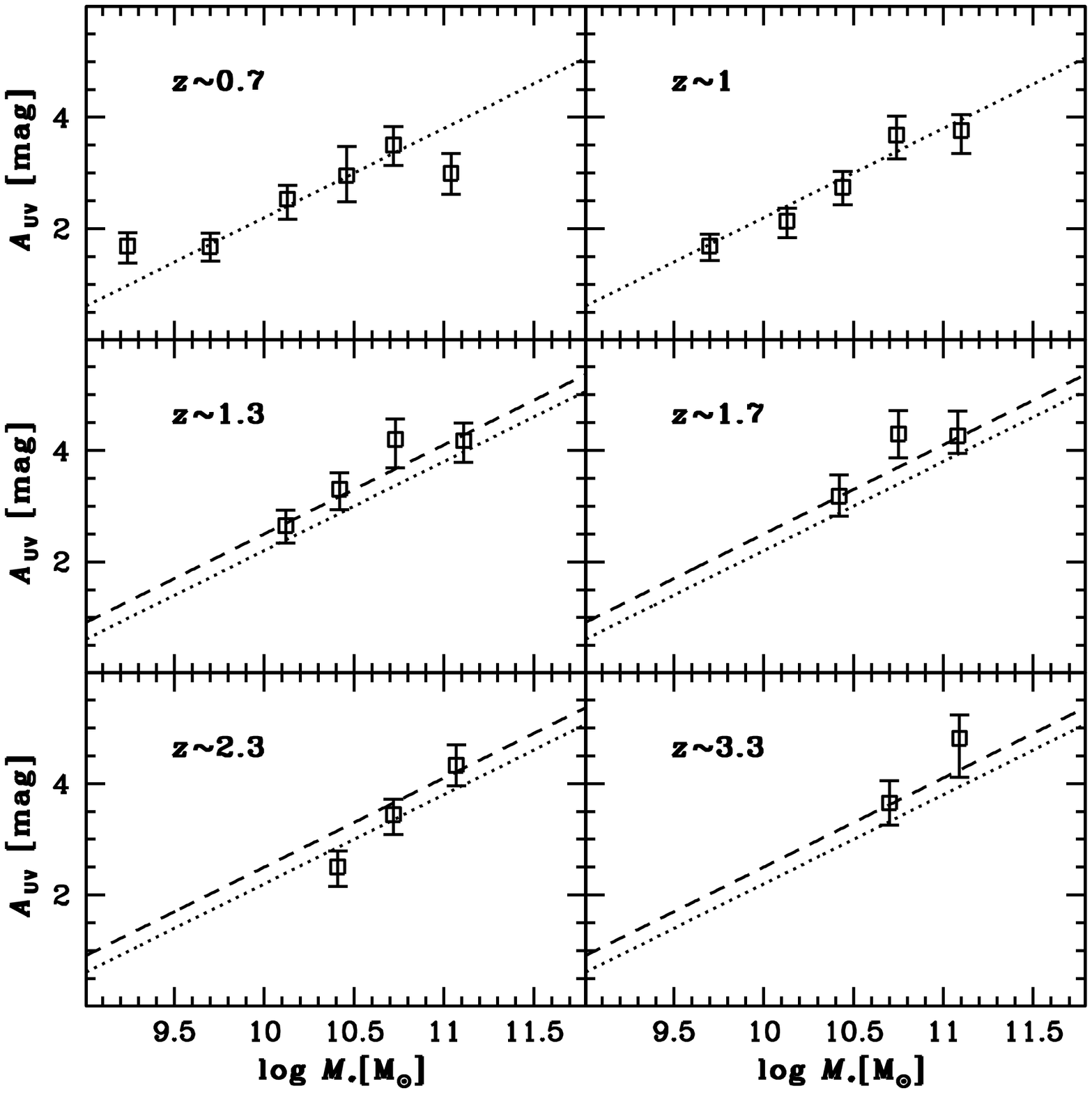}%
  \includegraphics[height=.460\textwidth]{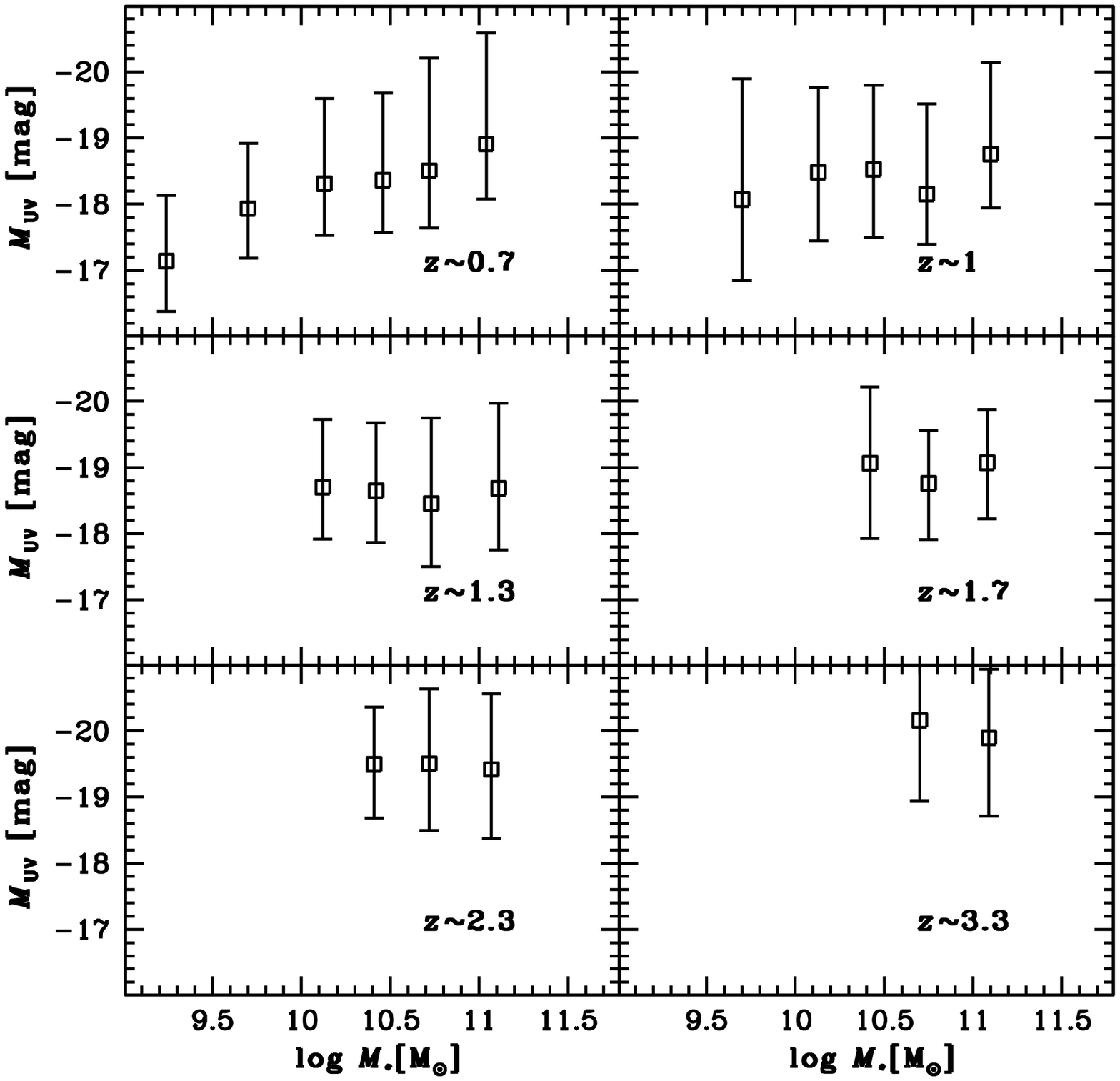}
  \includegraphics[height=.460\textwidth]{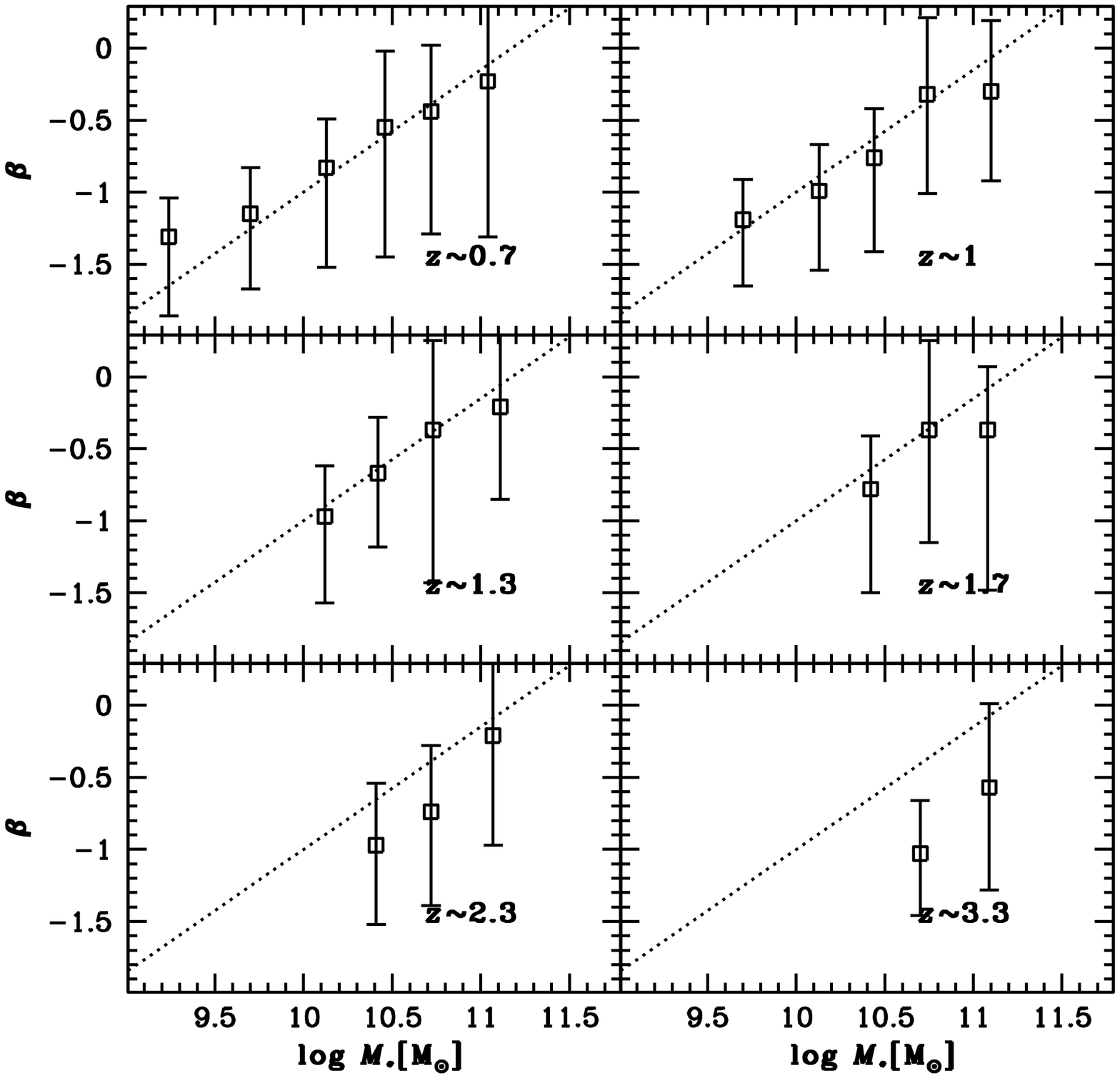}
  \includegraphics[height=.460\textwidth]{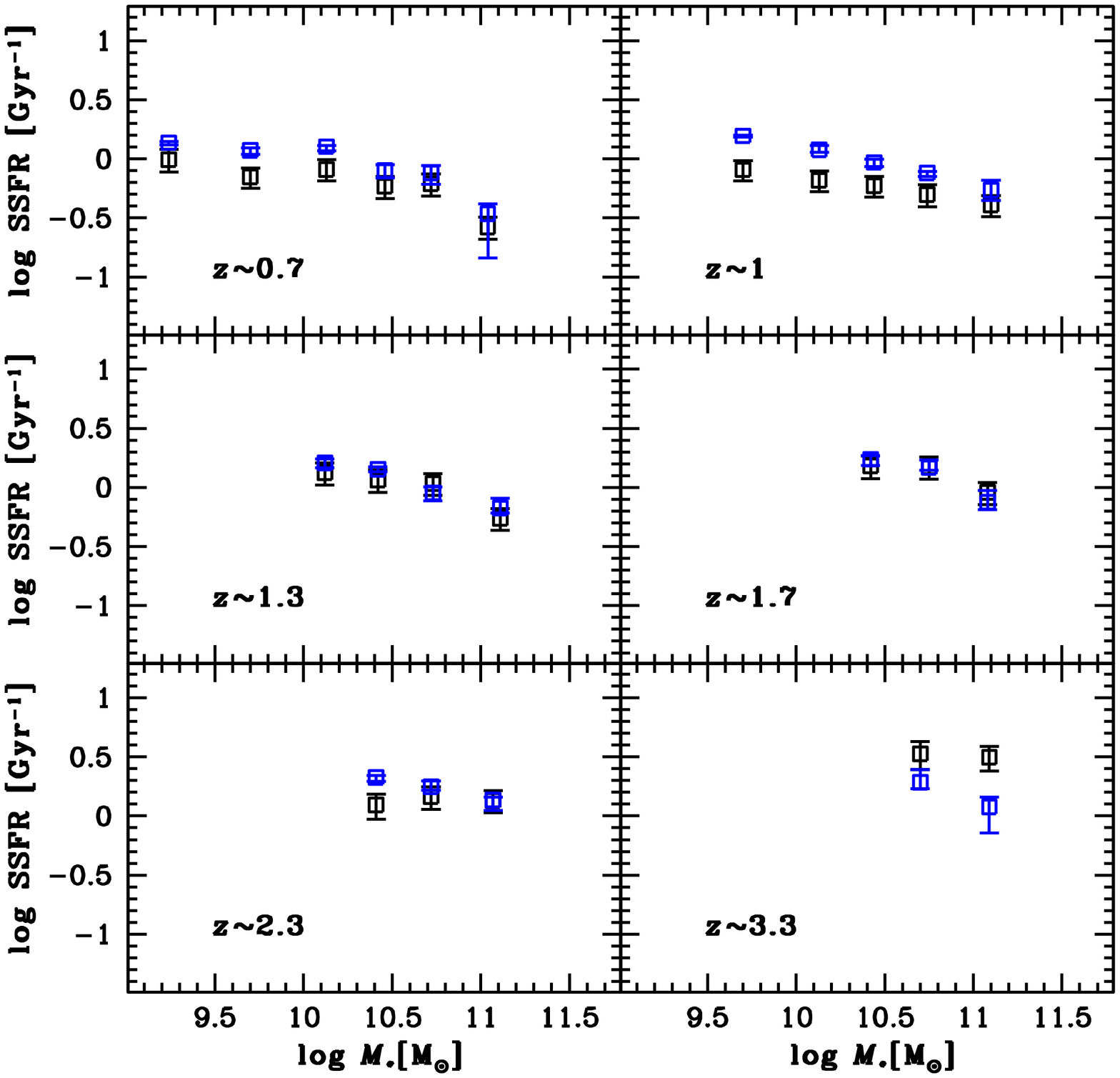}
  \caption{{\bf Top-left:}~Measured $A_{\rm UV}$ vs. $M_*$ correlation at different redshifts and for mass-complete bins. The dotted (dashed) lines are the linear fit to the measured values in the redshift range 0.5--1.2 (1.2--4.0).~As described in the text, above redshift 1 we see a higher normalization by only $\simeq$0.3 mag of attenuation at a fixed stellar mass. Using the \h~detections from Elbaz et al.~(2011) we estimate a dispersion in the correlation of $\simeq$1 magnitude~{\bf Top-right:} Median absolute magnitudes $M_{\rm UV}$ for each bin of stellar mass where stacking has been performed. Galaxies with very different stellar masses, and hence very different star formation rates, emit about the same dust attenuated UV light in average. Error bars show the rms of UV absolute magnitude distributions rather than the error on the median values.~{\bf Bottom-left:} Median UV spectral slope, $\beta$, for each bin of stellar mass where stacking has been performed. The dotted line is the best fit to the lowest redshift bin and is kept constant in all panels in order to show the evolution of the UV slope at a fixed stellar mass: the median $\beta$ slope becomes bluer with redshift by about 0.5 over the whole redshift range, but most of the evolution seems to happen at $z\ge$2. Error bars represent the rms of $\beta$ distributions in the stellar mass bins.~{\bf Bottom-right:} Comparison between the IR (black) and UV-corrected (blue) derived SSFRs as a function of redshift and stellar mass. At $z\ge3$ SFRs derived from the UV corrected emission underestimate the real SFR by a factor of around 2.}
\label{mass1}
\end{center}
\end{figure*}

\subsubsection{$A_{\rm UV}$ vs. $M_*$}
\label{auvmsec} 

In the top-left panel of Figure~ \ref{mass1} we show the correlation between dust attenuation and galaxy stellar mass. The correlation evolves only marginally, less than 0.3 magnitude in $A_{\rm UV}$, over the redshift range explored:

%\begin{equation}
% A_{\rm UV} = 1.6 \times \log M_* - \begin{cases} 13.9~~\textrm{mag}~~\textrm{for}~z=0.5-1; \\
% 13.3~~\textrm{mag}~~\textrm{for}~z=1-4. \end{cases}
%\label{auvm}                   
%\end{equation}

\begin{equation}
 A_{\rm UV} = 1.6 \times \log M_* - 13.5~~\textrm{mag}~~\textrm{for}~z=1.2-4.
\label{auvm}                   
\end{equation}

The stellar mass content of a star-forming galaxy turns out to be a robust proxy of the dust attenuation affecting its UV emission \citep[see, e.g.,][]{PP09,schaerer10,buat12,p13,kashino13,heinis14,oteo14}. Previous results focused on a relatively small redshift range, and sometimes on a UV selected sample, which, by construction, tends to be biased against the most massive and most obscured star-forming galaxies. Here we test this correlation over a wide redshift range and in a mass-complete way. Testing the real dispersion of a correlation obtained through stacking is not trivial and often impractical. To at least obtain an estimate of the correlation scatter we have used all \h~detected sources in the field (Elbaz et al.~2011) up to $z\simeq$1.3, where the GOODS-\h~data are deep enough to allow a good statistical description of the parent sample, and  estimate a dispersion of $\simeq$1 magnitude in $A_{\rm UV}$.

The mild evolution of this correlation lends support to a number of earlier studies claiming that the same amount of star formation suffered less dust extinction at high redshift compared to the local Universe \citep[e.g.,][]{reddy12,buat07}. Because of the evolution of the MS with redshift, the same amount of star formation is hosted in galaxies which are less and less massive as redshift increases, and hence they will suffer a correspondingly lower UV dust extinction. 

On the other hand, dust attenuation is known to correlate with both the ISM metallicity and the mode and geometry of star formation in the host galaxy \citep[i.e., normal vs. starbursting galaxies, e.g.,][]{cortese06,heck98}. The ISM metallicity is expected to decrease with increasing redshift at a fixed mass, according to the well-studied evolution of the mass-metallicity relation \citep[e.g.,][]{tremonti04,savaglio05,erb2006,zz13, steidel14,wuyts14}.By assuming that the ISM conditions of star-forming galaxies do not change with redshift, one would see an effective decrease with redshift of UV attenuation at a fixed stellar mass: if anything we measure a weak trend going in the opposite direction. This suggests that the ISM conditions of MS galaxies are evolving, becoming somehow more extreme with redshift, which might be understood as a combination of both geometrical effect (star-forming galaxies are becoming smaller with redshift) and physical effects~(gas fractions and SSFRs are also increasing), leading to an overall enhanced volume density of the UV radiation field, as also suggested by their higher dust temperature \citep[e.g.,][]{hoseong10,magdis12,symeo13}. We will come back to this topic in Section 5. 

In the top-right panel of Figure~\ref{mass1} we plot the median UV luminosities for the same stacking samples as in the top-left panel. We show that at all redshifts explored, there is basically no correlation between stellar mass - a robust proxy of the ongoing star formation and UV dust attenuation - and the emerging UV photons. The amount of UV radiation which escapes dust absorption rises systematically with redshift, mimicking the rise of the general star formation rate level. This is consistent with a number of previous studies \citep[e.g.,][]{PP09,buat12,heinis13}. Here we explore a wider redshift range and find basically the same result: the emerging UV light is not (or only marginally in the lowest redshift bin) correlated to the actual SFR present in a galaxy, because the tight correlation between  stellar mass (a proxy for the SFR, i.e., the intrinsic UV luminosity) and the dust attenuation conspires so that the emerging, average UV light is basically the same for very different SFR levels. 

In the bottom-left panel of Figure~\ref{mass1} 
we plot the median values of the galaxy UV spectral slope, $\beta$, per bin of stellar mass at different redshifts. The two quantities, $\beta$ and $M_*$, clearly correlate at all redshifts, but while the attenuation is fairly constant, or slightly increasing with redshift,  $\beta$ values are becoming systematically bluer with redshift. The fact that high mass galaxies at high redshift have a similar dust attenuation compared to similar mass galaxies at lower redshift, but have bluer UV slopes, has important implications for UV-derived SFRs in the high redshift Universe. Using standard recipes to correct the observed UV emission by means of the $\beta$ slope would systematically underpredict galaxy star formation rates. This is better quantified in the bottom-right panel of Figure~\ref{mass1} where we compare the MS evolution plot derived from \h~data~(black squares, as in Figure. 4) to the one derived from UV data which have been corrected by dust attenuation according to equation 9. The comparison shows three main features mainly dependent on redshift: 1) at redshift lower than $\simeq$1, the SFR$_{\rm UV_{\rm}^{\beta}}$ tends to overpredict the real SFR, possibly because old stellar populations present in star-forming galaxies contribute to create redder UV slopes, which is in agreement with previous results (see, e.g., Kong et al.~2004 and Overzier et al.~2011); in redshift range 1.5--2.5 the SFR$_{\rm UV_{\rm}^{\beta}}$ provides a nice match to the real SFR which is in agreement with the results of Daddi et al.~(2007), Pannella et al.~(2009a) and more recently Rodighiero et al.~(2014); at redshift $z \ge 3$, SFR$_{\rm UV_{\rm}^{\beta}}$  are systematically lower than the real SFR and underpredicting the SFR by a factor of around 2. Intriguingly enough, this underprediction of SFR conspire to create a plateau of the UV derived SSFR values at $z\ge2$ which was indeed observed in previous studies~\citep[see, e.g.,][]{daddi2009,stark09,gonzales10}. Lastly, the underestimate of SFR has an impact on the estimates of the global star formation rate density (SFRD) at $z\ge3$, which were obtained by applying standard recipes as in equation 9, that should be revised upward by a factor of around 2. A similar conclusion was already presented in de Barros et al. (2014) and Castellano et al.~(2014) for a sample of low mass Lyman-break selected galaxies. We extend their result to the high mass end of the star-forming galaxy population at high redshift, putting forward a more compelling case for an upward revision of the SFRD estimates obtained so far at redshift $z\ge3$ .  We will further discuss this point in the next section.

\subsubsection{$A_{\rm UV}$ vs. $\beta$ slope}

In this section we explore in more detail the correlation between UV dust attenuation and UV spectral slope, by repeating the stacking procedure in bins of $\beta$ slope. This correlation is possibly the most widely used method in the literature to derive UV dust attenuation and hence SFR from UV emission. Daddi et al.~(2004) calibrated it for a sample of star-forming galaxies at $z\simeq$2 using the observed $B-z$ color, which is a proxy for the UV spectral slope, and this calibration has been shown to behave well in a number of studies (e.g., Daddi et al  2007, Pannella et al 2009a, Rodighiero et al. 2014). Previously, \citet{meurer99} had shown that local starburst galaxies and $z\simeq$3 Lyman Break Galaxies (LBGs) were following the same tight correlation between dust attenuation and UV spectral slope. After Meurer et al.~(1999), numerous investigations have tried to understand if the so-called Meurer law would be followed by galaxies at all redshifts. Different studies have reported contradicting results on this topic. One thing that has become clear within a few years after the Meurer et al.~(1999) study was, for example, that local normal star-forming galaxies do not follow the Meurer relation but have ``redder'' spectra for the amount of dust attenuation they suffer (e.g., Kong et al.~2004, Buat et al.~2005, Overzier et al.~2011, \citealt{boquien12} and \citealt{grasha13}).

\begin{figure*}
\begin{center}
  \includegraphics[height=.460\textwidth]{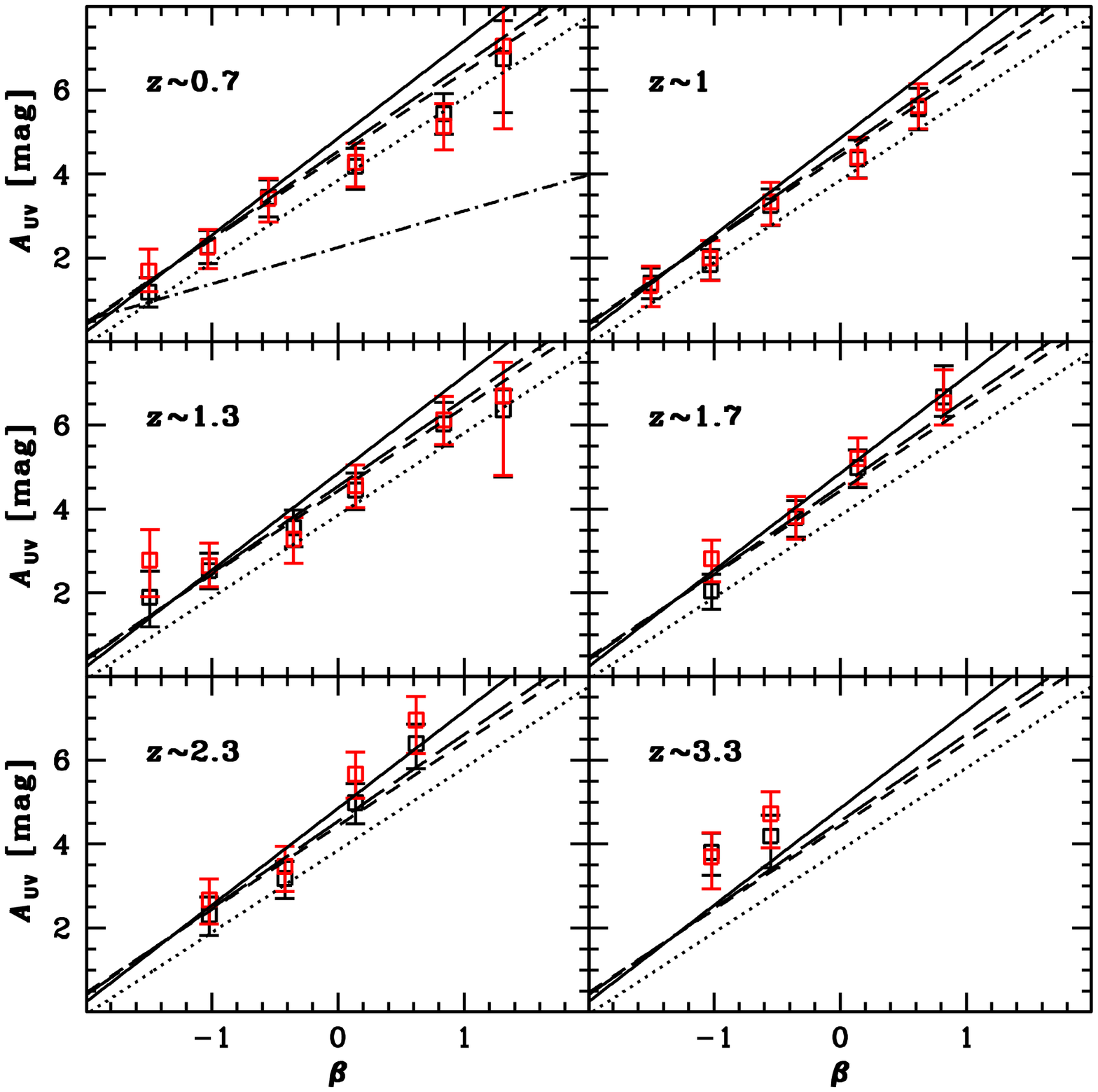}%
  \includegraphics[height=.460\textwidth]{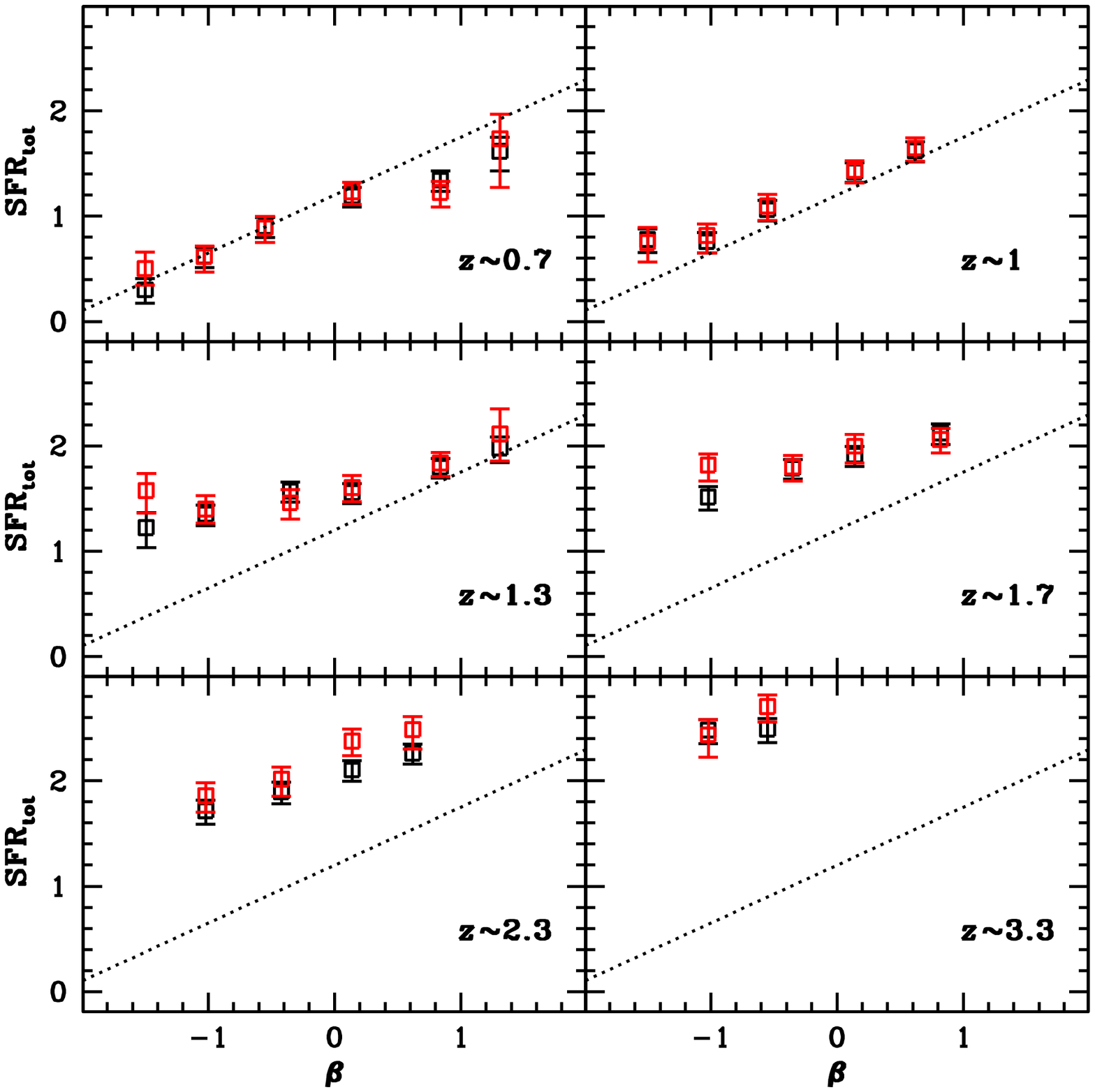}
  \includegraphics[height=.460\textwidth]{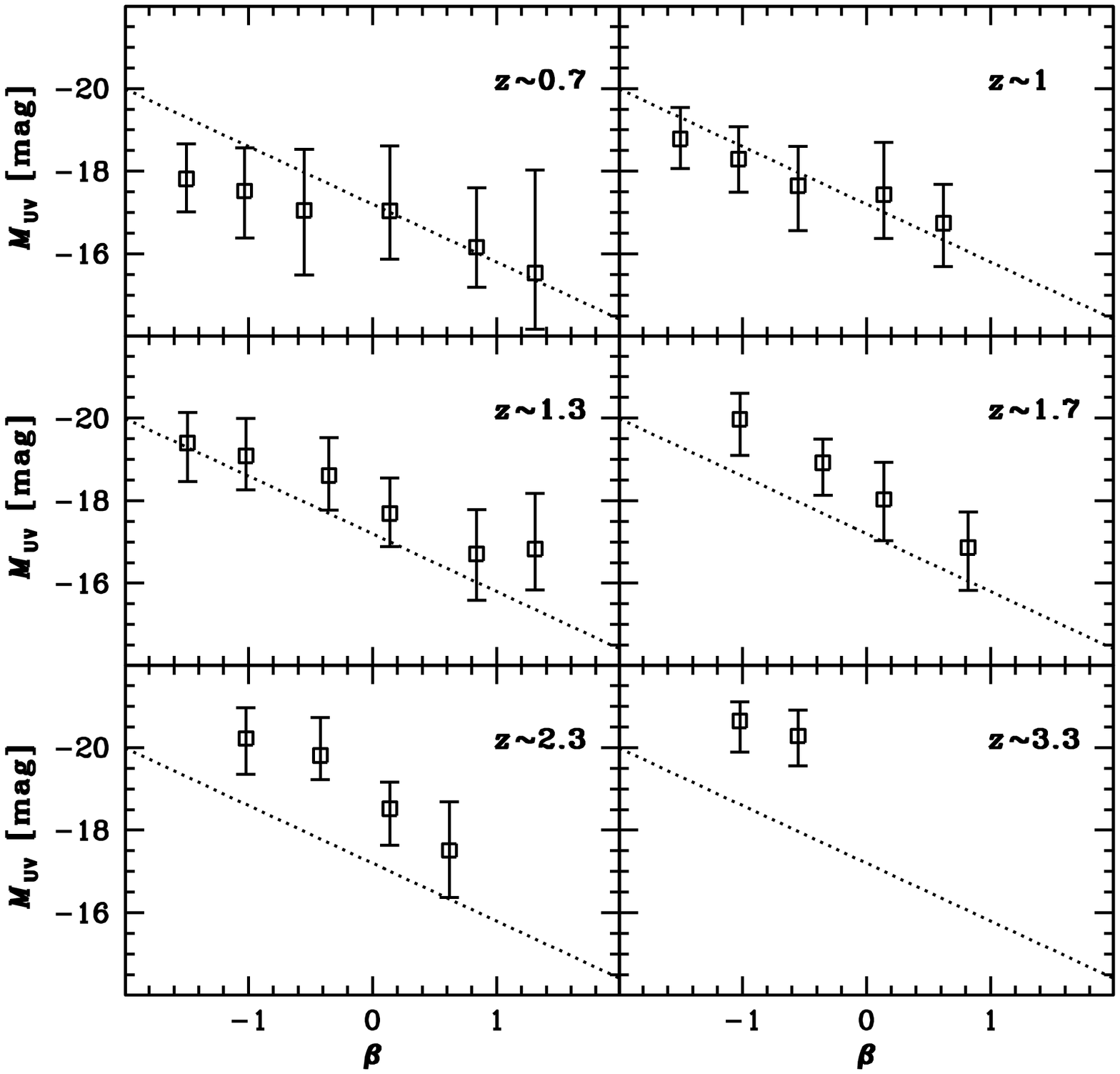}
  \caption{{\bf Top-left:}~ Measured $A_{\rm UV}$ vs. $\beta$ correlation at different redshifts for the whole sample of star-forming galaxies in the GOODS-N field. The $A_{\rm UV}$ measurements obtained from the stacked IR GOODS-H and radio VLA data are plotted with black and red squares. The predictions of different attenuation laws from Calzetti et al.~(2000), Meurer et al.~(1999), Daddi et al.~(2004), and Overzier et al.~(2011), are shown with solid, long dashed, dashed, and dotted lines, respectively. Only in the lowest redshift bin, at $z\sim$0.7, we show with a dot-dashed line the relation found for local star-forming galaxies by Boquien et al.~2012, see text for more details.~{\bf Top-right:} Total SFR from stacking the IR and radio data in bins of UV slope and redshifts. The dotted line is the fit to the lowest redshift bin stacking result and is kept identical in all panels to show the evolution in the correlation between the two quantities. At a fixed value of $\beta$, i.e. at fixed UV dust attenuation (to within a factor of 2 accuracy), the star formation rate grows by more than an order of magnitude.~{\bf Bottom:} Median UV absolute magnitude in the bins of $\beta$ used to produce the stacking results. A negative correlation is found, which points toward higher attenuation for galaxies with a fainter measured UV. Error bars show the rms values of the UV luminosities in the bins of $\beta$.}
\label{beta1}
\end{center}
\end{figure*}

We now investigate how the median UV spectral slope relates to dust attenuation. We plot at all redshifts the predicted correlations between dust attenuation and $\beta$ slope according to Meurer et al.~(1999), Calzetti et al.~(2000), Daddi et al.~(2004) and Overzier et al.~(2011)  in the top-left panel of Figure~ \ref{beta1}. The first three determinations overlap to a large extent at blue UV slopes, while they spread out at red slopes. The determination of Overzier stands out, clearly predicting redder slopes at a fixed attenuation. However, we remind the reader that the sample used in Overzier et al.~(2011) was strictly based on blue star-forming objects with no data-points having a UV slope redder than beta -0.5. For this reason the derived relation~(R. Overzier, private communication) should not be used for very attenuated galaxies.

When binned in $\beta$ values, the median dust attenuation of the MS galaxy population follows the prediction of the Calzetti law already in our lowest redshift bin with a possible hint of a dust attenuation overprediction only in the reddest slope bins and up to redshift $z\simeq$1.  This latter feature might be due to old stellar populations of massive galaxies already contributing partially to the measured red slope. The departure from the Meurer relation of normal star-forming galaxies in the local Universe has been thoroughly investigated in the last years and it is now well accepted to explain it as an age effect of the stellar populations hosted in local galaxies (see e.g., Kong et al.~2004, \citealt{boquien12} and \citealt{grasha13}). In the lowest redshift bin, we show, for illustrative purpose, the relation obtained by Boquien et al.~(2012) for a sample of 7, face-on and spatially resolved, local star-forming galaxies drawn from the \h~Reference Survey.

Already at $z\simeq$1.3 and up to $z\le$3 the correlation matches well the observed data points, agreeing with the results of Daddi et al (2007), who found the UV slope derived SFR to be in good agreement with those derived from 24\,$\mu$m and radio continuum data. In the last redshift bin, between redshift 3 and 4, we see that UV dust attenuation tends to overshoot the prediction of the Calzetti law, so that  galaxies tend to have bluer UV slope (an offset of about -0.5), compared to that predicted by the Calzetti law at a fixed amount of dust attenuation. These results are overall in very good agreement with the ones discussed in the previous subsection.

There are several possible explanations for such a finding. At face value, our result seems to support the suggestion of \citet{maiolino04} that high redshift galaxies have a grayer attenuation law than predicted by the Calzetti law. More recently, \citet{cast12,cast14}, \citet{alavi14} and \citet{debarros14}  have suggested that the bluer slopes can be explained by galaxy stellar populations with young ages and/or low metallicities. On the other hand, \citet{buat12,kriek13} have suggested that the presence of a dust bump, the feature at 2175\angstrom, in the spectra of high redshift galaxies, may result in underestimated values for the UV slope. Finally, blue spectral slopes at high redshift have been reported previously by a number of authors as the possible outcome of a measurement bias when estimating slopes directly from the observed photometry rather than, as we actually do in this study, using the best fit SED \citep[see, e.g.,][]{buat11,bouwens12,finkelstein12}. 

 We remind the reader that our $\beta$ estimate is directly proportional to the difference between two rest-frame magnitudes and as such one might wonder about the possible impact of the shifts applied to the photometric catalog on our conclusions. The only sensible difference we find by using the uncorrected photometry is in the first redshift bin. The uncorrected photometry would produce redder slopes ($\Delta \beta \sim$0.3)  that would possibly move our dust attenuation further away from the Meurer relation and possibly closer to the correlation found in the local Universe (see e.g., Boquien et al. 2012). All the other redshift bins remain unchanged with possibly a weak trend, but well within the estimated $\beta$ uncertainty, to obtain slightly bluer values compared to the ones we used. If anything this would reinforce our conclusions.

In the top-right panel of Figure~\ref{beta1} we show the evolution of the correlation between $\beta$ slope and total star formation rate at different redshifts. In all panels the dotted line is the best fit to the correlation in the lowest redshift bin, and is kept constant at all redshifts to aid comparison. As redshift increases the amount of star formation rate at fixed UV slope increases steadily. This can be interpreted as further evidence that at a fixed star formation rate there is less and less UV dust attenuation with increasing redshift.

In the bottom panel of Figure~ \ref{beta1} we plot the median measured UV absolute magnitudes from stacks in bins of $\beta$. 
%The analysis of the correlation between the measured UV luminosity and the dust attenuation is 
%is a basic diagnostic plot in 
%This can be see a new insight in the long after-sought correlation between the measured UV luminosity and dust attenuation, i.e. in trying to understand if it is possible to correct the observed UV light for dust attenuation and derived intrinsic UV luminosities and to finally build meaningful UV luminosity functions and cosmic star formation densities. 
Binning in $\beta$ values, again basically in UV dust attenuation, we find an anti-correlation between measured UV luminosities and UV slopes: galaxies with redder UV slopes, i.e., that are more dust attenuated, tend to have fainter UV measured luminosities and vice versa galaxies that are less attenuated have brighter luminosities. Pannella et al.~(2009a) found a similar result at $z\simeq$1.7, here we can extend this at all probed redshifts (see also Heinis et al.~2013 and 2014 for similar conclusions). 

Overall, the anti-correlation is consistent with an identical slope at all redshifts, but with an increasing normalization, with the UV-observed luminosity increasing by about 2 magnitudes (more than a factor 6 in luminosity) over the redshift range probed. 

Reddy et al.~(2009) found that low luminosity objects are less attenuated, statistically speaking, than more UV luminous ones. This goes in the opposite direction to that we found here.
%The anti-correlation found here seems indeed at odd with the positive correlation between dust attenuation and stellar mass discussed in the previous section. More massive galaxies are intrinsically more UV bright and also more dust attenuated, because of this it is reasonable to expect these objects to be very faint in the UV and even fainter than a much lower mass galaxy. 
%The point here is at the other end of the stellar mass function where galaxies are low mass, low star-forming and hence with a faint UV output and also  very low dust attenuation. These latter objects outnumber by order of magnitudes the high mass galaxies so
The difference between our result and the study of Reddy et al.~(2009) might easily be explained by the different selection bands used to build the galaxy samples. We use a $K$ band selection, which is very close to a stellar mass selection at these redshifts and is therefore sampling the mass function in a fairly complete way at the high mass end. Massive galaxies are intrinsically UV-bright but also extremely dust attenuated, so it is reasonable to expect these objects to be very faint in the UV and even fainter than lower mass galaxies. On the other hand, the LBG-like selection of the Reddy et al.~(2009) sample, which is mostly an observed frame UV selection, will sample the mass function in a sparse way, giving more weight toward the low-to-intermediate range of the stellar mass function, thereby missing almost completely the high mass end as well as very obscured galaxies, a very well known drawback of UV-selected samples. These UV-selected low mass galaxies preferentially have low star formation rates, low dust attenuation and faint UV output and they outnumber by several orders of magnitude the high mass galaxies.

\section{The evolving ISM of star-forming galaxies\\ over cosmic time}

In this section we will discuss our results and investigate how the interstellar medium of star-forming galaxies and  the physical conditions of star formation  evolved with redshift. The main ingredient we will use in this section is the fact that the correlation between UV dust attenuation and galaxy stellar mass remains fairly constant with redshift. Notwithstanding a 10-fold increase in star formation rate, indeed over more than 6 Gyrs (0.5$\le z \le$4), on average a galaxy of a given stellar mass suffers only a slight increase of dust attenuation. We measure at most an increase of about 0.3 magnitudes which corresponds to a factor 1.3 in UV flux attenuation. We will complement this non evolving $A_{\rm UV}$-$M_*$ relation with other measured scaling relations that describe the conditions of the star-forming galaxy ISM and finally  derive some evolutionary trends and conclusions. 

\begin{figure}
\begin{center}
  \includegraphics[height=.460\textwidth]{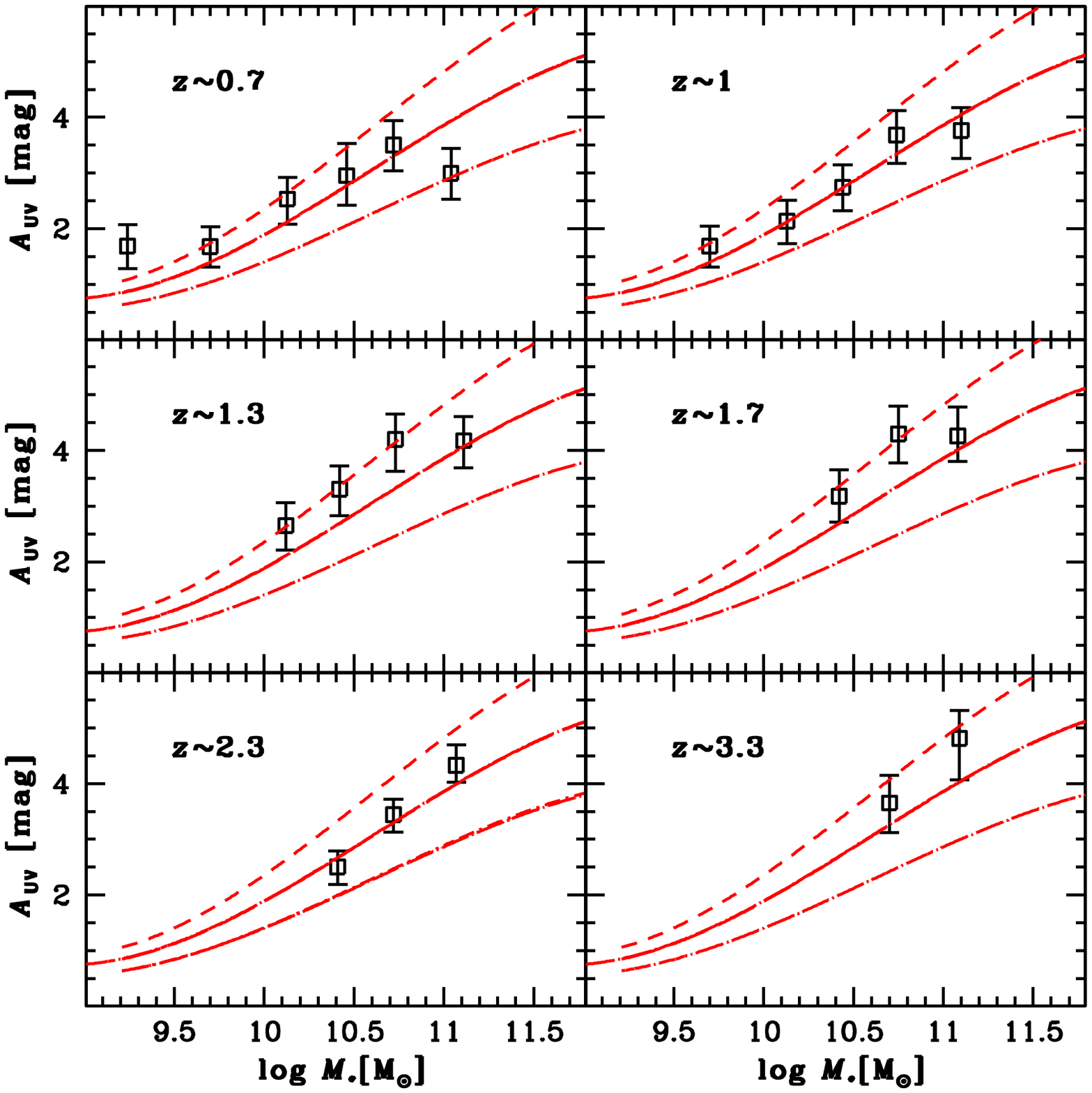}%
  \caption{Comparison between the continuum and line attenuation as a function of stellar mass and redshift. Black squares show our stacking measurements of the UV dust attenuation as a function of mass and redshift, while the red lines (dashed, solid, dot-dashed) represent the \citet{garn10} relation between H$_\alpha$ attenuation and stellar mass, which has been extrapolated to the UV rest-frame assuming the Calzetti et al.~(2000) attenuation law and scaled by 1,1.3 and 1.7 respectively. The latter value (1.7) is the factor found by \citet{calzetti2001} in the local Universe. }
\label{line}
\end{center}
\end{figure}

%and hence the dot-dashed line would represent our presently best guess for where the continuum $A_{\rm UV}$-$\log$$M_*$ correlation in the local Universe should be.

\subsection{The attenuation of continuum vs. line emission}

\ha~line emission is a well known and robust SFR tracer (e.g., Kennicutt 1998). Lying at 6563\angstrom, \ha~suffers much less dust extinction  compared to the UV rest-frame of a galaxy spectrum, and hence the uncertainty on SFR due to the poorly known dust attenuation can be an order of magnitude less severe as compared to UV based estimates. On the other hand, \ha~is mainly produced by extremely massive ($\ge$10\,M$_\odot$), short lived ($\simeq$ 10\,Myr) stars that are deeply embedded in the giant molecular cloud (GMC) \hii~regions, while stars producing the UV rest-frame stellar continuum are less massive ($<$10M$_\odot$), shine over time scales ten times longer, and have time to migrate out of the dense \hii~regions. The net outcome of this process is that  \ha~emission suffers from an extra attenuation that has, as we just described, an origin driven by the distribution and density of \hii~regions within the galaxy itself. 

%The extra amount of reddening suffered by \ha~emission has been quantified in the local Universe by Calzetti et al.~(2000) to be a factor of 1.7 that can be easily derived by using the equation 13 in Calzetti et al.~(2001)\footnote{The canonical value quoted in Calzetti et al.~(2000) for the extra reddening suffered by the nebular emission is in fact 2.3 (1/0.44) as opposed to the 1.7 we quote. The difference between the two values comes from the fact that Calzetti et al.~(2000) used two different extinction curves for the nebular~\citep{fitz1999} and continuum~(Calzetti et al. 2000) emission which have very similar shapes but different normalizations, i.e., $R_{\rm V}$=3.1 and 4.0, respectively, while we use the same Calzetti et al.~(2000) extinction curve for both emissions.}.  

 The extra amount of attenuation suffered by nebular emission has been quantified in the local Universe by Calzetti et al.~(2000) to be a factor of about 1.7. This latter can be derived by accounting for the fact that the stellar continuum suffers roughly one-half of the reddening suffered by the ionized gas, encoded into E(B-V)$_{star}$=0.44\,E(B-V)$_{gas}$, and for the fact that Calzetti et al.~(2000) used two different extinction curves for the nebular~\citep{fitz1999} and continuum~(Calzetti et al. 2000) emission\footnote{We refer the reader to the review paper of Calzetti~(2001) where all the details about the actual assumptions are extensively discussed.} which have similar shapes but different normalizations, i.e., $R_{\rm V}$=3.1 and 4.0, respectively and thus for example:

\begin{equation}
\frac{{A_{\rm V}}_{gas}}{{A_{\rm V}}_{star}} = \frac{{R_{\rm V}}_{gas}}{{R_{\rm V}}_{star}} \frac{E[B-V]_{\rm gas}}{E[B-V]_{\rm star}} \sim 1.7
\end{equation}

We will be assuming here that this ratio is constant in wavelength. Altough this is obviously not strictly true, because it assumes that the two mentioned extinction curves have exactly the same shape and only differ in their normalizations, it turns out not to be a bad approximation at the wavelengths of interest for this study, namely at the FUV~(1500\angstrom) and \ha~(6563\angstrom) wavelengths.

At higher redshift, things are much less constrained. \citet{erb2006} and \citet{reddy2010} claimed a vanishing diference, i.e. a factor close to 1, at $z\simeq$2 but this has not been confirmed by more recent studies \citep[e.g.,][]{fs09,mancini11,wuyts13}. 

The main problem for  high redshift studies is the lack of a robust star formation rate indicator to be directly compared to the \ha~and UV stellar continuum measurements. Most of the quoted studies rely in fact on indirect arguments, by estimating the star formation rate  from SED-fitting or from the $\beta$-corrected UV luminosity. This approach is, by construction, extremely uncertain and model dependent, and the end product of the comparison is closer to a correction factor, that produces in the end a consistent result, rather than a real physical measure. 

In this study we are able to look at the redshift evolution of the extra attenuation affecting \hii~regions in a way that is relatively model-independent. The idea is to compare the $A_{\rm UV}$-$M_*$ relation we have found in section \ref{auvmsec} to the one that similarly relates $A_{\rm \ha}$ to $M_*$ as a function of redshift. The latter relation was first established in the local Universe by \citet{garn10}, and more recently confirmed by \citet{zahid13}, by using the observed Balmer decrement, i.e. $H_\alpha$/$H_\beta$, measured in SDSS spectra as previously done in \citet{brinchmann04}.   

In the last few years several different studies have looked at the redshift evolution, at least up to $z\simeq$1.5, of the $A_{\rm \ha}$--$\log$$M_*$ relation \citep[e.g.,][]{sobral12,doming12,ibar13,kashino13,momcheva13,price13}. Compared to the SDSS studies mentioned above, these studies are affected by additional uncertainties and systematics mainly due to the fact that the \ha~line shifts into the observed near-infrared for z$\ge$0.5 and this often prevents a robust measurement of the Balmer decrement. Despite the diverse caveats, most of the results obtained so far, up to $z\simeq$1.5, suggest that there is no redshift evolution of the $A_{\rm H_\alpha}$-$\log$$M_*$ correlation \citep[see, e.g., Figure~ 6 in ][for a compilation of available results]{price13}. 

In order to investigate the nebular exinction we start from the $A_{\rm H_\alpha}$-$\log$$M_*$ correlation defined in \citet{garn10}:
\begin{equation}
 A_{\rm H_\alpha} = 0.91 + 0.77\cdot X + 0.11\cdot X^2 -0.09\cdot X^3
\end{equation}
\noindent
where the quantity X = ($\log$$M_*$ - 10 + 0.23) is the stellar mass term in units of 10$^{10}$M$_\odot$, which has been rescaled from the \citet{chabrier} IMF used in \citet{garn10} to the \citet{salpeter} IMF we are using in this study. We then assume a Calzetti et al.~(2000) attenuation law to derive the line reddening $E[B-V]_{\rm gas}$ as
\begin{equation}
 E[B-V]_{\rm gas} = A_{\rm \ha}/3.33
\end{equation}
\noindent
and the dust attenuation for nebular emission in the rest-frame UV as 
\begin{equation}
 A_{\rm UV - gas} = 10.4\,E[B-V]_{gas}~.
\end{equation}
\noindent

In Figure~ \ref{line} we overplot the rescaled \citet{garn10} relation divided by different scaling factors (1.7, 1.3 and 1) to the measured $A_{\rm UV}$-$\log$$M_*$ relation at different redshifts. The factor 1.7 corresponds to the value found by Calzetti et al.~(2000) in the local Universe. We find that this factor is clearly smaller at higher redshift, being 1.3 at $z\simeq$1 and becomes close to unity at higher $z$, i.e. there is a vanishing difference between the reddening of the \hii~regions and that affecting the young massive, but slightly older stars responsible for the rest-frame continuum UV emission. This can be understood through the fact that, on the one hand, galaxies are becoming slightly smaller in physical size (e.g., \citealt{ferguson2004}, Elbaz et al. 2011, Pannella et al. in preparation) and at the same time their star formation rate density dramatically increases from the local Universe out to $z\simeq$4, by more than a factor of 40 according to the evolution of the normalization of the $\log$$M_*$-$\log$SFR correlation. For this reason the volume density of \hii~regions is becoming so high as to almost entirely fill up the available galaxy surface. This suggests that the ISM of star-forming galaxies was more and more dense and opaque to UV radiation with increasing redshift, eventually resembling, at $z\ge$1.5, the physical properties of local \hii~regions, as it has already been suggested by \citet{reddy2010}. More recently \citet{wild11}, by analysing a sample of 15000 galaxies, drawn from the Sloan Digital Sky Survey, have found a correlation between the galaxy SSFR and the ratio between line and continuum reddening such that the ratio becomes smaller with increasing SSFR, i.e., as galaxies move toward the starburst regime locus. The correlation found by \citet{wild11} nicely explain our findings and also support our conclusions. 

 On the other hand \citet{kreckel13} have shown, for a sample of 8 nearby spatially resolved star-forming galaxies, that the ratio between line and continuum reddening increases with increasing SFR density which seems in contradiction with the Wild et al.~(2011) result and hence our conclusions. Very recently, \citet{boquien15} obtain results on M33 which are in nice agreement with the ones presented in our study and suggest possible causes explaining the controversial results of Kreckel et al.~(2013).

\begin{figure}
\begin{center}
  \includegraphics[height=.460\textwidth]{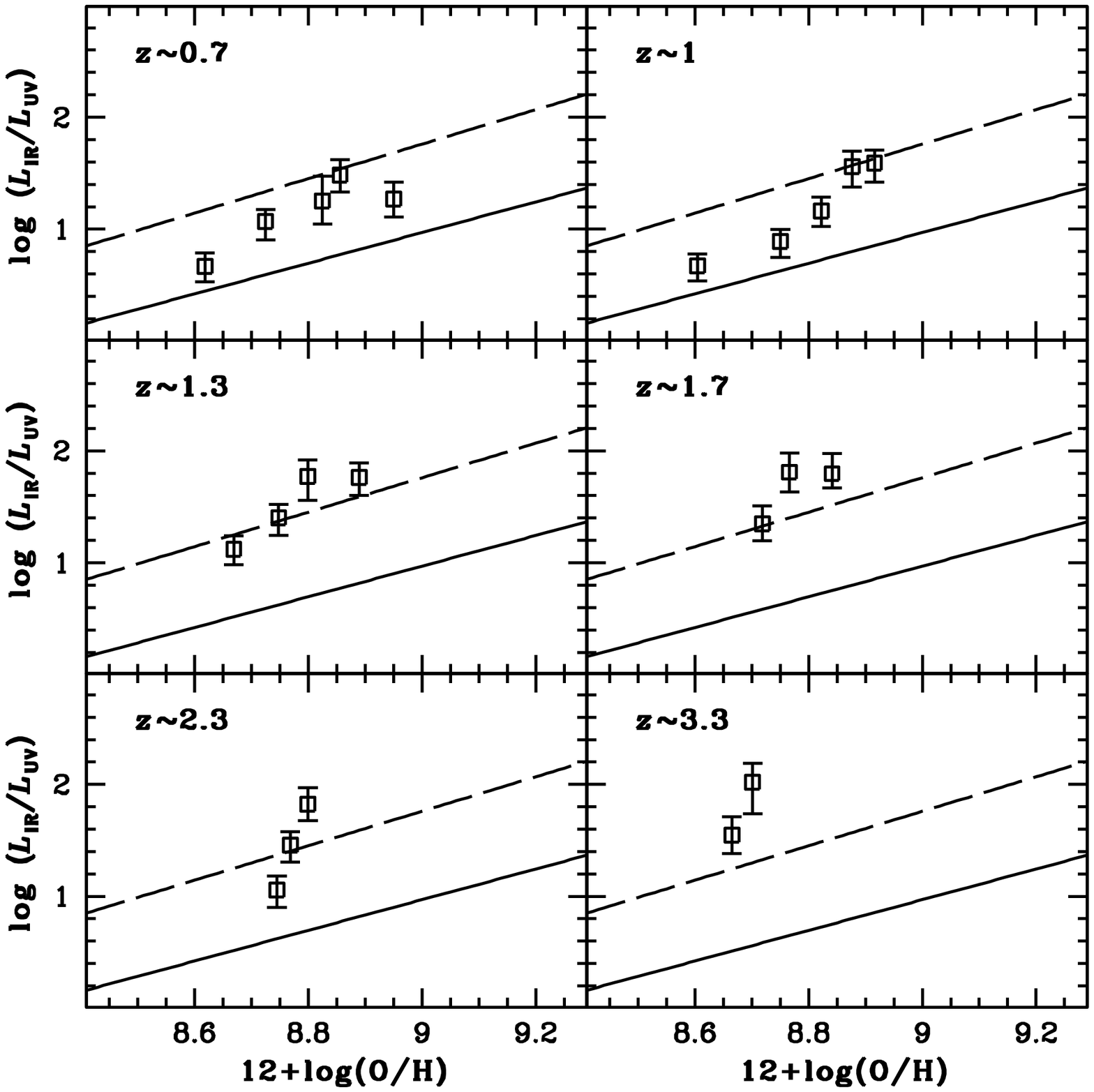}%
  \caption{ISM metallicity vs.~$\log$(\lir/$L_{\rm UV}$) at different redshifts. Here we estimate the ISM metal content using the fundamental metallicity relation (FMR) defined in Mannucci et al.~(2010). We show at all redshifts the correlations found in the local Universe for starburst \citep{heck98} and normal star-forming galaxies \citep{cortese06} with long-dashed and solid lines, respectively. MS galaxies tend to catch up at high redshift with the correlation followed by starburst galaxies in the local Universe, suggesting an evolution of the median star-forming galaxy ISM.}
\label{metal}
\end{center}
\end{figure}

\subsection{The relation between dust attenuation and ISM metallicity}

A more physical way to look at ISM conditions in star-forming galaxies and their possible evolution with redshift 
is to consider the correlation between dust attenuation and ISM metal content for star-forming galaxies. In the local Universe the correlation was first explored by \citet{heck98} for a sample of starburst galaxies, and found to yield a tight correlation for star-forming galaxies, with more dusty galaxies being more metal rich. The correlation is expected, since metals and more complex molecules are needed to build up the actual dust that absorbs the UV radiation. 

In a more recent study,  \citet{cortese06} looked at the properties of normal star-forming galaxies in the local Universe, a sample of objects much closer to what in this paper we are calling MS galaxies. They found that these objects were also following a tight correlation between dust attenuation and ISM metallicity, but that the correlation was offset with respect to the one of starburst galaxies: at a fixed metallicity, starburst galaxies are more obscured compared to normal star-forming galaxies. The offset in the correlation can be interpreted in terms of the different ISM conditions found in normal star-forming and starburst galaxies. The latter host star formation rate densities much higher than the former, both because star formation happens on smaller scales and because of the higher overall levels of star formation, and they can enrich most of their ISM much earlier and more quickly than normal smoothly star-forming MS galaxies.   
 
Assembling mass-complete samples of galaxies at high redshift with measured metallicities is practically impossible with the presently available facilities. Most of the existing data in the high redshift Universe faces the unavoidable bias of being limited by line sensitivity, and hence selected against the more dusty, more massive but also more metal rich systems. As already discussed, dust attenuation has an impact on the source detection itself, which becomes more and more severe with increasing redshift. This bias can in principle become extreme when it comes to metallicity measurements, where the easiest, and possibly only, objects for which it is possible to measure metallicity will be the ones with the lowest metallicities. This effect, which seems to nicely fit and explain some recent observations of metallicity and gas masses at high redshift \citep[see, e.g.,][]{ouchi13,tan13,troncoso13}, should be carefully taken into account when dealing with high redshift sources and deriving more general conclusions on the global galaxy population.

Here we estimate the median ISM metallicity for our samples of star-forming galaxies in bins of stellar mass and redshift, by applying the fundamental metallicity relation, FMR, found by Mannucci et al.~(2010). The relation is defined as a tri-dimensional surface linking SFR, metallicity and stellar mass of star-forming galaxies. It was first defined in a sample of local SDSS galaxies and then found to be essentially identical at higher redshift at least out to $z\simeq$2. The relation implies that galaxies obey an anti-correlation between SFR and metallicity: at a fixed stellar mass, the drop in metal content of star-forming galaxies at higher redshift \citep[e.g.,][]{tremonti04,savaglio05,erb2006,zz13} is compensated by the increase of SFR. 

As already stated, the FMR has been verified to remain unchanged up to $z\simeq$2 \citep[but see also, e.g.,][]{cullen14,steidel14,wuyts14}, while a sudden change has been claimed recently by the same authors in Troncoso et al.~(2013) for a sample of $z\simeq$3 galaxies. Of course, we have no way to test their claim here, because of how the mass-completeness of their sample could possibly affect their conclusions according to what discussed in the previous paragraph. Very recently, \citet{maier14} have shown that, when using the formalism put forward in \citet{lilly13} or equivalently the equation 2 of Mannucci et al.~(2010), as opposed to the equation 5 of the same paper, to describe the FMR, measured high-redshift metallicities are overall well described by the FMR predictions. Here we will assume that the FMR is still valid in our last redshift bin at z$\ge$3. 

We show in Figure~ \ref{metal} the evolution of dust attenuation (here shown as the classical quantity $\log$(\lir/$L_{\rm UV}$) instead of $A_{\rm UV}$ as in the previous figures, to be consistent with the studies of Heckman et al.~1998 and Cortese et al.~2006) vs. the ISM metallicity at different redshifts. We see a clear evolution in redshift, with MS galaxies moving progressively from the locus occupied locally by normal star-forming galaxies (solid line), to overshooting the one occupied locally by starburst galaxies (long-dashed line). We stress here that our result does not depend dramatically on the use of the FMR but that a qualitatively similar outcome would have been obtained by simply using the evolution of the mass-metallicity relation. In other words, the shift in redshift we are seeing is mostly driven by the fact that galaxies of a fix stellar mass are becoming less and less metal-rich with increasing time while keeping the same level of UV dust attenuation. 

Galaxies are becoming overall more compact in sizes at higher redshift and at the same time they are also significantly increasing their star formation rates and this trend makes them more and more similar to local starbursts in terms of ISM conditions. This similarity is also confirmed by the reported evolution of the IR SED shape of Main Sequence galaxies up to redshift $z\simeq$2 by Magdis et al.~(2012) and more recently by \citet{mag14}, and extended at higher redshift by Schreiber et al.~(in preparation), with the median dust temperature becoming higher with increasing redshift and indeed more similar to the dust temperatures of highly star-forming galaxies seen in the local Universe.

%Also, given the measured evolution of the stellar mass-metallicity evolution, $M_*$-Z, with star-forming galaxies being less metal rich at higher redshift \citep[e.g.,][]{tremonti04,savaglio05,erb2006}, we can predict that the local relation between dust attenuation and metal content \citep[see, e.g.,][]{cortese06} has instead to evolve with redshift: at a fixed metallicity galaxies were more dusty at higher redshift, simply because they were more massive.

\section{Conclusions}
We have presented first results of an ongoing project aimed at a better understanding
of the physics of star formation across cosmic time. We take advantage of one
of the deepest panchromatic data-sets available at present to select a star-forming galaxy sample
in the GOODS-N field and to obtain a  complete sampling of galaxy redshifts,
 star formation rates, stellar masses, and UV rest-frame properties. 
We quantitatively explore, with a mass-complete sample, the evolution of star formation and dust attenuation properties up to $z\simeq$4 as a function of the host galaxy properties. Our main results can be summarized as:

$\bullet$~We find that the slope of the SFR--$M_*$ correlation is equal to 0.8, and is consistent with being identical at all redshifts up to $z\simeq$1.5, while the normalization increases continously up to the highest redshift we are able to probe.   

$\bullet$~We have explored  the FIR-Radio correlation for a mass-selected sample of star-forming galaxies and found that the correlation does not evolve up to $z\simeq$4.    

$ \bullet$~We confirm that galaxy stellar mass is a robust proxy for the UV dust attenuation in star-forming galaxies: more massive galaxies are more dust-attenuated than less massive ones.   

$ \bullet$~We find that the correlation between UV attenuation and mass evolves very weakly with redshift: the amount of dust attenuation increases by less than 0.3 magnitudes over the redshift range 0.5--4 for fixed stellar mass. This finding explains the already reported evolution of the SFR--$A_{\rm UV}$ relation: the same amount of star formation is less attenuated at higher redshift, because it is hosted in less massive, and less metal rich, galaxies   

$\bullet$~We explored the correlation between UV dust attenuation and UV spectral slope, a widely used proxy for dust attenuation.
At all redshifts the two quantities correlate, as already shown in many previous studies, but the correlation evolves with redshift with star-forming  galaxies having bluer UV spectral slope for the same amount of dust attenuation, reaching values inconsistent with the ones predicted by the Calzetti law at z$\ge$3. At these redshifts the SFRs derived from the UV corrected emission underestimate the real SFR by a factor of around 2. Consequently, this should lead, as already suggested in Castellano et al.~(2014), to an upward revision, by the same factor, of the SFRD estimates at redshift $z\ge3$.  

$\bullet$~Combining our findings with previously published results from \ha~line emission surveys, we find that at $z\simeq$1 the line reddening is larger than the continuum reddening by a factor $\simeq$1.3 and becomes closer to a factor 1 by redshift $z\simeq$3. This is  substantially lower than the value  of around 1.7, found in the local Universe, and points toward a more compact and more dense star formation distribution in high redshift star-forming galaxies.    

$\bullet$~Finally, using the fundamental metallicity relation to estimate ISM metallicities, we find that the amount of dust attenuation at a fixed ISM metallicity increases with redshift and reaches the local value for highly star-forming galaxies. 

We speculate that our results point toward an evolution of the ISM conditions of the median Main Sequence star-forming galaxy, such that at $z\ge$1.5, Main Sequence galaxies have ISM properties more similar to those found in local starbursts.\\

\acknowledgements
We thank the referee for thorough and constructive comments that helped clarifying and improving this paper. We are grateful to N. Drory for sharing the SED-fitting code used to estimate galaxy stellar masses, to Masami Ouchi for kindly sharing the Subaru SuprimeCam deep $z$ and $ZR$ imaging of the GOODS-N field and to Gabe Brammer for kindly answering (too) many questions about {\it EAZY}. This research was supported by the French Agence Nationale de la Recherche (ANR) projects ANR-09-BLAN-0224 and ANR-08-JCJC-0008 and by the ERC-StG grant UPGAL 240039. We aknowledge the contribution of the FP7 SPACE project ``ASTRODEEP''~(Ref.No.~312725), supported by the European Commission. SJ acknowledges support from the EU through grant ERC-StG-257720. RJI acknowledges support from the European Research Council in the form of the Advanced Investigator Programme,
321302, COSMICISM. This research has made use of the NASA/ IPAC Infrared Science Archive, which is operated by the Jet Propulsion Laboratory, California Institute of Techno$\log$y, under contract with the National Aeronautics and Space Administration. This work is based in part on observations obtained with WIRCAM, a joint project of CFHT, Taiwan, Korea, Canada, France, at the Canada-France-Hawaii Telescope (CFHT) which is operated by the National Research Council (NRC) of Canada, the Institut National des Sciences de l'Univers of the Centre National de la Recherche Scientifique of France, and the University of Hawaii. This work is also based in part on data collected at Subaru Telescope, which is operated by the National Astronomical Observatory of Japan and on observations made with the NASA Galaxy Evolution Explorer. GALEX is operated for NASA by the California Institute of Technology under NASA contract NAS5-98034. PACS has been developed by a consortium of institutes led by MPE (Germany) and including UVIE (Austria); KU Leuven, CSL, IMEC (Belgium); CEA, LAM (France); MPIA (Germany); INAFIFSI/OAA/OAP/OAT, LENS, SISSA (Italy); IAC (Spain). This development has been supported by the funding agencies BMVIT (Austria), ESA-PRODEX (Belgium), CEA/CNES (France), DLR (Germany), ASI/INAF (Italy), and CICYT/MCYT (Spain). SPIRE has been developed by a consortium of institutes led by Cardiff University (UK) and including Univ. Lethbridge (Canada); NAOC (China); CEA, LAM (France); IFSI, Univ. Padua (Italy); IAC (Spain); Stockholm Observatory (Sweden); Imperial College London, RAL, UCL-MSSL, UKATC, Univ. Sussex (UK); and Caltech, JPL, NHSC, Univ. Colorado (USA). This development has been supported by national funding agencies: CSA (Canada); NAOC (China); CEA, CNES, CNRS (France); ASI (Italy); MCINN (Spain); Stockholm Observatory (Sweden); STFC (UK); and NASA (USA).

%\bibliography{tpp}

\begin{thebibliography}{171}
\expandafter\ifx\csname natexlab\endcsname\relax\def\natexlab#1{#1}\fi

\bibitem[{{Alavi} {et~al.}(2014){Alavi}, {Siana}, {Richard}, {Stark},
  {Scarlata}, {Teplitz}, {Freeman}, {Dominguez}, {Rafelski}, {Robertson}, \&
  {Kewley}}]{alavi14}
{Alavi}, A., {Siana}, B., {Richard}, J., {et~al.} 2014, \apj, 780, 143

\bibitem[{{Alexander} {et~al.}(2003){Alexander}, {Bauer}, {Brandt},
  {Schneider}, {Hornschemeier}, {Vignali}, {Barger}, {Broos}, {Cowie},
  {Garmire}, {Townsley}, {Bautz}, {Chartas}, \& {Sargent}}]{alexander03}
{Alexander}, D.~M., {Bauer}, F.~E., {Brandt}, W.~N., {et~al.} 2003, \aj, 126,
  539

\bibitem[{{Barger} {et~al.}(2014){Barger}, {Cowie}, {Chen}, {Owen}, {Wang},
  {Casey}, {Lee}, {Sanders}, \& {Williams}}]{barger14}
{Barger}, A.~J., {Cowie}, L.~L., {Chen}, C.-C., {et~al.} 2014, \apj, 784, 9

\bibitem[{{Barger} {et~al.}(2008){Barger}, {Cowie}, \& {Wang}}]{barger08}
{Barger}, A.~J., {Cowie}, L.~L., \& {Wang}, W.-H. 2008, \apj, 689, 687

\bibitem[{{Barger} {et~al.}(2012){Barger}, {Wang}, {Cowie}, {Owen}, {Chen}, \&
  {Williams}}]{barger12}
{Barger}, A.~J., {Wang}, W.-H., {Cowie}, L.~L., {et~al.} 2012, \apj, 761, 89

\bibitem[{{Bauer} {et~al.}(2004){Bauer}, {Alexander}, {Brandt}, {Schneider},
  {Treister}, {Hornschemeier}, \& {Garmire}}]{bauer04}
{Bauer}, F.~E., {Alexander}, D.~M., {Brandt}, W.~N., {et~al.} 2004, \aj, 128,
  2048

\bibitem[{{Bell}(2003)}]{bell03}
{Bell}, E.~F. 2003, \apj, 586, 794

\bibitem[{{Bell} {et~al.}(2012){Bell}, {van der Wel}, {Papovich}, {Kocevski},
  {Lotz}, {McIntosh}, {Kartaltepe}, {Faber}, {Ferguson}, {Koekemoer}, {Grogin},
  {Wuyts}, {Cheung}, {Conselice}, {Dekel}, {Dunlop}, {Giavalisco},
  {Herrington}, {Koo}, {McGrath}, {de Mello}, {Rix}, {Robaina}, \&
  {Williams}}]{bell12}
{Bell}, E.~F., {van der Wel}, A., {Papovich}, C., {et~al.} 2012, \apj, 753, 167

\bibitem[{{Bertin} \& {Arnouts}(1996)}]{bertin}
{Bertin}, E. \& {Arnouts}, S. 1996, A\&AS, 117, 393

\bibitem[{{B{\'e}thermin} {et~al.}(2012{\natexlab{a}}){B{\'e}thermin}, {Daddi},
  {Magdis}, {Sargent}, {Hezaveh}, {Elbaz}, {Le Borgne}, {Mullaney}, {Pannella},
  {Buat}, {Charmandaris}, {Lagache}, \& {Scott}}]{beth12}
{B{\'e}thermin}, M., {Daddi}, E., {Magdis}, G., {et~al.} 2012{\natexlab{a}},
  \apjl, 757, L23

\bibitem[{{B{\'e}thermin} {et~al.}(2010){B{\'e}thermin}, {Dole}, {Beelen}, \&
  {Aussel}}]{bethermin10}
{B{\'e}thermin}, M., {Dole}, H., {Beelen}, A., \& {Aussel}, H. 2010, \aap, 512,
  A78

\bibitem[{{B{\'e}thermin} {et~al.}(2012{\natexlab{b}}){B{\'e}thermin}, {Le
  Floc'h}, {Ilbert}, {Conley}, {Lagache}, {Amblard}, {Arumugam}, {Aussel},
  {Berta}, {Bock}, {Boselli}, {Buat}, {Casey}, {Castro-Rodr{\'{\i}}guez},
  {Cava}, {Clements}, {Cooray}, {Dowell}, {Eales}, {Farrah}, {Franceschini},
  {Glenn}, {Griffin}, {Hatziminaoglou}, {Heinis}, {Ibar}, {Ivison},
  {Kartaltepe}, {Levenson}, {Magdis}, {Marchetti}, {Marsden}, {Nguyen},
  {O'Halloran}, {Oliver}, {Omont}, {Page}, {Panuzzo}, {Papageorgiou},
  {Pearson}, {P{\'e}rez-Fournon}, {Pohlen}, {Rigopoulou}, {Roseboom},
  {Rowan-Robinson}, {Salvato}, {Schulz}, {Scott}, {Seymour}, {Shupe}, {Smith},
  {Symeonidis}, {Trichas}, {Tugwell}, {Vaccari}, {Valtchanov}, {Vieira},
  {Viero}, {Wang}, {Xu}, \& {Zemcov}}]{bethermin12}
{B{\'e}thermin}, M., {Le Floc'h}, E., {Ilbert}, O., {et~al.}
  2012{\natexlab{b}}, \aap, 542, A58

\bibitem[{{Bongiorno} {et~al.}(2012){Bongiorno}, {Merloni}, {Brusa},
  {Magnelli}, {Salvato}, {Mignoli}, {Zamorani}, {Fiore}, {Rosario}, {Mainieri},
  {Hao}, {Comastri}, {Vignali}, {Balestra}, {Bardelli}, {Berta}, {Civano},
  {Kampczyk}, {Le Floc'h}, {Lusso}, {Lutz}, {Pozzetti}, {Pozzi}, {Riguccini},
  {Shankar}, \& {Silverman}}]{bongiorno12}
{Bongiorno}, A., {Merloni}, A., {Brusa}, M., {et~al.} 2012, \mnras, 427, 3103

\bibitem[{{Boquien} {et~al.}(2012){Boquien}, {Buat}, {Boselli}, {Baes},
  {Bendo}, {Ciesla}, {Cooray}, {Cortese}, {Eales}, {Gavazzi}, {Gomez},
  {Lebouteiller}, {Pappalardo}, {Pohlen}, {Smith}, \& {Spinoglio}}]{boquien12}
{Boquien}, M., {Buat}, V., {Boselli}, A., {et~al.} 2012, \aap, 539, A145

\bibitem[{{Boquien} {et~al.}(2015){Boquien}, {Calzetti}, {Aalto}, {Boselli},
  {Braine}, {Buat}, {Combes}, {Israel}, {Kramer}, {Lord}, {Relano},
  {Rosolowsky}, {Stacey}, {Tabatabaei}, {van der Tak}, {van der Werf},
  {Verley}, \& {Xilouris}}]{boquien15}
{Boquien}, M., {Calzetti}, D., {Aalto}, S., {et~al.} 2015, ArXiv e-prints:
  1502.01347

\bibitem[{{Bourne} {et~al.}(2011){Bourne}, {Dunne}, {Ivison}, {Maddox},
  {Dickinson}, \& {Frayer}}]{bourne11}
{Bourne}, N., {Dunne}, L., {Ivison}, R.~J., {et~al.} 2011, \mnras, 410, 1155

\bibitem[{{Bouwens} {et~al.}(2009){Bouwens}, {Illingworth}, {Franx}, {Chary},
  {Meurer}, {Conselice}, {Ford}, {Giavalisco}, \& {van Dokkum}}]{bouwens09}
{Bouwens}, R.~J., {Illingworth}, G.~D., {Franx}, M., {et~al.} 2009, \apj, 705,
  936

\bibitem[{{Bouwens} {et~al.}(2012){Bouwens}, {Illingworth}, {Oesch}, {Franx},
  {Labb{\'e}}, {Trenti}, {van Dokkum}, {Carollo}, {Gonz{\'a}lez}, {Smit}, \&
  {Magee}}]{bouwens12}
{Bouwens}, R.~J., {Illingworth}, G.~D., {Oesch}, P.~A., {et~al.} 2012, \apj,
  754, 83

\bibitem[{{Brammer} {et~al.}(2008){Brammer}, {van Dokkum}, \&
  {Coppi}}]{brammer08}
{Brammer}, G.~B., {van Dokkum}, P.~G., \& {Coppi}, P. 2008, \apj, 686, 1503

\bibitem[{{Brammer} {et~al.}(2011){Brammer}, {Whitaker}, {van Dokkum},
  {Marchesini}, {Franx}, {Kriek}, {Labb{\'e}}, {Lee}, {Muzzin}, {Quadri},
  {Rudnick}, \& {Williams}}]{Brammer11}
{Brammer}, G.~B., {Whitaker}, K.~E., {van Dokkum}, P.~G., {et~al.} 2011, \apj,
  739, 24

\bibitem[{{Brinchmann} {et~al.}(2004){Brinchmann}, {Charlot}, {White},
  {Tremonti}, {Kauffmann}, {Heckman}, \& {Brinkmann}}]{brinchmann04}
{Brinchmann}, J., {Charlot}, S., {White}, S.~D.~M., {et~al.} 2004, \mnras, 351,
  1151

\bibitem[{{Bruzual} \& {Charlot}(2003)}]{bc03}
{Bruzual}, G. \& {Charlot}, S. 2003, \mnras, 344, 1000

\bibitem[{{Buat} {et~al.}(2011){Buat}, {Giovannoli}, {Heinis}, {Charmandaris},
  {Coia}, {Daddi}, {Dickinson}, {Elbaz}, {Hwang}, {Morrison}, {Dasyra},
  {Aussel}, {Altieri}, {Dannerbauer}, {Kartaltepe}, {Leiton}, {Magdis},
  {Magnelli}, \& {Popesso}}]{buat11}
{Buat}, V., {Giovannoli}, E., {Heinis}, S., {et~al.} 2011, \aap, 533, A93

\bibitem[{{Buat} {et~al.}(2014){Buat}, {Heinis}, {Boquien}, {Burgarella},
  {Charmandaris}, {Boissier}, {Boselli}, {Le Borgne}, \& {Morrison}}]{buat14}
{Buat}, V., {Heinis}, S., {Boquien}, M., {et~al.} 2014, \aap, 561, A39

\bibitem[{{Buat} {et~al.}(2005){Buat}, {Iglesias-P{\'a}ramo}, {Seibert},
  {Burgarella}, {Charlot}, {Martin}, {Xu}, {Heckman}, {Boissier}, {Boselli},
  {Barlow}, {Bianchi}, {Byun}, {Donas}, {Forster}, {Friedman}, {Jelinski},
  {Lee}, {Madore}, {Malina}, {Milliard}, {Morissey}, {Neff}, {Rich},
  {Schiminovitch}, {Siegmund}, {Small}, {Szalay}, {Welsh}, \& {Wyder}}]{buat05}
{Buat}, V., {Iglesias-P{\'a}ramo}, J., {Seibert}, M., {et~al.} 2005, \apjl,
  619, L51

\bibitem[{{Buat} {et~al.}(2007){Buat}, {Marcillac}, {Burgarella}, {Le Floc'h},
  {Takeuchi}, {Iglesias-Par{\`a}mo}, \& {Xu}}]{buat07}
{Buat}, V., {Marcillac}, D., {Burgarella}, D., {et~al.} 2007, \aap, 469, 19

\bibitem[{{Buat} {et~al.}(2012){Buat}, {Noll}, {Burgarella}, {Giovannoli},
  {Charmandaris}, {Pannella}, {Hwang}, {Elbaz}, {Dickinson}, {Magdis}, {Reddy},
  \& {Murphy}}]{buat12}
{Buat}, V., {Noll}, S., {Burgarella}, D., {et~al.} 2012, \aap, 545, A141

\bibitem[{{Burgarella} {et~al.}(2013){Burgarella}, {Buat}, {Gruppioni},
  {Cucciati}, {Heinis}, {Berta}, {B{\'e}thermin}, {Bock}, {Cooray}, {Dunlop},
  {Farrah}, {Franceschini}, {Le Floc'h}, {Lutz}, {Magnelli}, {Nordon},
  {Oliver}, {Page}, {Popesso}, {Pozzi}, {Riguccini}, {Vaccari}, \&
  {Viero}}]{burga13}
{Burgarella}, D., {Buat}, V., {Gruppioni}, C., {et~al.} 2013, \aap, 554, A70

\bibitem[{{Calzetti}(2001)}]{calzetti2001}
{Calzetti}, D. 2001, \pasp, 113, 1449

\bibitem[{{Calzetti} {et~al.}(2000){Calzetti}, {Armus}, {Bohlin}, {Kinney},
  {Koornneef}, \& {Storchi-Bergmann}}]{Calzetti00}
{Calzetti}, D., {Armus}, L., {Bohlin}, R.~C., {et~al.} 2000, \apj, 533, 682

\bibitem[{{Capak} {et~al.}(2007){Capak}, {Aussel}, {Ajiki}, {McCracken},
  {Mobasher}, {Scoville}, {Shopbell}, {Taniguchi}, {Thompson}, {Tribiano},
  {Sasaki}, {Blain}, {Brusa}, {Carilli}, {Comastri}, {Carollo}, {Cassata},
  {Colbert}, {Ellis}, {Elvis}, {Giavalisco}, {Green}, {Guzzo}, {Hasinger},
  {Ilbert}, {Impey}, {Jahnke}, {Kartaltepe}, {Kneib}, {Koda}, {Koekemoer},
  {Komiyama}, {Leauthaud}, {Lefevre}, {Lilly}, {Liu}, {Massey}, {Miyazaki},
  {Murayama}, {Nagao}, {Peacock}, {Pickles}, {Porciani}, {Renzini}, {Rhodes},
  {Rich}, {Salvato}, {Sanders}, {Scarlata}, {Schiminovich}, {Schinnerer},
  {Scodeggio}, {Sheth}, {Shioya}, {Tasca}, {Taylor}, {Yan}, \&
  {Zamorani}}]{capak2007}
{Capak}, P., {Aussel}, H., {Ajiki}, M., {et~al.} 2007, \apjs, 172, 99

\bibitem[{{Capak} {et~al.}(2004){Capak}, {Cowie}, {Hu}, {Barger}, {Dickinson},
  {Fernandez}, {Giavalisco}, {Komiyama}, {Kretchmer}, {McNally}, {Miyazaki},
  {Okamura}, \& {Stern}}]{Capak2004}
{Capak}, P., {Cowie}, L.~L., {Hu}, E.~M., {et~al.} 2004, AJ, 127, 180

\bibitem[{{Cardamone} {et~al.}(2010){Cardamone}, {Urry}, {Schawinski},
  {Treister}, {Brammer}, \& {Gawiser}}]{cardamone10}
{Cardamone}, C.~N., {Urry}, C.~M., {Schawinski}, K., {et~al.} 2010, \apjl, 721,
  L38

\bibitem[{{Carilli} {et~al.}(2008){Carilli}, {Lee}, {Capak}, {Schinnerer},
  {Lee}, {McCraken}, {Yun}, {Scoville}, {Smol{\v c}i{\'c}}, {Giavalisco},
  {Datta}, {Taniguchi}, \& {Urry}}]{carilli08}
{Carilli}, C.~L., {Lee}, N., {Capak}, P., {et~al.} 2008, \apj, 689, 883

\bibitem[{{Casey} {et~al.}(2012){Casey}, {Berta}, {B{\'e}thermin}, {Bock},
  {Bridge}, {Budynkiewicz}, {Burgarella}, {Chapin}, {Chapman}, {Clements},
  {Conley}, {Conselice}, {Cooray}, {Farrah}, {Hatziminaoglou}, {Ivison}, {le
  Floc'h}, {Lutz}, {Magdis}, {Magnelli}, {Oliver}, {Page}, {Pozzi},
  {Rigopoulou}, {Riguccini}, {Roseboom}, {Sanders}, {Scott}, {Seymour},
  {Valtchanov}, {Vieira}, {Viero}, \& {Wardlow}}]{casey12}
{Casey}, C.~M., {Berta}, S., {B{\'e}thermin}, M., {et~al.} 2012, \apj, 761, 140

\bibitem[{{Castellano} {et~al.}(2012){Castellano}, {Fontana}, {Grazian},
  {Pentericci}, {Santini}, {Koekemoer}, {Cristiani}, {Galametz}, {Gallerani},
  {Vanzella}, {Boutsia}, {Gallozzi}, {Giallongo}, {Maiolino}, {Menci}, \&
  {Paris}}]{cast12}
{Castellano}, M., {Fontana}, A., {Grazian}, A., {et~al.} 2012, \aap, 540, A39

\bibitem[{{Castellano} {et~al.}(2014){Castellano}, {Sommariva}, {Fontana},
  {Pentericci}, {Santini}, {Grazian}, {Amorin}, {Donley}, {Dunlop}, {Ferguson},
  {Fiore}, {Galametz}, {Giallongo}, {Guo}, {Huang}, {Koekemoer}, {Maiolino},
  {McLure}, {Paris}, {Schaerer}, {Troncoso}, \& {Vanzella}}]{cast14}
{Castellano}, M., {Sommariva}, V., {Fontana}, A., {et~al.} 2014, ArXiv~e-prints:~1403.0743

\bibitem[{{Chabrier}(2003)}]{chabrier}
{Chabrier}, G. 2003, \pasp, 115, 763

\bibitem[{{Chary} \& {Elbaz}(2001)}]{ce01}
{Chary}, R. \& {Elbaz}, D. 2001, \apj, 556, 562

\bibitem[{{Chary} \& {Pope}(2010)}]{charyandpope10}
{Chary}, R.-R. \& {Pope}, A. 2010, ArXiv~e-prints:~1003.1731

\bibitem[{{Cimatti} {et~al.}(2006){Cimatti}, {Daddi}, \& {Renzini}}]{cim06}
{Cimatti}, A., {Daddi}, E., \& {Renzini}, A. 2006, \aap, 453, L29

\bibitem[{{Cisternas} {et~al.}(2011){Cisternas}, {Jahnke}, {Bongiorno},
  {Inskip}, {Impey}, {Koekemoer}, {Merloni}, {Salvato}, \&
  {Trump}}]{cisternas11}
{Cisternas}, M., {Jahnke}, K., {Bongiorno}, A., {et~al.} 2011, \apjl, 741, L11

\bibitem[{{Condon}(1992)}]{condon92}
{Condon}, J.~J. 1992, \araa, 30, 575

\bibitem[{{Cortese} {et~al.}(2006){Cortese}, {Boselli}, {Buat}, {Gavazzi},
  {Boissier}, {Gil de Paz}, {Seibert}, {Madore}, \& {Martin}}]{cortese06}
{Cortese}, L., {Boselli}, A., {Buat}, V., {et~al.} 2006, \apj, 637, 242

\bibitem[{{Cullen} {et~al.}(2014){Cullen}, {Cirasuolo}, {McLure}, {Dunlop}, \&
  {Bowler}}]{cullen14}
{Cullen}, F., {Cirasuolo}, M., {McLure}, R.~J., {Dunlop}, J.~S., \& {Bowler},
  R.~A.~A. 2014, \mnras, 440, 2300

\bibitem[{{Daddi} {et~al.}(2004){Daddi}, {Cimatti}, {Renzini}, {Fontana},
  {Mignoli}, {Pozzetti}, {Tozzi}, \& {Zamorani}}]{daddi2004}
{Daddi}, E., {Cimatti}, A., {Renzini}, A., {et~al.} 2004, \apj, 617, 746

\bibitem[{{Daddi} {et~al.}(2009){Daddi}, {Dannerbauer}, {Stern}, {Dickinson},
  {Morrison}, {Elbaz}, {Giavalisco}, {Mancini}, {Pope}, \&
  {Spinrad}}]{daddi2009}
{Daddi}, E., {Dannerbauer}, H., {Stern}, D., {et~al.} 2009, \apj, 694, 1517

\bibitem[{{Daddi} {et~al.}(2007){Daddi}, {Dickinson}, {Morrison}, {Chary},
  {Cimatti}, {Elbaz}, {Frayer}, {Renzini}, {Pope}, {Alexander}, {Bauer},
  {Giavalisco}, {Huynh}, {Kurk}, \& {Mignoli}}]{daddi071}
{Daddi}, E., {Dickinson}, M., {Morrison}, G., {et~al.} 2007, \apj, 670, 156

\bibitem[{{de Barros} {et~al.}(2014){de Barros}, {Schaerer}, \&
  {Stark}}]{debarros14}
{de Barros}, S., {Schaerer}, D., \& {Stark}, D.~P. 2014, \aap, 563, A81

\bibitem[{{Del Moro} {et~al.}(2013){Del Moro}, {Alexander}, {Mullaney},
  {Daddi}, {Pannella}, {Bauer}, {Pope}, {Dickinson}, {Elbaz}, {Barthel},
  {Garrett}, {Brandt}, {Charmandaris}, {Chary}, {Dasyra}, {Gilli}, {Hickox},
  {Hwang}, {Ivison}, {Juneau}, {Le Floc'h}, {Luo}, {Morrison}, {Rovilos},
  {Sargent}, \& {Xue}}]{delmoro13}
{Del Moro}, A., {Alexander}, D.~M., {Mullaney}, J.~R., {et~al.} 2013, \aap,
  549, A59

\bibitem[{{Dickinson} {et~al.}(2003){Dickinson}, {Papovich}, {Ferguson}, \&
  {Budav{\' a}ri}}]{dickinson2003}
{Dickinson}, M., {Papovich}, C., {Ferguson}, H.~C., \& {Budav{\' a}ri}, T.
  2003, \apj, 587, 25

\bibitem[{{Dom{\'{\i}}nguez} {et~al.}(2013){Dom{\'{\i}}nguez}, {Siana},
  {Henry}, {Scarlata}, {Bedregal}, {Malkan}, {Atek}, {Ross}, {Colbert},
  {Teplitz}, {Rafelski}, {McCarthy}, {Bunker}, {Hathi}, {Dressler}, {Martin},
  \& {Masters}}]{doming12}
{Dom{\'{\i}}nguez}, A., {Siana}, B., {Henry}, A.~L., {et~al.} 2013, \apj, 763,
  145

\bibitem[{{Drory} {et~al.}(2004){Drory}, {Bender}, \& {Hopp}}]{drorymass}
{Drory}, N., {Bender}, R., \& {Hopp}, U. 2004, \apjl, 616, L103

\bibitem[{{Drory} {et~al.}(2009){Drory}, {Bundy}, {Leauthaud}, {Scoville},
  {Capak}, {Ilbert}, {Kartaltepe}, {Kneib}, {McCracken}, {Salvato}, {Sanders},
  {Thompson}, \& {Willott}}]{drory09}
{Drory}, N., {Bundy}, K., {Leauthaud}, A., {et~al.} 2009, \apj, 707, 1595

\bibitem[{{Drory} {et~al.}(2005){Drory}, {Salvato}, {Gabasch}, {Bender},
  {Hopp}, {Feulner}, \& {Pannella}}]{drory2005}
{Drory}, N., {Salvato}, M., {Gabasch}, A., {et~al.} 2005, \apjl, 619, L131

\bibitem[{{Elbaz} {et~al.}(2007){Elbaz}, {Daddi}, {Le Borgne}, {Dickinson},
  {Alexander}, {Chary}, {Starck}, {Brandt}, {Kitzbichler}, {MacDonald},
  {Nonino}, {Popesso}, {Stern}, \& {Vanzella}}]{elbaz07}
{Elbaz}, D., {Daddi}, E., {Le Borgne}, D., {et~al.} 2007, \aap, 468, 33

\bibitem[{{Elbaz} {et~al.}(2011){Elbaz}, {Dickinson}, {Hwang},
  {D{\'{\i}}az-Santos}, {Magdis}, {Magnelli}, {Le Borgne}, {Galliano},
  {Pannella}, {Chanial}, {Armus}, {Charmandaris}, {Daddi}, {Aussel}, {Popesso},
  {Kartaltepe}, {Altieri}, {Valtchanov}, {Coia}, {Dannerbauer}, {Dasyra},
  {Leiton}, {Mazzarella}, {Alexander}, {Buat}, {Burgarella}, {Chary}, {Gilli},
  {Ivison}, {Juneau}, {Le Floc'h}, {Lutz}, {Morrison}, {Mullaney}, {Murphy},
  {Pope}, {Scott}, {Brodwin}, {Calzetti}, {Cesarsky}, {Charlot}, {Dole},
  {Eisenhardt}, {Ferguson}, {F{\"o}rster Schreiber}, {Frayer}, {Giavalisco},
  {Huynh}, {Koekemoer}, {Papovich}, {Reddy}, {Surace}, {Teplitz}, {Yun}, \&
  {Wilson}}]{Elbaz11}
{Elbaz}, D., {Dickinson}, M., {Hwang}, H.~S., {et~al.} 2011, \aap, 533, A119

\bibitem[{{Elbaz} {et~al.}(2010){Elbaz}, {Hwang}, {Magnelli}, {Daddi},
  {Aussel}, {Altieri}, {Amblard}, {Andreani}, {Arumugam}, {Auld}, {Babbedge},
  {Berta}, {Blain}, {Bock}, {Bongiovanni}, {Boselli}, {Buat}, {Burgarella},
  {Castro-Rodriguez}, {Cava}, {Cepa}, {Chanial}, {Chary}, {Cimatti},
  {Clements}, {Conley}, {Conversi}, {Cooray}, {Dickinson}, {Dominguez},
  {Dowell}, {Dunlop}, {Dwek}, {Eales}, {Farrah}, {F{\"o}rster Schreiber},
  {Fox}, {Franceschini}, {Gear}, {Genzel}, {Glenn}, {Griffin}, {Gruppioni},
  {Halpern}, {Hatziminaoglou}, {Ibar}, {Isaak}, {Ivison}, {Lagache}, {Le
  Borgne}, {Le Floc'h}, {Levenson}, {Lu}, {Lutz}, {Madden}, {Maffei}, {Magdis},
  {Mainetti}, {Maiolino}, {Marchetti}, {Mortier}, {Nguyen}, {Nordon},
  {O'Halloran}, {Okumura}, {Oliver}, {Omont}, {Page}, {Panuzzo},
  {Papageorgiou}, {Pearson}, {Perez Fournon}, {P{\'e}rez Garc{\'{\i}}a},
  {Poglitsch}, {Pohlen}, {Popesso}, {Pozzi}, {Rawlings}, {Rigopoulou},
  {Riguccini}, {Rizzo}, {Rodighiero}, {Roseboom}, {Rowan-Robinson},
  {Saintonge}, {Sanchez Portal}, {Santini}, {Sauvage}, {Schulz}, {Scott},
  {Seymour}, {Shao}, {Shupe}, {Smith}, {Stevens}, {Sturm}, {Symeonidis},
  {Tacconi}, {Trichas}, {Tugwell}, {Vaccari}, {Valtchanov}, {Vieira},
  {Vigroux}, {Wang}, {Ward}, {Wright}, {Xu}, \& {Zemcov}}]{elbaba10}
{Elbaz}, D., {Hwang}, H.~S., {Magnelli}, B., {et~al.} 2010, \aap, 518, L29

\bibitem[{{Erb} {et~al.}(2006){Erb}, {Steidel}, {Shapley}, {Pettini}, {Reddy},
  \& {Adelberger}}]{erb2006}
{Erb}, D.~K., {Steidel}, C.~C., {Shapley}, A.~E., {et~al.} 2006, \apj, 647, 128

\bibitem[{{Ferguson} {et~al.}(2004){Ferguson}, {Dickinson}, {Giavalisco},
  {Kretchmer}, {Ravindranath}, {Idzi}, {Taylor}, {Conselice}, {Fall},
  {Gardner}, {Livio}, {Madau}, {Moustakas}, {Papovich}, {Somerville},
  {Spinrad}, \& {Stern}}]{ferguson2004}
{Ferguson}, H.~C., {Dickinson}, M., {Giavalisco}, M., {et~al.} 2004, \apjl,
  600, L107

\bibitem[{{Finkelstein} {et~al.}(2012){Finkelstein}, {Papovich}, {Salmon},
  {Finlator}, {Dickinson}, {Ferguson}, {Giavalisco}, {Koekemoer}, {Reddy},
  {Bassett}, {Conselice}, {Dunlop}, {Faber}, {Grogin}, {Hathi}, {Kocevski},
  {Lai}, {Lee}, {McLure}, {Mobasher}, \& {Newman}}]{finkelstein12}
{Finkelstein}, S.~L., {Papovich}, C., {Salmon}, B., {et~al.} 2012, \apj, 756,
  164

\bibitem[{{Fitzpatrick}(1999)}]{fitz1999}
{Fitzpatrick}, E.~L. 1999, \pasp, 111, 63

\bibitem[{{Fontana} {et~al.}(2006){Fontana}, {Salimbeni}, {Grazian},
  {Giallongo}, {Pentericci}, {Nonino}, {Fontanot}, {Menci}, {Monaco},
  {Cristiani}, {Vanzella}, {de Santis}, \& {Gallozzi}}]{font06}
{Fontana}, A., {Salimbeni}, S., {Grazian}, A., {et~al.} 2006, \aap, 459, 745

\bibitem[{{F{\"o}rster Schreiber} {et~al.}(2009){F{\"o}rster Schreiber},
  {Genzel}, {Bouch{\'e}}, {Cresci}, {Davies}, {Buschkamp}, {Shapiro},
  {Tacconi}, {Hicks}, {Genel}, {Shapley}, {Erb}, {Steidel}, {Lutz},
  {Eisenhauer}, {Gillessen}, {Sternberg}, {Renzini}, {Cimatti}, {Daddi},
  {Kurk}, {Lilly}, {Kong}, {Lehnert}, {Nesvadba}, {Verma}, {McCracken},
  {Arimoto}, {Mignoli}, \& {Onodera}}]{fs09}
{F{\"o}rster Schreiber}, N.~M., {Genzel}, R., {Bouch{\'e}}, N., {et~al.} 2009,
  \apj, 706, 1364

\bibitem[{{Gabasch} {et~al.}(2008){Gabasch}, {Goranova}, {Hopp}, {Noll}, \&
  {Pannella}}]{gab2008}
{Gabasch}, A., {Goranova}, Y., {Hopp}, U., {Noll}, S., \& {Pannella}, M. 2008,
  \mnras, 383, 1319

\bibitem[{{Garn} \& {Best}(2010)}]{garn10}
{Garn}, T. \& {Best}, P.~N. 2010, \mnras, 409, 421

\bibitem[{{Garn} {et~al.}(2010){Garn}, {Sobral}, {Best}, {Geach}, {Smail},
  {Cirasuolo}, {Dalton}, {Dunlop}, {McLure}, \& {Farrah}}]{garn2010}
{Garn}, T., {Sobral}, D., {Best}, P.~N., {et~al.} 2010, \mnras, 402, 2017

\bibitem[{{Gilbank} {et~al.}(2010){Gilbank}, {Balogh}, {Glazebrook}, {Bower},
  {Baldry}, {Davies}, {Hau}, {Li}, \& {McCarthy}}]{gilbank10}
{Gilbank}, D.~G., {Balogh}, M.~L., {Glazebrook}, K., {et~al.} 2010, \mnras,
  405, 2419

\bibitem[{{Gonz{\'a}lez} {et~al.}(2010){Gonz{\'a}lez}, {Labb{\'e}}, {Bouwens},
  {Illingworth}, {Franx}, {Kriek}, \& {Brammer}}]{gonzales10}
{Gonz{\'a}lez}, V., {Labb{\'e}}, I., {Bouwens}, R.~J., {et~al.} 2010, \apj,
  713, 115

\bibitem[{{Grasha} {et~al.}(2013){Grasha}, {Calzetti}, {Andrews}, {Lee}, \&
  {Dale}}]{grasha13}
{Grasha}, K., {Calzetti}, D., {Andrews}, J.~E., {Lee}, J.~C., \& {Dale}, D.~A.
  2013, \apj, 773, 174

\bibitem[{{Heckman} {et~al.}(1998){Heckman}, {Robert}, {Leitherer}, {Garnett},
  \& {van der Rydt}}]{heck98}
{Heckman}, T.~M., {Robert}, C., {Leitherer}, C., {Garnett}, D.~R., \& {van der
  Rydt}, F. 1998, \apj, 503, 646

\bibitem[{{Heinis} {et~al.}(2013){Heinis}, {Buat}, {B{\'e}thermin}, {Aussel},
  {Bock}, {Boselli}, {Burgarella}, {Conley}, {Cooray}, {Farrah}, {Ibar},
  {Ilbert}, {Ivison}, {Magdis}, {Marsden}, {Oliver}, {Page}, {Rodighiero},
  {Roehlly}, {Schulz}, {Scott}, {Smith}, {Viero}, {Wang}, \&
  {Zemcov}}]{heinis13}
{Heinis}, S., {Buat}, V., {B{\'e}thermin}, M., {et~al.} 2013, \mnras, 429, 1113

\bibitem[{{Heinis} {et~al.}(2014){Heinis}, {Buat}, {B{\'e}thermin}, {Bock},
  {Burgarella}, {Conley}, {Cooray}, {Farrah}, {Ilbert}, {Magdis}, {Marsden},
  {Oliver}, {Rigopoulou}, {Roehlly}, {Schulz}, {Symeonidis}, {Viero}, {Xu}, \&
  {Zemcov}}]{heinis14}
{Heinis}, S., {Buat}, V., {B{\'e}thermin}, M., {et~al.} 2014, \mnras, 437, 1268

\bibitem[{{Helou} {et~al.}(1985){Helou}, {Soifer}, \&
  {Rowan-Robinson}}]{helou85}
{Helou}, G., {Soifer}, B.~T., \& {Rowan-Robinson}, M. 1985, \apjl, 298, L7

\bibitem[{{Hopkins} \& {Beacom}(2006)}]{hop06}
{Hopkins}, A.~M. \& {Beacom}, J.~F. 2006, \apj, 651, 142

\bibitem[{{Hwang} {et~al.}(2010){Hwang}, {Elbaz}, {Magdis}, {Daddi},
  {Symeonidis}, {Altieri}, {Amblard}, {Andreani}, {Arumugam}, {Auld}, {Aussel},
  {Babbedge}, {Berta}, {Blain}, {Bock}, {Bongiovanni}, {Boselli}, {Buat},
  {Burgarella}, {Castro-Rodr{\'{\i}}guez}, {Cava}, {Cepa}, {Chanial}, {Chapin},
  {Chary}, {Cimatti}, {Clements}, {Conley}, {Conversi}, {Cooray},
  {Dannerbauer}, {Dickinson}, {Dominguez}, {Dowell}, {Dunlop}, {Dwek}, {Eales},
  {Farrah}, {Schreiber}, {Fox}, {Franceschini}, {Gear}, {Genzel}, {Glenn},
  {Griffin}, {Gruppioni}, {Halpern}, {Hatziminaoglou}, {Ibar}, {Isaak},
  {Ivison}, {Jeong}, {Lagache}, {Le Borgne}, {Le Floc'h}, {Lee}, {Lee}, {Lee},
  {Levenson}, {Lu}, {Lutz}, {Madden}, {Maffei}, {Magnelli}, {Mainetti},
  {Maiolino}, {Marchetti}, {Mortier}, {Nguyen}, {Nordon}, {O'Halloran},
  {Okumura}, {Oliver}, {Omont}, {Page}, {Panuzzo}, {Papageorgiou}, {Pearson},
  {P{\'e}rez-Fournon}, {Garc{\'{\i}}a}, {Poglitsch}, {Pohlen}, {Popesso},
  {Pozzi}, {Rawlings}, {Rigopoulou}, {Riguccini}, {Rizzo}, {Rodighiero},
  {Roseboom}, {Rowan-Robinson}, {Saintonge}, {Portal}, {Santini}, {Sauvage},
  {Schulz}, {Scott}, {Seymour}, {Shao}, {Shupe}, {Smith}, {Stevens}, {Sturm},
  {Tacconi}, {Trichas}, {Tugwell}, {Vaccari}, {Valtchanov}, {Vieira},
  {Vigroux}, {Wang}, {Ward}, {Wright}, {Xu}, \& {Zemcov}}]{hoseong10}
{Hwang}, H.~S., {Elbaz}, D., {Magdis}, G., {et~al.} 2010, \mnras, 409, 75

\bibitem[{{Ibar} {et~al.}(2009){Ibar}, {Ivison}, {Biggs}, {Lal}, {Best}, \&
  {Green}}]{ibar09}
{Ibar}, E., {Ivison}, R.~J., {Biggs}, A.~D., {et~al.} 2009, \mnras, 397, 281

\bibitem[{{Ibar} {et~al.}(2013){Ibar}, {Sobral}, {Best}, {Ivison}, {Smail},
  {Arumugam}, {Berta}, {B{\'e}thermin}, {Bock}, {Cava}, {Conley}, {Farrah},
  {Geach}, {Ikarashi}, {Kohno}, {Le Floc'h}, {Lutz}, {Magdis}, {Magnelli},
  {Marsden}, {Oliver}, {Page}, {Pozzi}, {Riguccini}, {Schulz}, {Seymour},
  {Smith}, {Symeonidis}, {Wang}, {Wardlow}, \& {Zemcov}}]{ibar13}
{Ibar}, E., {Sobral}, D., {Best}, P.~N., {et~al.} 2013, \mnras, 434, 3218

\bibitem[{{Ilbert} {et~al.}(2013){Ilbert}, {McCracken}, {Le F{\`e}vre},
  {Capak}, {Dunlop}, {Karim}, {Renzini}, {Caputi}, {Boissier}, {Arnouts},
  {Aussel}, {Comparat}, {Guo}, {Hudelot}, {Kartaltepe}, {Kneib}, {Krogager},
  {Le Floc'h}, {Lilly}, {Mellier}, {Milvang-Jensen}, {Moutard}, {Onodera},
  {Richard}, {Salvato}, {Sanders}, {Scoville}, {Silverman}, {Taniguchi},
  {Tasca}, {Thomas}, {Toft}, {Tresse}, {Vergani}, {Wolk}, \& {Zirm}}]{ilbert13}
{Ilbert}, O., {McCracken}, H.~J., {Le F{\`e}vre}, O., {et~al.} 2013, \aap, 556,
  A55

\bibitem[{{Ilbert} {et~al.}(2010){Ilbert}, {Salvato}, {Le Floc'h}, {Aussel},
  {Capak}, {McCracken}, {Mobasher}, {Kartaltepe}, {Scoville}, {Sanders},
  {Arnouts}, {Bundy}, {Cassata}, {Kneib}, {Koekemoer}, {Le F{\`e}vre}, {Lilly},
  {Surace}, {Taniguchi}, {Tasca}, {Thompson}, {Tresse}, {Zamojski}, {Zamorani},
  \& {Zucca}}]{ilbert10}
{Ilbert}, O., {Salvato}, M., {Le Floc'h}, E., {et~al.} 2010, \apj, 709, 644

\bibitem[{{Ivison} {et~al.}(2010){Ivison}, {Magnelli}, {Ibar}, {Andreani},
  {Elbaz}, {Altieri}, {Amblard}, {Arumugam}, {Auld}, {Aussel}, {Babbedge},
  {Berta}, {Blain}, {Bock}, {Bongiovanni}, {Boselli}, {Buat}, {Burgarella},
  {Castro-Rodr{\'{\i}}guez}, {Cava}, {Cepa}, {Chanial}, {Cimatti}, {Cirasuolo},
  {Clements}, {Conley}, {Conversi}, {Cooray}, {Daddi}, {Dominguez}, {Dowell},
  {Dwek}, {Eales}, {Farrah}, {F{\"o}rster Schreiber}, {Fox}, {Franceschini},
  {Gear}, {Genzel}, {Glenn}, {Griffin}, {Gruppioni}, {Halpern},
  {Hatziminaoglou}, {Isaak}, {Lagache}, {Levenson}, {Lu}, {Lutz}, {Madden},
  {Maffei}, {Magdis}, {Mainetti}, {Maiolino}, {Marchetti}, {Morrison},
  {Mortier}, {Nguyen}, {Nordon}, {O'Halloran}, {Oliver}, {Omont}, {Owen},
  {Page}, {Panuzzo}, {Papageorgiou}, {Pearson}, {P{\'e}rez-Fournon}, {P{\'e}rez
  Garc{\'{\i}}a}, {Poglitsch}, {Pohlen}, {Popesso}, {Pozzi}, {Rawlings},
  {Raymond}, {Rigopoulou}, {Riguccini}, {Rizzo}, {Rodighiero}, {Roseboom},
  {Rowan-Robinson}, {Saintonge}, {Sanchez Portal}, {Santini}, {Schulz},
  {Scott}, {Seymour}, {Shao}, {Shupe}, {Smith}, {Stevens}, {Sturm},
  {Symeonidis}, {Tacconi}, {Trichas}, {Tugwell}, {Vaccari}, {Valtchanov},
  {Vieira}, {Vigroux}, {Wang}, {Ward}, {Wright}, {Xu}, \& {Zemcov}}]{ivison10}
{Ivison}, R.~J., {Magnelli}, B., {Ibar}, E., {et~al.} 2010, \aap, 518, L31

\bibitem[{{Juneau} {et~al.}(2013){Juneau}, {Dickinson}, {Bournaud},
  {Alexander}, {Daddi}, {Mullaney}, {Magnelli}, {Kartaltepe}, {Hwang},
  {Willner}, {Coil}, {Rosario}, {Trump}, {Weiner}, {Willmer}, {Cooper},
  {Elbaz}, {Faber}, {Frayer}, {Kocevski}, {Laird}, {Monkiewicz}, {Nandra},
  {Newman}, {Salim}, \& {Symeonidis}}]{juneau13}
{Juneau}, S., {Dickinson}, M., {Bournaud}, F., {et~al.} 2013, \apj, 764, 176

\bibitem[{{Kajisawa} {et~al.}(2011){Kajisawa}, {Ichikawa}, {Tanaka}, {Yamada},
  {Akiyama}, {Suzuki}, {Tokoku}, {Katsuno Uchimoto}, {Konishi}, {Yoshikawa},
  {Nishimura}, {Omata}, {Ouchi}, {Iwata}, {Hamana}, \& {Onodera}}]{kajisawa11}
{Kajisawa}, M., {Ichikawa}, T., {Tanaka}, I., {et~al.} 2011, \pasj, 63, 379

\bibitem[{{Karim} {et~al.}(2011){Karim}, {Schinnerer},
  {Mart{\'{\i}}nez-Sansigre}, {Sargent}, {van der Wel}, {Rix}, {Ilbert},
  {Smol{\v c}i{\'c}}, {Carilli}, {Pannella}, {Koekemoer}, {Bell}, \&
  {Salvato}}]{karim11}
{Karim}, A., {Schinnerer}, E., {Mart{\'{\i}}nez-Sansigre}, A., {et~al.} 2011,
  \apj, 730, 61

\bibitem[{{Kashino} {et~al.}(2013){Kashino}, {Silverman}, {Rodighiero},
  {Renzini}, {Arimoto}, {Daddi}, {Lilly}, {Sanders}, {Kartaltepe}, {Zahid},
  {Nagao}, {Sugiyama}, {Capak}, {Carollo}, {Chu}, {Hasinger}, {Ilbert},
  {Kajisawa}, {Kewley}, {Koekemoer}, {Kova{\v c}}, {Le F{\`e}vre}, {Masters},
  {McCracken}, {Onodera}, {Scoville}, {Strazzullo}, {Symeonidis}, \&
  {Taniguchi}}]{kashino13}
{Kashino}, D., {Silverman}, J.~D., {Rodighiero}, G., {et~al.} 2013, \apjl, 777,
  L8

\bibitem[{{Kennicutt}(1998)}]{kennicutt98}
{Kennicutt}, Jr., R.~C. 1998, \araa, 36, 189

\bibitem[{{Kirkpatrick} {et~al.}(2012){Kirkpatrick}, {Pope}, {Alexander},
  {Charmandaris}, {Daddi}, {Dickinson}, {Elbaz}, {Gabor}, {Hwang}, {Ivison},
  {Mullaney}, {Pannella}, {Scott}, {Altieri}, {Aussel}, {Bournaud}, {Buat},
  {Coia}, {Dannerbauer}, {Dasyra}, {Kartaltepe}, {Leiton}, {Lin}, {Magdis},
  {Magnelli}, {Morrison}, {Popesso}, \& {Valtchanov}}]{kirk12}
{Kirkpatrick}, A., {Pope}, A., {Alexander}, D.~M., {et~al.} 2012, \apj, 759,
  139

\bibitem[{{Kong} {et~al.}(2004){Kong}, {Charlot}, {Brinchmann}, \&
  {Fall}}]{kong2004}
{Kong}, X., {Charlot}, S., {Brinchmann}, J., \& {Fall}, S.~M. 2004, \mnras,
  349, 769

\bibitem[{{Kreckel} {et~al.}(2013){Kreckel}, {Groves}, {Schinnerer}, {Johnson},
  {Aniano}, {Calzetti}, {Croxall}, {Draine}, {Gordon}, {Crocker}, {Dale},
  {Hunt}, {Kennicutt}, {Meidt}, {Smith}, \& {Tabatabaei}}]{kreckel13}
{Kreckel}, K., {Groves}, B., {Schinnerer}, E., {et~al.} 2013, \apj, 771, 62

\bibitem[{{Kriek} \& {Conroy}(2013)}]{kriek13}
{Kriek}, M. \& {Conroy}, C. 2013, \apjl, 775, L16

\bibitem[{{Kriek} {et~al.}(2009){Kriek}, {van Dokkum}, {Labb{\'e}}, {Franx},
  {Illingworth}, {Marchesini}, \& {Quadri}}]{fast}
{Kriek}, M., {van Dokkum}, P.~G., {Labb{\'e}}, I., {et~al.} 2009, \apj, 700,
  221

\bibitem[{{Kurczynski} \& {Gawiser}(2010)}]{Kurcz10}
{Kurczynski}, P. \& {Gawiser}, E. 2010, \aj, 139, 1592

\bibitem[{{Law} {et~al.}(2011){Law}, {Gordon}, \& {Misselt}}]{law11}
{Law}, K.-H., {Gordon}, K.~D., \& {Misselt}, K.~A. 2011, \apj, 738, 124

\bibitem[{{Lilly} {et~al.}(2013){Lilly}, {Carollo}, {Pipino}, {Renzini}, \&
  {Peng}}]{lilly13}
{Lilly}, S.~J., {Carollo}, C.~M., {Pipino}, A., {Renzini}, A., \& {Peng}, Y.
  2013, \apj, 772, 119

\bibitem[{{Lin} {et~al.}(2012){Lin}, {Dickinson}, {Jian}, {Merson}, {Baugh},
  {Scott}, {Foucaud}, {Wang}, {Yan}, {Yan}, {Cheng}, {Guo}, {Helly}, {Kirsten},
  {Koo}, {Lagos}, {Meger}, {Messias}, {Pope}, {Simard}, {Grogin}, \&
  {Wang}}]{lin12}
{Lin}, L., {Dickinson}, M., {Jian}, H.-Y., {et~al.} 2012, \apj, 756, 71

\bibitem[{{Madau} \& {Dickinson}(2014)}]{mad14}
{Madau}, P. \& {Dickinson}, M. 2014, ArXiv~e-prints:~1403.0007

\bibitem[{{Magdis} {et~al.}(2012){Magdis}, {Daddi}, {B{\'e}thermin}, {Sargent},
  {Elbaz}, {Pannella}, {Dickinson}, {Dannerbauer}, {da Cunha}, {Walter},
  {Rigopoulou}, {Charmandaris}, {Hwang}, \& {Kartaltepe}}]{magdis12}
{Magdis}, G.~E., {Daddi}, E., {B{\'e}thermin}, M., {et~al.} 2012, \apj, 760, 6

\bibitem[{{Magdis} {et~al.}(2010){Magdis}, {Elbaz}, {Daddi}, {Morrison},
  {Dickinson}, {Rigopoulou}, {Gobat}, \& {Hwang}}]{magdis10}
{Magdis}, G.~E., {Elbaz}, D., {Daddi}, E., {et~al.} 2010, \apj, 714, 1740

\bibitem[{{Magnelli} {et~al.}(2014){Magnelli}, {Lutz}, {Saintonge}, {Berta},
  {Santini}, {Symeonidis}, {Altieri}, {Andreani}, {Aussel}, {B{\'e}thermin},
  {Bock}, {Bongiovanni}, {Cepa}, {Cimatti}, {Conley}, {Daddi}, {Elbaz},
  {F{\"o}rster Schreiber}, {Genzel}, {Ivison}, {Le Floc'h}, {Magdis},
  {Maiolino}, {Nordon}, {Oliver}, {Page}, {P{\'e}rez Garc{\'{\i}}a},
  {Poglitsch}, {Popesso}, {Pozzi}, {Riguccini}, {Rodighiero}, {Rosario},
  {Roseboom}, {Sanchez-Portal}, {Scott}, {Sturm}, {Tacconi}, {Valtchanov},
  {Wang}, \& {Wuyts}}]{mag14}
{Magnelli}, B., {Lutz}, D., {Saintonge}, A., {et~al.} 2014, \aap, 561, A86

\bibitem[{{Magnelli} {et~al.}(2012){Magnelli}, {Lutz}, {Santini}, {Saintonge},
  {Berta}, {Albrecht}, {Altieri}, {Andreani}, {Aussel}, {Bertoldi},
  {B{\'e}thermin}, {Bongiovanni}, {Capak}, {Chapman}, {Cepa}, {Cimatti},
  {Cooray}, {Daddi}, {Danielson}, {Dannerbauer}, {Dunlop}, {Elbaz}, {Farrah},
  {F{\"o}rster Schreiber}, {Genzel}, {Hwang}, {Ibar}, {Ivison}, {Le Floc'h},
  {Magdis}, {Maiolino}, {Nordon}, {Oliver}, {P{\'e}rez Garc{\'{\i}}a},
  {Poglitsch}, {Popesso}, {Pozzi}, {Riguccini}, {Rodighiero}, {Rosario},
  {Roseboom}, {Salvato}, {Sanchez-Portal}, {Scott}, {Smail}, {Sturm},
  {Swinbank}, {Tacconi}, {Valtchanov}, {Wang}, \& {Wuyts}}]{magnelli12}
{Magnelli}, B., {Lutz}, D., {Santini}, P., {et~al.} 2012, \aap, 539, A155

\bibitem[{{Maier} {et~al.}(2014){Maier}, {Lilly}, {Ziegler}, {Contini}, {Perez
  Montero}, {Peng}, \& {Balestra}}]{maier14}
{Maier}, C., {Lilly}, S.~J., {Ziegler}, B., {et~al.} 2014, ArXiv~e-prints:~1406.6069

\bibitem[{{Maiolino} {et~al.}(2004){Maiolino}, {Schneider}, {Oliva}, {Bianchi},
  {Ferrara}, {Mannucci}, {Pedani}, \& {Roca Sogorb}}]{maiolino04}
{Maiolino}, R., {Schneider}, R., {Oliva}, E., {et~al.} 2004, \nat, 431, 533

\bibitem[{{Mancini} {et~al.}(2011){Mancini}, {F{\"o}rster Schreiber},
  {Renzini}, {Cresci}, {Hicks}, {Peng}, {Vergani}, {Lilly}, {Carollo},
  {Pozzetti}, {Zamorani}, {Daddi}, {Genzel}, {Maraston}, {McCracken},
  {Tacconi}, {Bouch{\'e}}, {Davies}, {Oesch}, {Shapiro}, {Mainieri}, {Lutz},
  {Mignoli}, \& {Sternberg}}]{mancini11}
{Mancini}, C., {F{\"o}rster Schreiber}, N.~M., {Renzini}, A., {et~al.} 2011,
  \apj, 743, 86

\bibitem[{{Maraston} {et~al.}(2010){Maraston}, {Pforr}, {Renzini}, {Daddi},
  {Dickinson}, {Cimatti}, \& {Tonini}}]{maraston10}
{Maraston}, C., {Pforr}, J., {Renzini}, A., {et~al.} 2010, \mnras, 407, 830

\bibitem[{{Marchesini} {et~al.}(2009){Marchesini}, {van Dokkum}, {F{\"o}rster
  Schreiber}, {Franx}, {Labb{\'e}}, \& {Wuyts}}]{mmm09}
{Marchesini}, D., {van Dokkum}, P.~G., {F{\"o}rster Schreiber}, N.~M., {et~al.}
  2009, \apj, 701, 1765

\bibitem[{{Meurer} {et~al.}(1999){Meurer}, {Heckman}, \& {Calzetti}}]{meurer99}
{Meurer}, G.~R., {Heckman}, T.~M., \& {Calzetti}, D. 1999, \apj, 521, 64

\bibitem[{{Momcheva} {et~al.}(2013){Momcheva}, {Lee}, {Ly}, {Salim}, {Dale},
  {Ouchi}, {Finn}, \& {Ono}}]{momcheva13}
{Momcheva}, I.~G., {Lee}, J.~C., {Ly}, C., {et~al.} 2013, \aj, 145, 47

\bibitem[{{Morrison} {et~al.}(2010){Morrison}, {Owen}, {Dickinson}, {Ivison},
  \& {Ibar}}]{morrison00}
{Morrison}, G.~E., {Owen}, F.~N., {Dickinson}, M., {Ivison}, R.~J., \& {Ibar},
  E. 2010, \apjs, 188, 178

\bibitem[{{Mullaney} {et~al.}(2012){Mullaney}, {Pannella}, {Daddi},
  {Alexander}, {Elbaz}, {Hickox}, {Bournaud}, {Altieri}, {Aussel}, {Coia},
  {Dannerbauer}, {Dasyra}, {Dickinson}, {Hwang}, {Kartaltepe}, {Leiton},
  {Magdis}, {Magnelli}, {Popesso}, {Valtchanov}, {Bauer}, {Brandt}, {Del Moro},
  {Hanish}, {Ivison}, {Juneau}, {Luo}, {Lutz}, {Sargent}, {Scott}, \&
  {Xue}}]{mullaney12}
{Mullaney}, J.~R., {Pannella}, M., {Daddi}, E., {et~al.} 2012, \mnras, 419, 95

\bibitem[{{Muzzin} {et~al.}(2013){Muzzin}, {Marchesini}, {Stefanon}, {Franx},
  {McCracken}, {Milvang-Jensen}, {Dunlop}, {Fynbo}, {Brammer}, {Labb{\'e}}, \&
  {van Dokkum}}]{muzzin13}
{Muzzin}, A., {Marchesini}, D., {Stefanon}, M., {et~al.} 2013, \apj, 777, 18

\bibitem[{{Noeske} {et~al.}(2007){Noeske}, {Weiner}, {Faber}, {Papovich},
  {Koo}, {Somerville}, {Bundy}, {Conselice}, {Newman}, {Schiminovich}, {Le
  Floc'h}, {Coil}, {Rieke}, {Lotz}, {Primack}, {Barmby}, {Cooper}, {Davis},
  {Ellis}, {Fazio}, {Guhathakurta}, {Huang}, {Kassin}, {Martin}, {Phillips},
  {Rich}, {Small}, {Willmer}, \& {Wilson}}]{noeske07}
{Noeske}, K.~G., {Weiner}, B.~J., {Faber}, S.~M., {et~al.} 2007, \apjl, 660,
  L43

\bibitem[{{Nordon} {et~al.}(2013){Nordon}, {Lutz}, {Saintonge}, {Berta},
  {Wuyts}, {F{\"o}rster Schreiber}, {Genzel}, {Magnelli}, {Poglitsch},
  {Popesso}, {Rosario}, {Sturm}, \& {Tacconi}}]{nord13}
{Nordon}, R., {Lutz}, D., {Saintonge}, A., {et~al.} 2013, \apj, 762, 125

\bibitem[{{Oteo} {et~al.}(2014){Oteo}, {Bongiovanni}, {Magdis},
  {P{\'e}rez-Garc{\'{\i}}a}, {Cepa}, {Dom{\'{\i}}nguez S{\'a}nchez},
  {Ederoclite}, {S{\'a}nchez-Portal}, \& {Pintos-Castro}}]{oteo14}
{Oteo}, I., {Bongiovanni}, {\'A}., {Magdis}, G., {et~al.} 2014, \mnras, 439,
  1337

\bibitem[{{Oteo} {et~al.}(2013){Oteo}, {Magdis}, {Bongiovanni},
  {P{\'e}rez-Garc{\'{\i}}a}, {Cepa}, {Cedr{\'e}s}, {Ederoclite},
  {S{\'a}nchez-Portal}, {Aguerri}, {Alfaro}, {Altieri}, {Andreani},
  {Aparicio-Villegas}, {Aussel}, {Ben{\'{\i}}tez}, {Berta}, {Broadhurst},
  {Cabrera-Ca{\~n}o}, {Castander}, {Cervi{\~n}o}, {Cimatti},
  {Cristobal-Hornillos}, {Daddi}, {Elbaz}, {Fernandez-Soto}, {Schreiber},
  {Genzel}, {Gonzalez-Delgado}, {Husillos}, {Infante}, {Le Floc'h}, {Lutz},
  {Magnelli}, {Maiolino}, {M{\'a}rquez}, {Mart{\'{\i}}nez}, {Masegosa},
  {Matute}, {Moles}, {Molino}, {Olmo}, {Perea}, {P{\'e}rez-Mart{\'{\i}}nez},
  {Pintos-Castro}, {Poglitsch}, {Polednikova}, {Popesso}, {Povi{\'c}}, {Pozzi},
  {Prada}, {Quintana}, {Riguccini}, {Sturm}, {Tacconi}, {Valtchanov}, \&
  {Viironen}}]{oteo13}
{Oteo}, I., {Magdis}, G., {Bongiovanni}, {\'A}., {et~al.} 2013, \mnras, 435,
  158

\bibitem[{{Ouchi} {et~al.}(2013){Ouchi}, {Ellis}, {Ono}, {Nakanishi}, {Kohno},
  {Momose}, {Kurono}, {Ashby}, {Shimasaku}, {Willner}, {Fazio}, {Tamura}, \&
  {Iono}}]{ouchi13}
{Ouchi}, M., {Ellis}, R., {Ono}, Y., {et~al.} 2013, \apj, 778, 102

\bibitem[{{Ouchi} {et~al.}(2009){Ouchi}, {Mobasher}, {Shimasaku}, {Ferguson},
  {Fall}, {Ono}, {Kashikawa}, {Morokuma}, {Nakajima}, {Okamura}, {Dickinson},
  {Giavalisco}, \& {Ohta}}]{ouchi09}
{Ouchi}, M., {Mobasher}, B., {Shimasaku}, K., {et~al.} 2009, \apj, 706, 1136

\bibitem[{{Overzier} {et~al.}(2011){Overzier}, {Heckman}, {Wang}, {Armus},
  {Buat}, {Howell}, {Meurer}, {Seibert}, {Siana}, {Basu-Zych}, {Charlot},
  {Gon{\c c}alves}, {Martin}, {Neill}, {Rich}, {Salim}, \&
  {Schiminovich}}]{overzier11}
{Overzier}, R.~A., {Heckman}, T.~M., {Wang}, J., {et~al.} 2011, \apjl, 726, L7

\bibitem[{{Pannella} {et~al.}(2009{\natexlab{a}}){Pannella}, {Carilli},
  {Daddi}, {McCracken}, {Owen}, {Renzini}, {Strazzullo}, {Civano}, {Koekemoer},
  {Schinnerer}, {Scoville}, {Smol{\v c}i{\'c}}, {Taniguchi}, {Aussel}, {Kneib},
  {Ilbert}, {Mellier}, {Salvato}, {Thompson}, \& {Willott}}]{PP09}
{Pannella}, M., {Carilli}, C.~L., {Daddi}, E., {et~al.} 2009{\natexlab{a}},
  \apjl, 698, L116

\bibitem[{{Pannella} {et~al.}(2013){Pannella}, {Elbaz}, \& {Daddi}}]{p13}
{Pannella}, M., {Elbaz}, D., \& {Daddi}, E. 2013, in IAU Symposium, Vol. 292,
  IAU Symposium, ed. T.~{Wong} \& J.~{Ott}, 289--289

\bibitem[{{Pannella} {et~al.}(2009{\natexlab{b}}){Pannella}, {Gabasch},
  {Goranova}, {Drory}, {Hopp}, {Noll}, {Saglia}, {Strazzullo}, \&
  {Bender}}]{P09}
{Pannella}, M., {Gabasch}, A., {Goranova}, Y., {et~al.} 2009{\natexlab{b}},
  \apj, 701, 787

\bibitem[{{Pannella} {et~al.}(2006){Pannella}, {Hopp}, {Saglia}, {Bender},
  {Drory}, {Salvato}, {Gabasch}, \& {Feulner}}]{P06}
{Pannella}, M., {Hopp}, U., {Saglia}, R.~P., {et~al.} 2006, \apjl, 639, L1

\bibitem[{{Papovich} {et~al.}(2011){Papovich}, {Finkelstein}, {Ferguson},
  {Lotz}, \& {Giavalisco}}]{papovich11}
{Papovich}, C., {Finkelstein}, S.~L., {Ferguson}, H.~C., {Lotz}, J.~M., \&
  {Giavalisco}, M. 2011, \mnras, 412, 1123

\bibitem[{{Papovich} {et~al.}(2006){Papovich}, {Moustakas}, {Dickinson}, {Le
  Floc'h}, {Rieke}, {Daddi}, {Alexander}, {Bauer}, {Brandt}, {Dahlen}, {Egami},
  {Eisenhardt}, {Elbaz}, {Ferguson}, {Giavalisco}, {Lucas}, {Mobasher},
  {P{\'e}rez-Gonz{\'a}lez}, {Stutz}, {Rieke}, \& {Yan}}]{papovich06}
{Papovich}, C., {Moustakas}, L.~A., {Dickinson}, M., {et~al.} 2006, \apj, 640,
  92

\bibitem[{{Peng} {et~al.}(2002){Peng}, {Ho}, {Impey}, \& {Rix}}]{peng2002}
{Peng}, C.~Y., {Ho}, L.~C., {Impey}, C.~D., \& {Rix}, H.-W. 2002, \aj, 124, 266

\bibitem[{{Peng} {et~al.}(2010){Peng}, {Lilly}, {Kova{\v c}}, {Bolzonella},
  {Pozzetti}, {Renzini}, {Zamorani}, {Ilbert}, {Knobel}, {Iovino}, {Maier},
  {Cucciati}, {Tasca}, {Carollo}, {Silverman}, {Kampczyk}, {de Ravel},
  {Sanders}, {Scoville}, {Contini}, {Mainieri}, {Scodeggio}, {Kneib}, {Le
  F{\`e}vre}, {Bardelli}, {Bongiorno}, {Caputi}, {Coppa}, {de la Torre},
  {Franzetti}, {Garilli}, {Lamareille}, {Le Borgne}, {Le Brun}, {Mignoli},
  {Perez Montero}, {Pello}, {Ricciardelli}, {Tanaka}, {Tresse}, {Vergani},
  {Welikala}, {Zucca}, {Oesch}, {Abbas}, {Barnes}, {Bordoloi}, {Bottini},
  {Cappi}, {Cassata}, {Cimatti}, {Fumana}, {Hasinger}, {Koekemoer},
  {Leauthaud}, {Maccagni}, {Marinoni}, {McCracken}, {Memeo}, {Meneux}, {Nair},
  {Porciani}, {Presotto}, \& {Scaramella}}]{peng10}
{Peng}, Y.-j., {Lilly}, S.~J., {Kova{\v c}}, K., {et~al.} 2010, \apj, 721, 193

\bibitem[{{Penner} {et~al.}(2012){Penner}, {Dickinson}, {Pope}, {Dey},
  {Magnelli}, {Pannella}, {Altieri}, {Aussel}, {Buat}, {Bussmann},
  {Charmandaris}, {Coia}, {Daddi}, {Dannerbauer}, {Elbaz}, {Hwang},
  {Kartaltepe}, {Lin}, {Magdis}, {Morrison}, {Popesso}, {Scott}, \&
  {Valtchanov}}]{penner12}
{Penner}, K., {Dickinson}, M., {Pope}, A., {et~al.} 2012, \apj, 759, 28

\bibitem[{{P{\'e}rez-Gonz{\'a}lez} {et~al.}(2008){P{\'e}rez-Gonz{\'a}lez},
  {Rieke}, {Villar}, {Barro}, {Blaylock}, {Egami}, {Gallego}, {Gil de Paz},
  {Pascual}, {Zamorano}, \& {Donley}}]{pgp08}
{P{\'e}rez-Gonz{\'a}lez}, P.~G., {Rieke}, G.~H., {Villar}, V., {et~al.} 2008,
  \apj, 675, 234

\bibitem[{{Popesso} {et~al.}(2012){Popesso}, {Magnelli}, {Buttiglione}, {Lutz},
  {Poglitsch}, {Berta}, {Nordon}, {Altieri}, {Aussel}, {Billot}, {Gastaud},
  {Ali}, {Balog}, {Cava}, {Feuchtgruber}, {Gonzalez Garcia}, {Geis}, {Kiss},
  {Klaas}, {Linz}, {Liu}, {Moor}, {Morin}, {Muller}, {Nielbock}, {Okumura},
  {Osterhage}, {Ottensamer}, {Paladini}, {Pezzuto}, {Dublier Pritchard},
  {Regibo}, {Rodighiero}, {Royer}, {Sauvage}, {Sturm}, {Wetzstein},
  {Wieprecht}, \& {Wiezorrek}}]{popesso12}
{Popesso}, P., {Magnelli}, B., {Buttiglione}, S., {et~al.} 2012, ArXiv~e-prints:~1211.4257

\bibitem[{{Price} {et~al.}(2014){Price}, {Kriek}, {Brammer}, {Conroy},
  {F{\"o}rster Schreiber}, {Franx}, {Fumagalli}, {Lundgren}, {Momcheva},
  {Nelson}, {Skelton}, {van Dokkum}, {Whitaker}, \& {Wuyts}}]{price13}
{Price}, S.~H., {Kriek}, M., {Brammer}, G.~B., {et~al.} 2014, \apj, 788, 86

\bibitem[{{Reddy} {et~al.}(2012){Reddy}, {Dickinson}, {Elbaz}, {Morrison},
  {Giavalisco}, {Ivison}, {Papovich}, {Scott}, {Buat}, {Burgarella},
  {Charmandaris}, {Daddi}, {Magdis}, {Murphy}, {Altieri}, {Aussel},
  {Dannerbauer}, {Dasyra}, {Hwang}, {Kartaltepe}, {Leiton}, {Magnelli}, \&
  {Popesso}}]{reddy12}
{Reddy}, N., {Dickinson}, M., {Elbaz}, D., {et~al.} 2012, \apj, 744, 154

\bibitem[{{Reddy} {et~al.}(2010){Reddy}, {Erb}, {Pettini}, {Steidel}, \&
  {Shapley}}]{reddy2010}
{Reddy}, N.~A., {Erb}, D.~K., {Pettini}, M., {Steidel}, C.~C., \& {Shapley},
  A.~E. 2010, \apj, 712, 1070

\bibitem[{{Renzini}(2009)}]{renzini09}
{Renzini}, A. 2009, \mnras, 398, L58

\bibitem[{{Rodighiero} {et~al.}(2010){Rodighiero}, {Cimatti}, {Gruppioni},
  {Popesso}, {Andreani}, {Altieri}, {Aussel}, {Berta}, {Bongiovanni},
  {Brisbin}, {Cava}, {Cepa}, {Daddi}, {Dominguez-Sanchez}, {Elbaz}, {Fontana},
  {F{\"o}rster Schreiber}, {Franceschini}, {Genzel}, {Grazian}, {Lutz},
  {Magdis}, {Magliocchetti}, {Magnelli}, {Maiolino}, {Mancini}, {Nordon},
  {Perez Garcia}, {Poglitsch}, {Santini}, {Sanchez-Portal}, {Pozzi},
  {Riguccini}, {Saintonge}, {Shao}, {Sturm}, {Tacconi}, {Valtchanov},
  {Wetzstein}, \& {Wieprecht}}]{rod10}
{Rodighiero}, G., {Cimatti}, A., {Gruppioni}, C., {et~al.} 2010, \aap, 518, L25

\bibitem[{{Rodighiero} {et~al.}(2011){Rodighiero}, {Daddi}, {Baronchelli},
  {Cimatti}, {Renzini}, {Aussel}, {Popesso}, {Lutz}, {Andreani}, {Berta},
  {Cava}, {Elbaz}, {Feltre}, {Fontana}, {F{\"o}rster Schreiber},
  {Franceschini}, {Genzel}, {Grazian}, {Gruppioni}, {Ilbert}, {Le Floch},
  {Magdis}, {Magliocchetti}, {Magnelli}, {Maiolino}, {McCracken}, {Nordon},
  {Poglitsch}, {Santini}, {Pozzi}, {Riguccini}, {Tacconi}, {Wuyts}, \&
  {Zamorani}}]{rod11}
{Rodighiero}, G., {Daddi}, E., {Baronchelli}, I., {et~al.} 2011, \apjl, 739,
  L40

\bibitem[{{Rodighiero} {et~al.}(2014){Rodighiero}, {Renzini}, {Daddi},
  {Baronchelli}, {Berta}, {Cresci}, {Franceschini}, {Gruppioni}, {Lutz},
  {Mancini}, {Santini}, {Zamorani}, {Silverman}, {Kashino}, {Andreani},
  {Cimatti}, {Dominguez Sanchez}, {Le Floch}, {Magnelli}, {Popesso}, \&
  {Pozzi}}]{rodighiero14}
{Rodighiero}, G., {Renzini}, A., {Daddi}, E., {et~al.} 2014, ArXiv~e-prints:~1406.1189

\bibitem[{{Rosario} {et~al.}(2013){Rosario}, {Mozena}, {Wuyts}, {Nandra},
  {Koekemoer}, {McGrath}, {Hathi}, {Dekel}, {Donley}, {Dunlop}, {Faber},
  {Ferguson}, {Giavalisco}, {Grogin}, {Guo}, {Kocevski}, {Koo}, {Laird},
  {Newman}, {Rangel}, \& {Somerville}}]{rosario13}
{Rosario}, D.~J., {Mozena}, M., {Wuyts}, S., {et~al.} 2013, \apj, 763, 59

\bibitem[{{Salim} {et~al.}(2005){Salim}, {Charlot}, {Rich}, {Kauffmann},
  {Heckman}, {Barlow}, {Bianchi}, {Byun}, {Donas}, {Forster}, {Friedman},
  {Jelinsky}, {Lee}, {Madore}, {Malina}, {Martin}, {Milliard}, {Morrissey},
  {Neff}, {Schiminovich}, {Seibert}, {Siegmund}, {Small}, {Szalay}, {Welsh}, \&
  {Wyder}}]{salim2005}
{Salim}, S., {Charlot}, S., {Rich}, R.~M., {et~al.} 2005, \apjl, 619, L39

\bibitem[{{Salim} {et~al.}(2007){Salim}, {Rich}, {Charlot}, {Brinchmann},
  {Johnson}, {Schiminovich}, {Seibert}, {Mallery}, {Heckman}, {Forster},
  {Friedman}, {Martin}, {Morrissey}, {Neff}, {Small}, {Wyder}, {Bianchi},
  {Donas}, {Lee}, {Madore}, {Milliard}, {Szalay}, {Welsh}, \& {Yi}}]{salim07}
{Salim}, S., {Rich}, R.~M., {Charlot}, S., {et~al.} 2007, \apjs, 173, 267

\bibitem[{{Salmi} {et~al.}(2012){Salmi}, {Daddi}, {Elbaz}, {Sargent},
  {Dickinson}, {Renzini}, {Bethermin}, \& {Le Borgne}}]{salmi12}
{Salmi}, F., {Daddi}, E., {Elbaz}, D., {et~al.} 2012, \apjl, 754, L14

\bibitem[{{Salpeter}(1955)}]{salpeter}
{Salpeter}, E.~E. 1955, \apj, 121, 161

\bibitem[{{Salvato} {et~al.}(2009){Salvato}, {Hasinger}, {Ilbert}, {Zamorani},
  {Brusa}, {Scoville}, {Rau}, {Capak}, {Arnouts}, {Aussel}, {Bolzonella},
  {Buongiorno}, {Cappelluti}, {Caputi}, {Civano}, {Cook}, {Elvis}, {Gilli},
  {Jahnke}, {Kartaltepe}, {Impey}, {Lamareille}, {Le Floc'h}, {Lilly},
  {Mainieri}, {McCarthy}, {McCracken}, {Mignoli}, {Mobasher}, {Murayama},
  {Sasaki}, {Sanders}, {Schiminovich}, {Shioya}, {Shopbell}, {Silverman},
  {Smol{\v c}i{\'c}}, {Surace}, {Taniguchi}, {Thompson}, {Trump}, {Urry}, \&
  {Zamojski}}]{salvato09}
{Salvato}, M., {Hasinger}, G., {Ilbert}, O., {et~al.} 2009, \apj, 690, 1250

\bibitem[{{Santini} {et~al.}(2009){Santini}, {Fontana}, {Grazian}, {Salimbeni},
  {Fiore}, {Fontanot}, {Boutsia}, {Castellano}, {Cristiani}, {de Santis},
  {Gallozzi}, {Giallongo}, {Menci}, {Nonino}, {Paris}, {Pentericci}, \&
  {Vanzella}}]{santini09}
{Santini}, P., {Fontana}, A., {Grazian}, A., {et~al.} 2009, \aap, 504, 751

\bibitem[{{Santini} {et~al.}(2012){Santini}, {Rosario}, {Shao}, {Lutz},
  {Maiolino}, {Alexander}, {Altieri}, {Andreani}, {Aussel}, {Bauer}, {Berta},
  {Bongiovanni}, {Brandt}, {Brusa}, {Cepa}, {Cimatti}, {Daddi}, {Elbaz},
  {Fontana}, {F{\"o}rster Schreiber}, {Genzel}, {Grazian}, {Le Floc'h},
  {Magnelli}, {Mainieri}, {Nordon}, {P{\'e}rez Garcia}, {Poglitsch}, {Popesso},
  {Pozzi}, {Riguccini}, {Rodighiero}, {Salvato}, {Sanchez-Portal}, {Sturm},
  {Tacconi}, {Valtchanov}, \& {Wuyts}}]{santini12}
{Santini}, P., {Rosario}, D.~J., {Shao}, L., {et~al.} 2012, \aap, 540, A109

\bibitem[{{Sargent} {et~al.}(2012){Sargent}, {B{\'e}thermin}, {Daddi}, \&
  {Elbaz}}]{sargent12}
{Sargent}, M.~T., {B{\'e}thermin}, M., {Daddi}, E., \& {Elbaz}, D. 2012, \apjl,
  747, L31

\bibitem[{{Sargent} {et~al.}(2010){Sargent}, {Schinnerer}, {Murphy}, {Aussel},
  {Le Floc'h}, {Frayer}, {Mart{\'{\i}}nez-Sansigre}, {Oesch}, {Salvato},
  {Smol{\v c}i{\'c}}, {Zamorani}, {Brusa}, {Cappelluti}, {Carilli}, {Carollo},
  {Ilbert}, {Kartaltepe}, {Koekemoer}, {Lilly}, {Sanders}, \&
  {Scoville}}]{sargent10}
{Sargent}, M.~T., {Schinnerer}, E., {Murphy}, E., {et~al.} 2010, \apjs, 186,
  341

\bibitem[{{Savaglio} {et~al.}(2005){Savaglio}, {Glazebrook}, {Le Borgne},
  {Juneau}, {Abraham}, {Chen}, {Crampton}, {McCarthy}, {Carlberg}, {Marzke},
  {Roth}, {J{\o}rgensen}, \& {Murowinski}}]{savaglio05}
{Savaglio}, S., {Glazebrook}, K., {Le Borgne}, D., {et~al.} 2005, \apj, 635,
  260

\bibitem[{{Schaerer} \& {de Barros}(2010)}]{schaerer10}
{Schaerer}, D. \& {de Barros}, S. 2010, \aap, 515, A73

\bibitem[{{Schreiber} {et~al.}(2015){Schreiber}, {Pannella}, {Elbaz},
  {B{\'e}thermin}, {Inami}, {Dickinson}, {Magnelli}, {Wang}, {Aussel}, {Daddi},
  {Juneau}, {Shu}, {Sargent}, {Buat}, {Faber}, {Ferguson}, {Giavalisco},
  {Koekemoer}, {Magdis}, {Morrison}, {Papovich}, {Santini}, \&
  {Scott}}]{schreiber15}
{Schreiber}, C., {Pannella}, M., {Elbaz}, D., {et~al.} 2015, \aap, 575, A74

\bibitem[{{Sobral} {et~al.}(2012){Sobral}, {Best}, {Matsuda}, {Smail}, {Geach},
  \& {Cirasuolo}}]{sobral12}
{Sobral}, D., {Best}, P.~N., {Matsuda}, Y., {et~al.} 2012, \mnras, 420, 1926

\bibitem[{{Stark} {et~al.}(2009){Stark}, {Ellis}, {Bunker}, {Bundy}, {Targett},
  {Benson}, \& {Lacy}}]{stark09}
{Stark}, D.~P., {Ellis}, R.~S., {Bunker}, A., {et~al.} 2009, \apj, 697, 1493

\bibitem[{{Steidel} {et~al.}(2014){Steidel}, {Rudie}, {Strom}, {Pettini},
  {Reddy}, {Shapley}, {Trainor}, {Erb}, {Turner}, {Konidaris}, {Kulas}, {Mace},
  {Matthews}, \& {McLean}}]{steidel14}
{Steidel}, C.~C., {Rudie}, G.~C., {Strom}, A.~L., {et~al.} 2014, ArXiv~e-prints:~1405.5473

\bibitem[{{Strazzullo} {et~al.}(2013){Strazzullo}, {Gobat}, {Daddi}, {Onodera},
  {Carollo}, {Dickinson}, {Renzini}, {Arimoto}, {Cimatti}, {Finoguenov}, \&
  {Chary}}]{strazzullo13}
{Strazzullo}, V., {Gobat}, R., {Daddi}, E., {et~al.} 2013, \apj, 772, 118

\bibitem[{{Strazzullo} {et~al.}(2010){Strazzullo}, {Pannella}, {Owen},
  {Bender}, {Morrison}, {Wang}, \& {Shupe}}]{strazzullo10}
{Strazzullo}, V., {Pannella}, M., {Owen}, F.~N., {et~al.} 2010, \apj, 714, 1305

\bibitem[{{Symeonidis} {et~al.}(2013){Symeonidis}, {Vaccari}, {Berta}, {Page},
  {Lutz}, {Arumugam}, {Aussel}, {Bock}, {Boselli}, {Buat}, {Capak}, {Clements},
  {Conley}, {Conversi}, {Cooray}, {Dowell}, {Farrah}, {Franceschini},
  {Giovannoli}, {Glenn}, {Griffin}, {Hatziminaoglou}, {Hwang}, {Ibar},
  {Ilbert}, {Ivison}, {Floc'h}, {Lilly}, {Kartaltepe}, {Magnelli}, {Magdis},
  {Marchetti}, {Nguyen}, {Nordon}, {O'Halloran}, {Oliver}, {Omont},
  {Papageorgiou}, {Patel}, {Pearson}, {P{\'e}rez-Fournon}, {Pohlen}, {Popesso},
  {Pozzi}, {Rigopoulou}, {Riguccini}, {Rosario}, {Roseboom}, {Rowan-Robinson},
  {Salvato}, {Schulz}, {Scott}, {Seymour}, {Shupe}, {Smith}, {Valtchanov},
  {Wang}, {Xu}, {Zemcov}, \& {Wuyts}}]{symeo13}
{Symeonidis}, M., {Vaccari}, M., {Berta}, S., {et~al.} 2013, \mnras, 431, 2317

\bibitem[{{Tan} {et~al.}(2013){Tan}, {Daddi}, {Sargent}, {Magdis}, {Hodge},
  {B{\'e}thermin}, {Bournaud}, {Carilli}, {Dannerbauer}, {Dickinson}, {Elbaz},
  {Gao}, {Morrison}, {Owen}, {Pannella}, {Riechers}, \& {Walter}}]{tan13}
{Tan}, Q., {Daddi}, E., {Sargent}, M., {et~al.} 2013, \apjl, 776, L24

\bibitem[{{Thomas} {et~al.}(2005){Thomas}, {Maraston}, {Bender}, \& {Mendes de
  Oliveira}}]{thomas2005}
{Thomas}, D., {Maraston}, C., {Bender}, R., \& {Mendes de Oliveira}, C. 2005,
  \apj, 621, 673

\bibitem[{{Tremonti} {et~al.}(2004){Tremonti}, {Heckman}, {Kauffmann},
  {Brinchmann}, {Charlot}, {White}, {Seibert}, {Peng}, {Schlegel}, {Uomoto},
  {Fukugita}, \& {Brinkmann}}]{tremonti04}
{Tremonti}, C.~A., {Heckman}, T.~M., {Kauffmann}, G., {et~al.} 2004, \apj, 613,
  898

\bibitem[{{Troncoso} {et~al.}(2014){Troncoso}, {Maiolino}, {Sommariva},
  {Cresci}, {Mannucci}, {Marconi}, {Meneghetti}, {Grazian}, {Cimatti},
  {Fontana}, {Nagao}, \& {Pentericci}}]{troncoso13}
{Troncoso}, P., {Maiolino}, R., {Sommariva}, V., {et~al.} 2014, \aap, 563, A58

\bibitem[{{Viero} {et~al.}(2013){Viero}, {Moncelsi}, {Quadri}, {Arumugam},
  {Assef}, {B{\'e}thermin}, {Bock}, {Bridge}, {Casey}, {Conley}, {Cooray},
  {Farrah}, {Glenn}, {Heinis}, {Ibar}, {Ikarashi}, {Ivison}, {Kohno},
  {Marsden}, {Oliver}, {Roseboom}, {Schulz}, {Scott}, {Serra}, {Vaccari},
  {Vieira}, {Wang}, {Wardlow}, {Wilson}, {Yun}, \& {Zemcov}}]{viero13}
{Viero}, M.~P., {Moncelsi}, L., {Quadri}, R.~F., {et~al.} 2013, \apj, 779, 32

\bibitem[{{Wang} {et~al.}(2010){Wang}, {Cowie}, {Barger}, {Keenan}, \&
  {Ting}}]{Wang10}
{Wang}, W.-H., {Cowie}, L.~L., {Barger}, A.~J., {Keenan}, R.~C., \& {Ting},
  H.-C. 2010, \apjs, 187, 251

\bibitem[{{Whitaker} {et~al.}(2011){Whitaker}, {Labb{\'e}}, {van Dokkum},
  {Brammer}, {Kriek}, {Marchesini}, {Quadri}, {Franx}, {Muzzin}, {Williams},
  {Bezanson}, {Illingworth}, {Lee}, {Lundgren}, {Nelson}, {Rudnick}, {Tal}, \&
  {Wake}}]{whita11}
{Whitaker}, K.~E., {Labb{\'e}}, I., {van Dokkum}, P.~G., {et~al.} 2011, \apj,
  735, 86

\bibitem[{{Whitaker} {et~al.}(2012){Whitaker}, {van Dokkum}, {Brammer}, \&
  {Franx}}]{whita12}
{Whitaker}, K.~E., {van Dokkum}, P.~G., {Brammer}, G., \& {Franx}, M. 2012,
  \apjl, 754, L29

\bibitem[{{Wild} {et~al.}(2011){Wild}, {Charlot}, {Brinchmann}, {Heckman},
  {Vince}, {Pacifici}, \& {Chevallard}}]{wild11}
{Wild}, V., {Charlot}, S., {Brinchmann}, J., {et~al.} 2011, \mnras, 417, 1760

\bibitem[{{Williams} {et~al.}(2009){Williams}, {Quadri}, {Franx}, {van Dokkum},
  \& {Labb{\'e}}}]{williams09}
{Williams}, R.~J., {Quadri}, R.~F., {Franx}, M., {van Dokkum}, P., \&
  {Labb{\'e}}, I. 2009, \apj, 691, 1879

\bibitem[{{Wuyts} {et~al.}(2014){Wuyts}, {Kurk}, {F{\"o}rster Schreiber},
  {Genzel}, {Wisnioski}, {Bandara}, {Wuyts}, {Beifiori}, {Bender}, {Brammer},
  {Burkert}, {Buschkamp}, {Carollo}, {Chan}, {Davies}, {Eisenhauer}, {Fossati},
  {Kulkarni}, {Lang}, {Lilly}, {Lutz}, {Mancini}, {Mendel}, {Momcheva}, {Naab},
  {Nelson}, {Renzini}, {Rosario}, {Saglia}, {Seitz}, {Sharples}, {Sternberg},
  {Tacchella}, {Tacconi}, {van Dokkum}, \& {Wilman}}]{wuyts14}
{Wuyts}, E., {Kurk}, J., {F{\"o}rster Schreiber}, N.~M., {et~al.} 2014, ArXiv
  e-prints: 1403.0007

\bibitem[{{Wuyts} {et~al.}(2013){Wuyts}, {F{\"o}rster Schreiber}, {Nelson},
  {van Dokkum}, {Brammer}, {Chang}, {Faber}, {Ferguson}, {Franx}, {Fumagalli},
  {Genzel}, {Grogin}, {Kocevski}, {Koekemoer}, {Lundgren}, {Lutz}, {McGrath},
  {Momcheva}, {Rosario}, {Skelton}, {Tacconi}, {van der Wel}, \&
  {Whitaker}}]{wuyts13}
{Wuyts}, S., {F{\"o}rster Schreiber}, N.~M., {Nelson}, E.~J., {et~al.} 2013,
  \apj, 779, 135

\bibitem[{{Wuyts} {et~al.}(2011){Wuyts}, {F{\"o}rster Schreiber}, {van der
  Wel}, {Magnelli}, {Guo}, {Genzel}, {Lutz}, {Aussel}, {Barro}, {Berta},
  {Cava}, {Graci{\'a}-Carpio}, {Hathi}, {Huang}, {Kocevski}, {Koekemoer},
  {Lee}, {Le Floc'h}, {McGrath}, {Nordon}, {Popesso}, {Pozzi}, {Riguccini},
  {Rodighiero}, {Saintonge}, \& {Tacconi}}]{wuyts11}
{Wuyts}, S., {F{\"o}rster Schreiber}, N.~M., {van der Wel}, A., {et~al.} 2011,
  \apj, 742, 96

\bibitem[{{Wuyts} {et~al.}(2007){Wuyts}, {Labb{\'e}}, {Franx}, {Rudnick}, {van
  Dokkum}, {Fazio}, {F{\"o}rster Schreiber}, {Huang}, {Moorwood}, {Rix},
  {R{\"o}ttgering}, \& {van der Werf}}]{wuyts07}
{Wuyts}, S., {Labb{\'e}}, I., {Franx}, M., {et~al.} 2007, \apj, 655, 51

\bibitem[{{Yun} {et~al.}(2001){Yun}, {Reddy}, \& {Condon}}]{yun2001}
{Yun}, M.~S., {Reddy}, N.~A., \& {Condon}, J.~J. 2001, \apj, 554, 803

\bibitem[{{Zahid} {et~al.}(2013{\natexlab{a}}){Zahid}, {Geller}, {Kewley},
  {Hwang}, {Fabricant}, \& {Kurtz}}]{zz13}
{Zahid}, H.~J., {Geller}, M.~J., {Kewley}, L.~J., {et~al.} 2013{\natexlab{a}},
  \apjl, 771, L19

\bibitem[{{Zahid} {et~al.}(2013{\natexlab{b}}){Zahid}, {Yates}, {Kewley}, \&
  {Kudritzki}}]{zahid13}
{Zahid}, H.~J., {Yates}, R.~M., {Kewley}, L.~J., \& {Kudritzki}, R.~P.
  2013{\natexlab{b}}, \apj, 763, 92

\end{thebibliography}

\appendix

\section{On the bolometric corrections for stacking results} 
\label{applir}
\begin{figure}
\begin{center}
  \includegraphics[height=.460\textwidth]{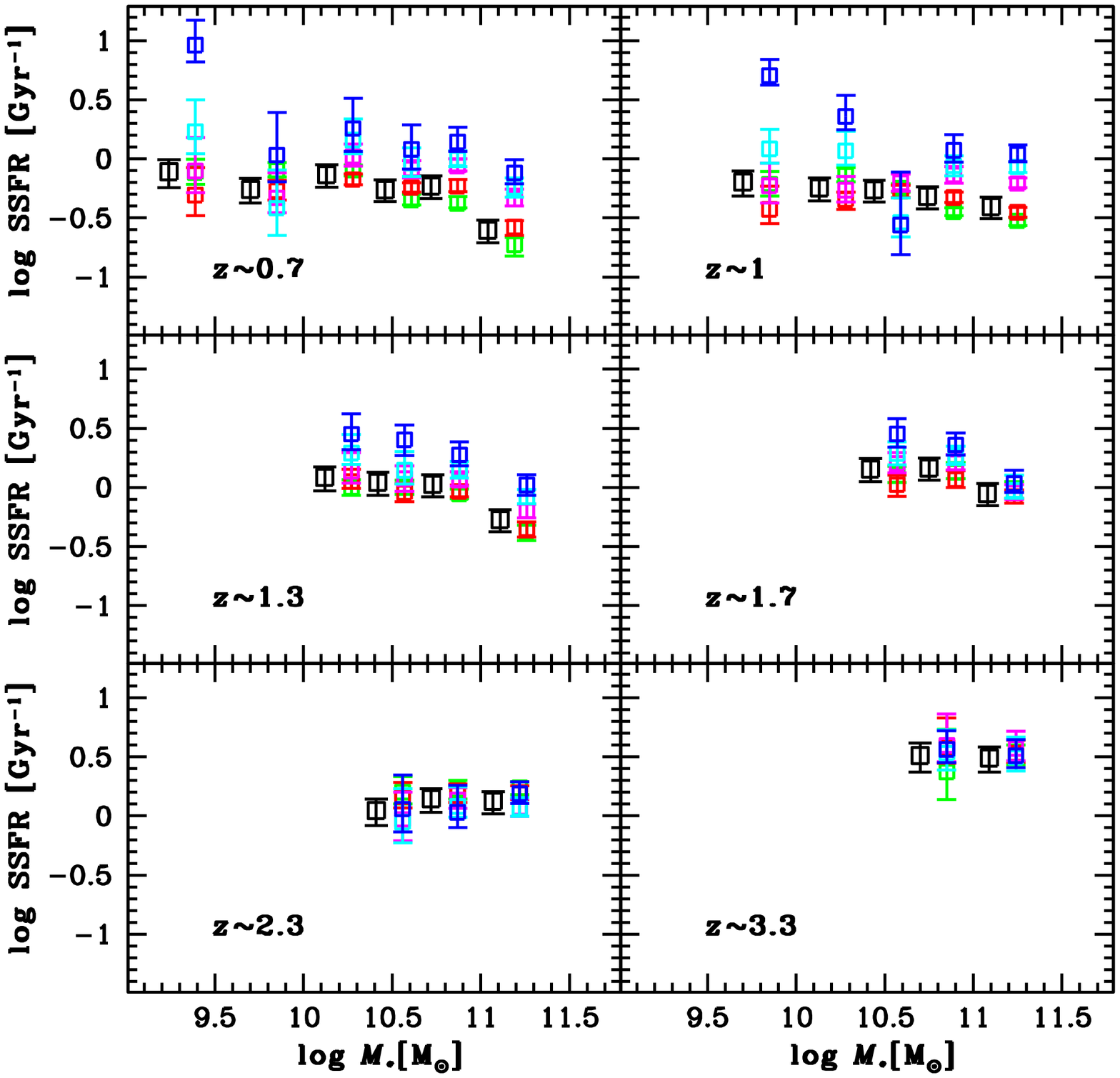}%
  \includegraphics[height=.460\textwidth]{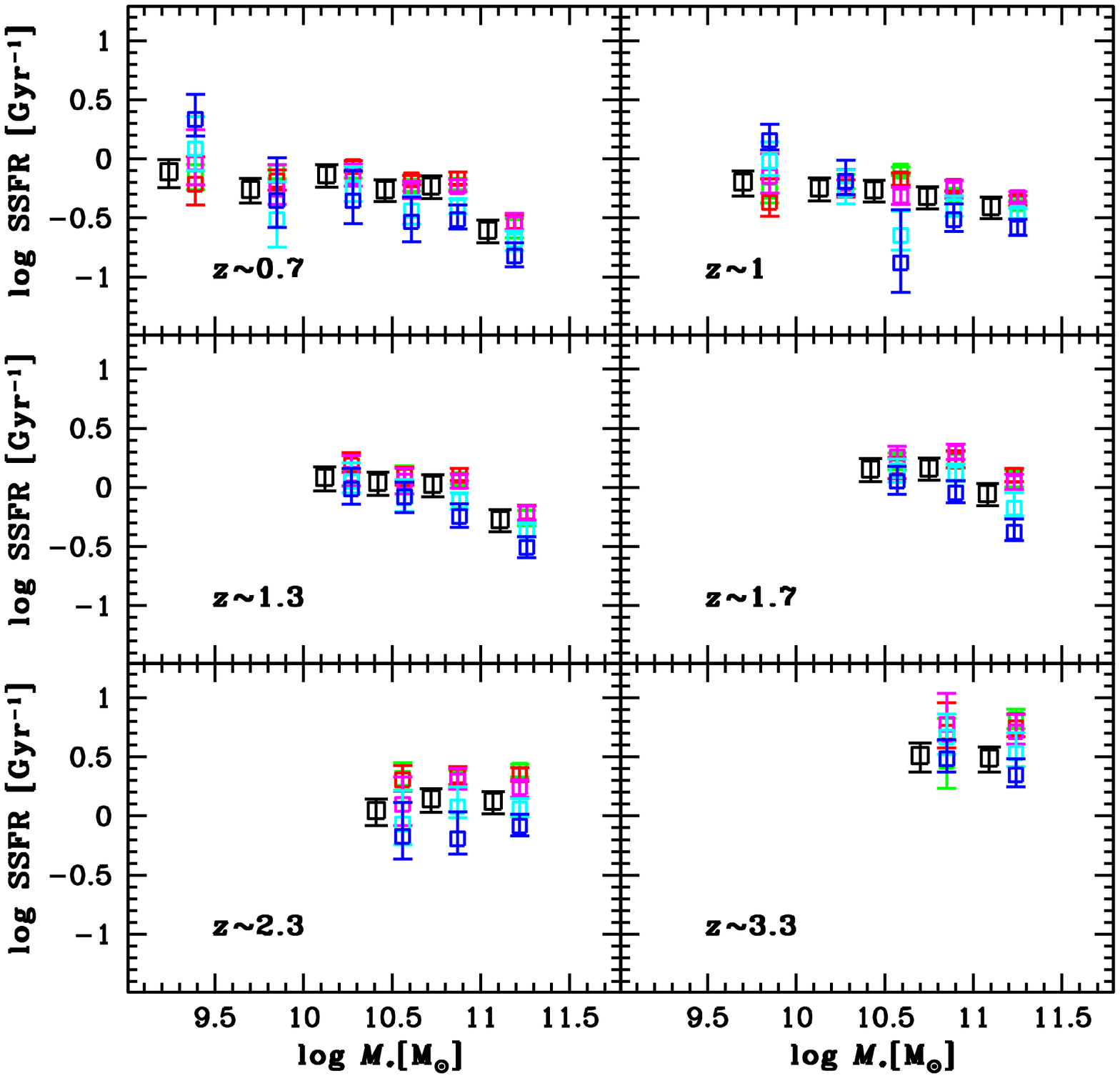}
  \caption{{\bf Left:} Single band derived SSFRs (green--100\,$\mu$m, red--160\,$\mu$m, magenta--250\,$\mu$m, cyan--350\,$\mu$m and blue--500\,$\mu$m) after applying a bolometric correction to each stacking result using the Chary\&Elbaz (2001) library. We also plot in black, horizontally offset for the sake of clarity, the SSFR derived from the multi-band IR fitting again using the same library but allowing for a free normalization of the SEDs.~{\bf Right:} The same information as in the other panel, with the only difference that we have this time used the MS SED described in Elbaz et al.~(2011) to derive bolometric corrections for all the IR band flux densities output from the stacking routine.}
\label{elbazerie}
\end{center}
\end{figure}

In this appendix we discuss the impact of bolometric corrections in the different \h~bands on the derivation of the total IR luminosity. The GOODS-N field is one of the few fields observed deeply in the IR \h~bands, which makes it special compared to other deep extragalactic fields. Previously only deep \s~24\,$\mu$m data were available to infer the total IR emission and hence an estimate of the SFR. Here we test this with all the PACS/SPIRE bands of \h~by comparing the outcome of the Chary~\&~Elbaz (2001) library\footnote{Publicly available at http://david.elbaz3.free.fr/astro\_codes/chary\_elbaz.html} and the new MS SED defined in Elbaz et al.~(2011), as we have shown in Figure~\ref{elbazerie}. A similar comparison has already been presented in \citet{elbaba10} and Elbaz et al.~(2011) but this was limited to \h~detections. Here we extend this to IR stacking results which could be useful in case that only relatively shallow data are available. By comparing to the results from the multi-band IR SED-fitting, we can draw the following conclusions:~a) the new MS template (Elbaz et al.~2011) significantly improves the bolometric corrections for the SPIRE bands only up to redshift $z\simeq$2;~and b) the Chary~\&~Elbaz (2001) recipe works best for the bolometric corrections of the PACS bands at all the redshifts explored.

\section{Testing radio stacking at different resolutions}
\label{apprad}

\begin{figure}
\begin{center}
  \includegraphics[height=.460\textwidth]{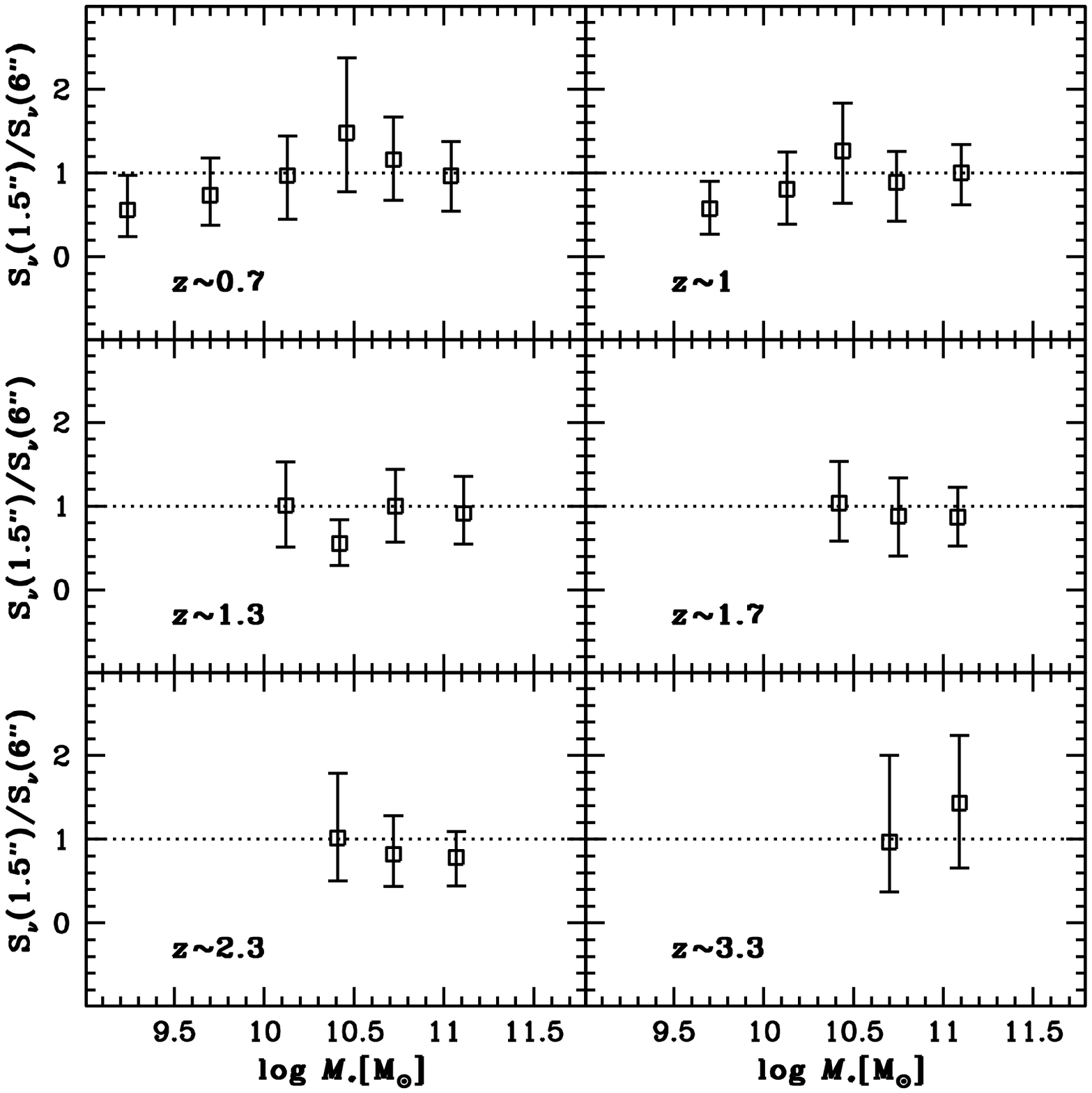}%
  \includegraphics[height=.46\textwidth, bb = 20 150 590 660, clip]{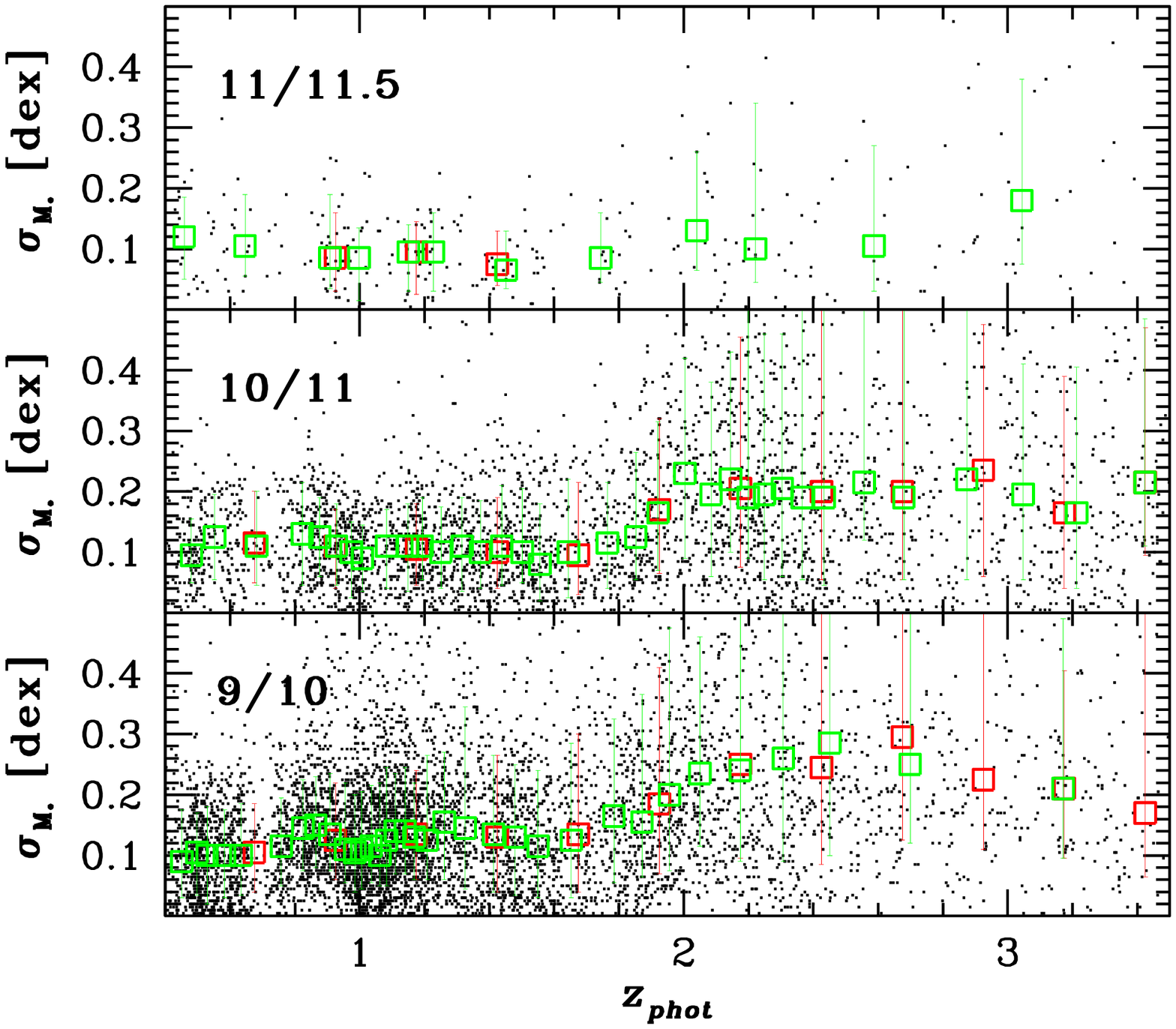}
  \caption{{\bf Left:}~Ratios between the total radio flux densities obtained from stacking at two different image resolutions, 1.5 and 6 arcsec. We note here that a Gaussian-convolved model is used to fit the higher resolution image to account for the physical sizes of sources, while a simple PSF fitting has been performed for the 6" resolution image.~{\bf Right:}~The stellar mass error (see text  for details) as a function of redshift and in three bins of stellar masses (indicated on the top-left of each panel) are plotted with black dots. There is an increase of the median error (green and red empty squares) with redshift, passing from $\sim$0.1~dex below redshift z=1.5 to $\sim$0.2~dex above redshift z=2, and with decreasing stellar mass.}
\label{radiomass}
\end{center}
\end{figure}

In this appendix we discuss the accuracy of the derived radio stacking results. Following Pannella et al.~(2009a), in order to derive the radio fluxes of the stacked images we fit a two-dimensional Gaussian, convolved with the image dirty beam, plus a background pedestal level using the {\it GALFIT} software (Peng et al.~2002). Allowing for a range of sizes, this kind of fitting naturally takes into account the physical sizes of the galaxies at different stellar masses and redshifts. One possible concern of such an approach could be a systematic bias introduced by the variable sizes, which might in principle, impact the results of our study, such as the constancy of the radio-FIR correlation. To test for possible biases, we used the VLA image at 6" resolution (G.E. Morrison, private communication), which has been produced with the same visibility dataset but with a different tapering of the data that gives more weight to short VLA baselines. The 6" resolution image is less deep than the natural 1.5" image and therefore not optimal for the science goals of this study. Still, we can fruitfully use the lower resolution image by assuming that flux densities of the stacking results in such image can be derived robustly by a simple PSF fitting, i.e., not allowing for any impact of the source sizes. We show the comparison in the left panel of Figure~ \ref{radiomass}. The results are in excellent agreement and usually within less than 10\% difference, that is a value close to the minimum statistical error dereived from our bootstrapping simulations.

\section{Testing galaxy stellar mass estimates with mock catalogs and a different SED-fitting code}
\label{appmass}

 In this appendix we try to assess the accuracy of our {\it FAST} derived stellar mass estimates by using both mock catalogs and the results from the SED-fitting
code described in detail in \citet{drorymass,drory09}. 

To test mass accuracies on mock catalogs we proceeded as follows. We have created a suite of 200 mock catalogs by randomly shifting the photometric points within the measured accuracies. We have then run FAST on each mock catalog and derived for each object a distribution of best fit stellar masses. In the right panel of Figure~\ref{radiomass}  we plot the stellar mass error as a function of redshift in three stellar mass bins. There is an increase of the median error with redshift, specifically passing from ~0.1 dex below redshift z=1.5 to ~0.2 above redshift z=2 which likely depend on the fact that going to hiher redshift the photometry gets more noisy in the optical rest-frame. Also worth noting is the increase of errors as stellar masses gets smaller, again likely driven by the global S/N decrease. This level of inaccuracy should be considered a good approximation to the real errors involved but still a lower limit as it does not factor in the discrepancies between different stellar population models.

Finally, we have also tested our FAST masses against the masses obtained by SED-fitting code described in \citet{drorymass,drory09}. This code has been already used in a number of published GOODS-\h~studies \citep{mullaney12,kirk12,penner12,magdis12,delmoro13}  making use of an earlier version of the photometric catalog in the GOODS-North field. For this reason a comparison between the two codes is useful both in terms of global accuracies but also to understand the possible impact on the published results. We parameterize the possible star
formation histories (SFHs) by a two-component model, consisting of a
main, smooth component, described by an exponentially declining star
formation rate ($\psi(t) \propto \exp(-t/\tau)$)s, linearly combined with
a secondary burst of star formation. The main component timescale
$\tau$~varies in the range 0.1--20~Gyr, and its metallicity is fixed to
solar. The age of the main component, $t$, is allowed to vary between
0.01~Gyr and the age of the Universe at the object's redshift.  The
secondary burst of star formation, which cannot contain more than 10\%
of the galaxy's total stellar mass, is modeled as a 100~Myr old
constant star formation rate episode of solar metallicity.  We adopt a
Salpeter~(1955) IMF for both components,
with lower and upper mass cutoffs of 0.1 and 100~\msun, respectively. Adopting the Calzetti et al.~(2000) attenuation law, both the main
component and the burst are allowed to exhibit a variable amount of
attenuation by dust with $A_V^m$~$\in [0,1.5]$~and~$A_V^b$~$\in$[0, 2] for
the main component and the burst, respectively. 

The results from the two runs with different codes and different SFHs, show no systematic differences in 
stellar mass estimates, except at the highest stellar masses ($\log M_* \ge 11.2$), where the two-component fitting predicts 
larger masses by 0.2~dex, with a scatter of about 0.2~dex.

\end{document}